\documentclass[aps,10pt,groupedaddress,prx,superscriptaddress,twocolumn,balancelastpage, tightenlines]{revtex4-2}

\usepackage{env} 
\usepackage{soul}
\usepackage{subcaption}
\usepackage{tocloft}
\usepackage{thmtools} 
\usepackage{thm-restate}
\usepackage[nolong,nosuper]{glossaries}
\usepackage{glossaries-prefix}
\makeglossaries
\glsdisablehyper

\usepackage{array}
\newcolumntype{C}[1]{>{\centering\arraybackslash}p{#1}}
\definecolor{lightgray}{gray}{0.85}

\theoremstyle{plain}  % Bold label, upright body (for definitions, examples)

\hypersetup{
    colorlinks=true,
    linkcolor=navy,
    citecolor=navy,
    urlcolor=navy,
    breaklinks=true,
}

\makeatletter
\renewcommand\@fnsymbol[1]{\ensuremath{\ifcase#1\or\dagger\or *\or\ddagger\or\mathsection\or\mathparagraph\else\fi}}
\makeatother

\usepackage{titletoc}
\newcommand\DoToC{%
  \startcontents
  \printcontents{}{1}{\textbf{Contents}\vskip3pt\hrule\vskip5pt}
  \vskip3pt\hrule\vskip5pt
}

\titlecontents{section}[2.9em]{\addvspace{4pt}\bfseries}
  {\hspace*{-2.9em}\makebox[2.9em][l]{\thecontentslabel}}
  {\hspace*{-2.9em}}
  {\mdseries\titlerule*[0.5pc]{.}\contentspage}

\begin{document}

\title{Hybrid Quantum Neural Networks: Theory, Implementations,
and Applications}

\author{Léo Monbroussou}
\affiliation{Terra Quantum AG, Kornhausstrasse 25, St.~Gallen, 9000, Switzerland} 
\affiliation{School of Informatics, University of Edinburgh, 10 Crichton Street, Edinburgh, EH8 9AB, United Kingdom}

\author{Maniraman Periyasamy}
\affiliation{Terra Quantum AG, Kornhausstrasse 25, St.~Gallen, 9000, Switzerland} 

\author{Viacheslav Kuzmin}
\affiliation{Terra Quantum AG, Kornhausstrasse 25, St.~Gallen, 9000, Switzerland} 

\author{Pavel Sekatski}
\affiliation{Terra Quantum AG, Kornhausstrasse 25, St.~Gallen, 9000, Switzerland}
\affiliation{Département de Physique Appliquée, Université de Genève, 24 Quai Ernest-Ansermet, Geneva, 1211, Switzerland}

\author{Viktoria Patapovich}
\affiliation{Terra Quantum AG, Kornhausstrasse 25, St.~Gallen, 9000, Switzerland}

\author{Asel Sagingalieva}
\affiliation{Terra Quantum AG, Kornhausstrasse 25, St.~Gallen, 9000, Switzerland}

\author{Alexey Melnikov}
\affiliation{Terra Quantum AG, Kornhausstrasse 25, St.~Gallen, 9000, Switzerland}

\begin{abstract}
    Artificial intelligence has been transformed by deep neural networks, yet the search for new learning architectures continues. Quantum machine learning offers one such direction, and hybrid quantum neural networks, which combine classical neural-network components with quantum information processing units, have emerged as a practical framework for near-term quantum technologies. However, the rapid development of the field across diverse architectures, benchmarks and hardware assumptions makes it difficult to assess the utility of various proposals, identify where genuine advantages may arise, and determine how practitioners can use these models. While recent benchmarks caution that such gains have not yet been demonstrated at scale, theoretical work has identified tasks on which quantum models hold provable advantages, and hybrid approaches have delivered promising results on practical problems using deliberately compact quantum components and substantially fewer trainable parameters. Here, we review hybrid quantum neural networks for the machine-learning and quantum-machine-learning communities. We summarize their main theoretical and methodological foundations, survey some of the most promising architectures developed so far, and examine their implementation challenges and reported performance. By consolidating these perspectives, this review provides a structured view of the state of the field and helps identify promising paths for future research and application-driven development.
\end{abstract}

\maketitle

% ##### start TOC stuff #####
\DoToC
% ##### end TOC stuff #####
\newacronym[shortplural=GMMs]{GMM}{GMM}{Gaussian Mixture Model}
\newacronym[shortplural=HMMs]{HMM}{HMM}{Hidden Markov Model}
\newacronym[shortplural=DNNs]{DNN}{DNN}{Deep Neural Network}
\newacronym[shortplural=SVDs]{SVD}{SVD}{Singular Value Decomposition}
\newacronym{MCTS}{MCTS}{Monte Carlo Tree Search}
\newacronym{MDP}{MDP}{Markov Decision Process}
\newacronym{CMDP}{CMDP}{Constrained Markov Decision Process}
\newacronym{RL}{RL}{Reinforcement Learning}
\newacronym[shortplural=DTs]{DT}{DT}{Decision Tree}
\newacronym{SMT}{SMT}{Satisfiability Modulo Theories}
\newacronym{IL}{IL}{Imitation Learning}
\newacronym[shortplural=CNNs]{CNN}{CNN}{Convolutional Neural Network}
\newacronym[shortplural=GANs]{GAN}{GAN}{Generative Adversarial Network}
\newacronym[shortplural=DQNs]{DQN}{DQN}{Deep Q-Network}
\newacronym{AI}{AI}{Artificial Intelligence}
\newacronym{PPO}{PPO}{Proximal Policy Optimization}
\newacronym{ML}{ML}{Machine Learning}
\newacronym{QML}{QML}{Quantum Machine Learning}
\newacronym{QNN}{QNN}{Quantum Neural Network}
\newacronym{NISQ}{NISQ}{Noisy Intermediate Scale Quantum}
\newacronym{QC}{QC}{Quantum Circuit}
\newacronym{VQC}{VQC}{Variational Quantum Circuit}
\newacronym{MNIST}{MNIST}{Modified National Institute Of Standards And Technology}
\newacronym{FIM}{FIM}{Fisher Information Matrix}
\newacronym{IDU}{IDU}{Incremental Data-Uploading}
\newacronym{DRU}{DRU}{Data Re-Uploading}
\newacronym{QRL}{QRL}{Quantum Reinforcement Learning}
\newacronym{VQA}{VQA}{Variational Quantum Algorithm}
\newacronym{SPSA}{SPSA}{Simultaneous Perturbation Stochastic Approximation}
\newacronym{SGD}{SGD}{Stochastic Gradient Descent}
\newacronym{QEM}{QEM}{Quantum Error Mitigation}
\newacronym{QEC}{QEC}{Quantum Error Correction}
\newacronym{VQ-DQN}{VQ-DQN}{Variational Quantum Deep Q-Networks}
\newacronym{BCQ}{BCQ}{Batch-Constraint Deep Q-Learning}
\newacronym{BCQQ}{BCQQ}{Batch-Constraint Quantum Q-Learning}
\newacronym{QPU}{QPU}{Quantum Processing Unit}
\newacronym{HQNN}{HQNN}{Hybrid Quantum Neural Network}
\newacronym{HQML}{HQML}{Hybrid Quantum Machine Learning}
\newacronym{PQS}{PQS}{Practical Quantum Software}
\newacronym[shortplural=MLPs]{MLP}{MLP}{Multi Layer Perceptron}
\newacronym[shortplural=FNOs]{FNO}{FNO}{Fourier Neural Operator}
\newacronym[shortplural=PDEs]{PDE}{PDE}{Partial Differential Equation}
\newacronym{RNN}{RNN}{Recurrent Neural Network}
\newacronym{LSTM}{LSTM}{Long Short-Term Memory}
\newacronym{PQN}{PQN}{Parallel Quantum Layers}
\newacronym{QDI}{QDI}{Quantum Depth Infused Layer}
\newacronym{PHN}{PHN}{Parallel Hybrid Network}
\newacronym{MAE}{MAE}{Mean Absolute Error}
\newacronym{MSE}{MSE}{Mean Squared Error}
\newacronym{FTQC}{FTQC}{Fault-Tolerant Quantum Computers}
\newacronym{POVM}{POVM}{Positive Operator-Valued Measure}
\newacronym[shortplural=BPs]{BP}{BP}{Barren Plateau}
\newacronym{DLA}{DLA}{Dynamical Lie Algebra}
\newacronym{QFIM}{QFIM}{Quantum Fisher Information Matrix}
\newacronym{DQFIM}{DQFIM}{Data Quantum Fisher Information Matrix}
\newacronym{LCU}{LCU}{Linear Combination Of Unitaries}
\newacronym[shortplural=QGANs]{QGAN}{QGAN}{Quantum Generative Adversarial Network}
\newacronym{CFD}{CFD}{Computational Fluid Dynamics}
\newacronym{IQP}{IQP}{Instantaneous Quantum Polynomial-Time}
\newacronym[shortplural=SIRENs]{SIREN}{SIREN}{Sinusoidal Representation Network}
\newacronym{RBS}{RBS}{Reconfigurable Beam Splitter}
\newacronym{FBS}{FBS}{Fermionic Beam Splitter}
\newacronym{JIT}{JIT}{Just-In-Time}
\newacronym{MNR}{MNR}{Metric--Noise--Resource}
\newacronym{GQML}{GQML}{Geometric Quantum Machine Learning}

\begin{figure*}[t!]
\centering
\input{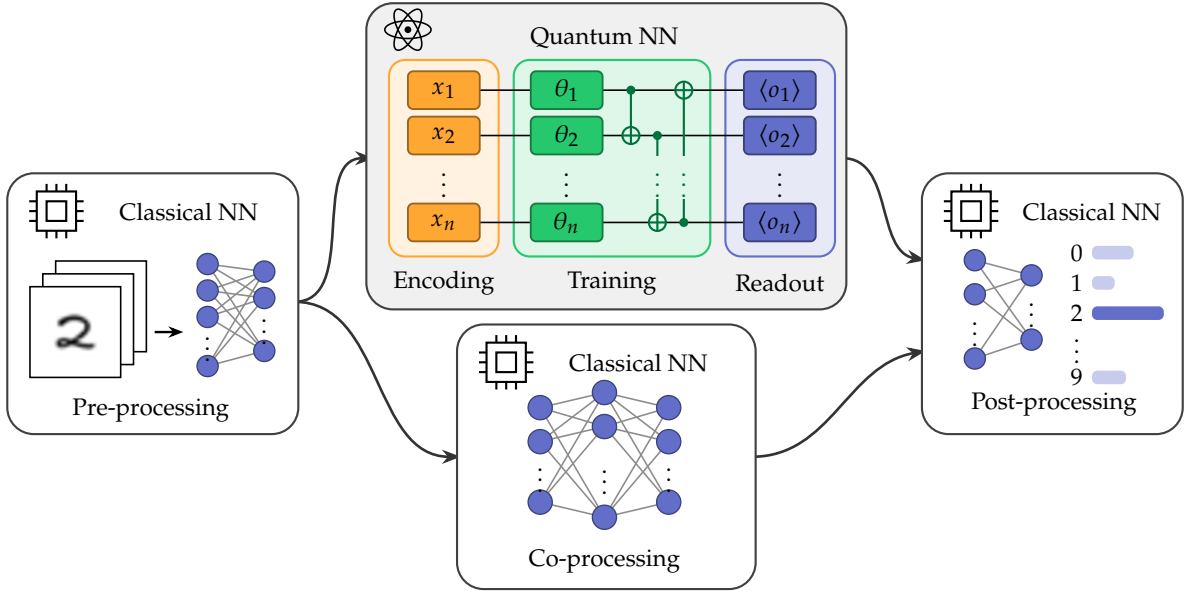}
\caption{Architecture example of a hybrid quantum neural network. A classical
computer handles data pre- and post-processing, parallel co-processing, and parameter optimization, while a quantum circuit (parameterized by trainable
angles) applies a unitary transformation in an exponentially large Hilbert space that, together with data encoding and measurement, realizes  a nonlinear input--output map which can be hard to reproduce classically.}
\label{fig:QNN}
\end{figure*}

\newpage

\begin{figure*}[t!]
    \centering
    \includegraphics[width=\linewidth]{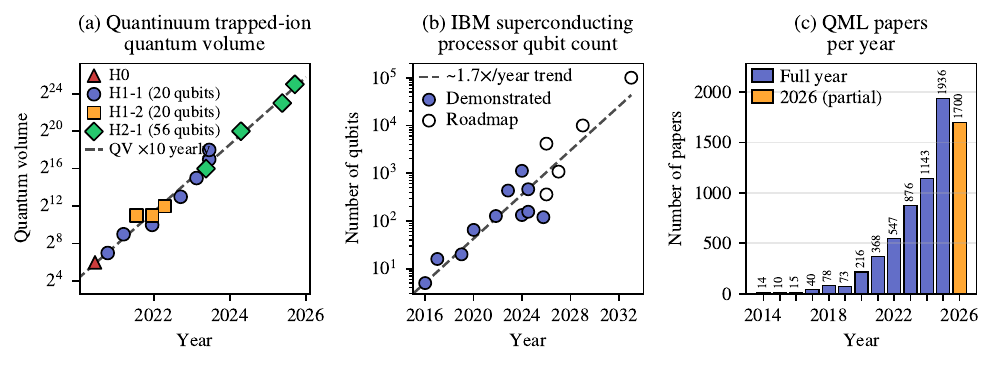}
    \caption{Exponential growth trends in quantum hardware and \gls{QML}
        research.
        (a)~Quantum Volume of Quantinuum's trapped-ion processors
        (H1 and H2 series)~\protect\cite{Quantinuum2025QV33M, Quantinuum2025Helios}.
        (b)~Physical qubit count of IBM's superconducting processors;
        filled symbols are realized devices, open symbols are published
        roadmap targets~\protect\cite{IBM2025Nighthawk, IBM2025roadmap,
        IBMQuantumEvolution2024}.
        (c)~Number of papers published per year containing the phrase
        ``Quantum Machine Learning'' in the title or abstract,
        as indexed by OpenAlex~\protect\cite{Priem2022OpenAlex} (covering arXiv
        preprints, journal articles, and conference proceedings).
        The orange bar denotes the most recent year (partial data from early July 2026).}
    \label{fig:hardware_plot_scaling}
\end{figure*}

\section{Introduction}
\label{sec:introduction}

\Gls{ML} has advanced across many areas of technology and research. \gls{ML} models such as LLMs have improved natural-language understanding and generation and are already used in healthcare~\cite{Ayers2023ChatGPTJAMA}, law~\cite{Liu2023LegalBench}, education~\cite{Carvallo2023ChatGPT}, finance~\cite{Cao2024QuantAI}, entertainment~\cite{Covington2016YouTube}, and scientific discovery~\cite{Jumper2021AlphaFold, yamada2025aiscientist}. Generative models, from DALL·E~2~\cite{Ramesh2022Dalle2} to text-to-video models such as Sora~\cite{openai2024sora}, are applied in art and design, and deep-learning perception systems are used in autonomous driving~\cite{Badue2021Survey}. Across these domains \gls{ML} has moved from a research tool to a widely deployed technology, from recommendation systems to industrial predictive analytics. In parallel, quantum computing is developing rapidly: by exploiting quantum-mechanical phenomena (superposition, entanglement, and interference), quantum algorithms can process information in fundamentally different ways from classical computing~\cite{Nielsen2011Quantum}, with algorithms such as Shor's~\cite{Shor1997Polynomial} and Grover's~\cite{grover1996fast} providing proven exponential and quadratic speedups for specific problem classes when the quantum hardware is mature enough \cite{Aaronson2015Complexity}. 
As quantum hardware~\cite{Preskill2018Quantum, arute2019quantum} and algorithms~\cite{Biamonte2017QML} continue to advance, this context has motivated researchers to explore how quantum information processing can enhance machine learning and vice versa; in this review we focus on \glspl{HQNN}: models that integrate classical neural network layers with quantum information processing units into a single trainable pipeline (\autoref{fig:QNN}).

The current success of \gls{ML} is underpinned by empirical scaling laws: model performance improves as a power law with model size, dataset size, and compute budget~\cite{kaplan2020scaling, nakkiran2021deep}. In the overparameterized regime where modern networks operate, larger models trained on more data consistently perform better. This trend is evident
in the evolution of neural networks, where state-of-the-art models now boast
billions of parameters and require substantial computational resources. For
comparison, OpenAI's GPT-1 had 0.12~billion parameters, GPT-2 expanded
to 1.5~billion, and GPT-3 grew to
175~billion~\cite{radford2018gpt1, radford2019gpt2, brown2020gpt3},
with subsequent models such as GPT-4 and GPT-5 continuing the
trend~\cite{openai2024gpt4, openai2025gpt5}. This rapid increase in model
complexity has proven effective but brings several challenges in the long
term. The escalating number of parameters dramatically increases the demand
for memory and computational power, raising the costs associated with training
and deploying these models. Moreover, the high demand for the required
hardware, primarily GPUs, has resulted in a shortage in the market, creating a significant barrier to development in the field.

This bottleneck prompts a critical question: is continuing down this path
sustainable? While classical algorithmic improvements, such as
sparsity~\cite{Cheng2024pruning}, quantization~\cite{Gholami2022quantization},
distillation~\cite{Hinton2015distilling}, and mixture-of-experts
architectures~\cite{Cai2024MoESurvey}, offer one direction for mitigating
these costs, an orthogonal and complementary approach lies in developing
fundamentally new information processing
paradigms~\cite{hernandez2020measuring}. Quantum algorithms emerge as a
particularly promising candidate in this respect, complementing
classical methods at two levels. First,
for certain classical problem classes, quantum routines can provide
provable computational speedups over the best-known classical
alternatives~\cite{Biamonte2017QML, liu2021rigorous}. Second, some tasks are intrinsically quantum-mechanical
and can be hard to address with classical machine learning
alone: learning properties of quantum
states~\cite{Huang2022Science}, simulating many-body quantum
systems, or recognizing quantum phases of
matter~\cite{cong2019quantum} all involve data that lives in an
exponentially large Hilbert space and can require processing that
preserves or exploits quantum coherence.

The hardware enabling this pursuit is maturing rapidly. Quantum
processors have exhibited exponential improvement in both quality and
scale over the past decade (\autoref{fig:hardware_plot_scaling}), a
trend sometimes framed as a ``Moore's Law'' for quantum
hardware~\cite{Neven2019law, Cho2021quantum, Google2024Willow}. However, early
demonstrations of quantum advantage~\cite{arute2019quantum,
kim2023evidence, King2025beyond, Google2025echoes} have so far been
vigorously contested by improved classical
algorithms~\cite{huang2020classical, tindall2306efficient,
Begusic2024fast, Mauron2025challenging}. Yet this ongoing interplay
has given rise to the era of \emph{quantum
utility}~\cite{kim2023evidence, IBM_utility}, in which quantum
processors deliver results beyond brute-force classical simulation.
This has motivated the search for a practical quantum advantage in
machine learning, with research interest in \gls{QML} growing in
step, as reflected in the annual publication volume
(\autoref{fig:hardware_plot_scaling}c).

Current devices nevertheless remain in the \gls{NISQ}
regime~\cite{Preskill2018Quantum}, with limited qubit counts and
residual gate noise. This constraint is precisely what motivates
hybrid quantum--classical architectures: rather than running
end-to-end quantum algorithms that presume fault-tolerant machines,
\glspl{HQNN} delegate the bulk of the computation to mature classical
networks and reserve the quantum component for the narrow subroutine
where it offers a principled advantage~\cite{mari2020transfer}.
Rigorous results establishing quantum advantages for specific learning
tasks~\cite{liu2021rigorous, Huang2022Science} (discussed in
\autoref{subsec:why_hybrid}) supply the theoretical grounding for
this division of labour; the remainder of this section lays out the
full set of motivations in detail.

\subsection{Why Hybrid? Motivations for Quantum-Classical Neural Networks}
\label{subsec:why_hybrid}

The case for \glspl{HQNN} rests not on a single argument but on a convergence
of motivations spanning theory, practice, and pragmatics. We organize them
from strongest to most speculative.

\paragraph{Processing quantum data.}
The most rigorous justification for quantum components in neural networks concerns learning from inherently quantum data. Huang et al.~\cite{Huang2022Science} proved that a quantum learner equipped with quantum memory can predict properties of physical systems from exponentially fewer experiments than any classical learner restricted to measuring each copy independently, a separation in sample complexity demonstrated experimentally on Google's Sycamore processor. Quantum convolutional neural networks were introduced to recognize quantum phases of matter directly from many-body states at low sampling cost and with robustness to gate noise~\cite{cong2019quantum}. Notably, on the datasets commonly used to benchmark them, such networks can be matched or outperformed by a purely classical surrogate~\cite{bermejo_quantum_2026}: under standard initialization their action is restricted to low-order (low-bodyness) measurements of the input, which are classically tractable, and the usual benchmarks turn out to be classifiable from exactly that information. The essential quantum resource is thus the acquisition of measurement data (e.g.\ classical shadows \cite{aaronson_shadow_2018}) from the physical system, after which classical post-processing suffices. This sharpens the hybrid case: the quantum device handles what is classically intractable (preparing and measuring the state), while conventional networks perform the learning. Such hybrid designs are a natural fit for tasks involving quantum experimental data, from quantum chemistry and materials science to quantum sensing.

\paragraph{Theoretical foundations for quantum advantage.}
Rigorous theoretical results establish that quantum advantage in machine learning is not merely hypothetical. Liu et al.~\cite{liu2021rigorous} constructed a classification task for which quantum kernel methods provably achieve an exponential speedup over any classical learner, assuming the hardness of the discrete logarithm problem. Huang et al.~\cite{huang2021power} showed that data availability can sometimes close the quantum--classical gap, but also identified settings where quantum models retain a provable prediction advantage. The common thread running through these results is that a provable separation arises whenever the learning task embeds a classically intractable problem, often one secured by a cryptographic hardness assumption, that a quantum model can evaluate efficiently while no classical learner can. Quantum advantage in this sense is therefore a property of the problem's structure rather than of quantum models in general: it is guaranteed only when the data-generating process carries the right kind of hidden structure. \autoref{sec:theory} develops this picture across the known separation results.

\paragraph{Fourier structure: a bridge between classical and quantum neural networks.} A key insight underpinning HQNNs is the connection between \gls{VQC} and Fourier analysis. Schuld et al.~\cite{Schuld_2021_Effect} showed that a broad class of data-encoding \gls{VQC}s naturally compute truncated Fourier series, whose accessible frequency spectrum is determined by the encoding gates (though, as discussed in \autoref{subsec:VQC_as_Fourier_Models}, this does not hold for every encoding strategy). The significance of this correspondence lies in the specific inductive bias it reveals: a spectral structure that gives direct control over the frequencies the model can represent. The same property underlies a family of classical architectures built to overcome the low-frequency (spectral) bias of standard networks and capture the fine-grained, high-frequency content they otherwise miss. \Glspl{SIREN}~\cite{sitzmann2020siren} use periodic activation functions to capture fine-grained signals, Fourier feature networks~\cite{tancik2020fourier} map inputs through random Fourier bases to overcome the spectral bias of standard \glspl{MLP}, and \glspl{FNO}~\cite{li2020fourier} learn \gls{PDE} solution operators in the frequency domain. All three approaches demonstrate that a Fourier-structured inductive bias is powerful for oscillatory, high-frequency, and physics-related functions. Crucially, this bias is shared by classical and quantum models alike: it is not in itself sufficient to distinguish the capacities of quantum and classical neural networks, since the classical architectures above realize the same spectral structure. Quantum circuits provide it natively, and the number of accessible Fourier frequencies can grow exponentially with the number of data-reuploading layers~\cite{Schuld_2021_Effect,
kordzanganeh2023exponentially, PerezSalinas2020datareuploading}.
However, the coefficients attached to these frequencies are correlated and constrained, so the achievable function class can be more limited than the raw spectrum count suggests~\cite{Schuld_2021_Effect, wiedmann2025fourier}. This shared structure motivates a hybrid division of labour: classical layers perform generic feature extraction, while a \gls{VQC} offers a compact, parameter-efficient representation of the periodic, high-frequency part of the target function. The advantage sought is therefore one of efficiency rather than capacity: an economical route to a spectral structure a classical network could also represent, but typically only with more parameters.

\paragraph{Complementary strengths and practical utility.}

Classical networks excel at processing large volumes of classical data, high-dimensional feature extraction, and benefit from decades of optimized infrastructure. Quantum circuits offer a fundamentally different computational primitive: operations in a Hilbert space whose dimension grows
exponentially with the number of qubits, parameterized by a number of trainable angles that scales only polynomially in qubit count and circuit depth. Rather than competing, these capabilities are complementary. In one common hybrid pipeline for classical data (\autoref{fig:QNN}), classical layers first compress high-dimensional raw data into a compact latent representation whose dimension matches the limited number of qubits available on current hardware; the quantum layer then maps this representation, via data encoding and a parameterized unitary, into the exponentially large Hilbert space spanned by those qubits; the measurement yields a nonlinear function of the input whose representational capacity would require a substantially wider or deeper classical network to
replicate. When the input is itself quantum data, the roles shift: the quantum device can perform the classically intractable state preparation and measurement (and may apply trainable quantum operations to the state before readout), while classical layers handle the remaining post-processing and learning. This particular allocation is not the only one: classical networks can handle scalable feature extraction while quantum circuits provide parameter-efficient transformations in structured function spaces, but the converse is equally natural: in quantum kernel methods and quantum reservoir computing, the quantum circuit itself serves as the feature extractor, embedding inputs into a classically intractable feature space from which a simple classical readout learns. What the hybrid setting offers is the freedom to assign each role where it is most effective.

\paragraph{Energy and resource efficiency prospects.}

As quantum hardware matures, mapping suitable computations to quantum processors may yield energy advantages. Different platforms (superconducting, photonic, and neutral-atom) offer distinct and still-evolving energy profiles, several of them favorable~\cite{auffeves2022quantum}. While current QPUs do not yet outperform GPUs in energy efficiency for typical ML workloads due to cryogenic overhead~\cite{jaschke2023quantum, fellous2023limitations}, the trajectory of hardware improvement suggests this balance may shift (see \autoref{subsubsec:Energy_Consumption} for a detailed analysis).

\paragraph{Current empirical benchmarks.}
These theoretical guarantees notwithstanding, the current empirical evidence remains mixed. Several systematic benchmarking studies paint a nuanced picture. Across large benchmark suites, out-of-the-box classical baselines often match or outperform quantum classifiers, with entanglement frequently neutral or even detrimental at small scales~\cite{bowles2024better}. Independent time-series benchmarks~\cite{fellner_quantum_2026} comparing variational quantum algorithms against classical baselines found that, on generic benchmarks, variational models do not yet surpass classical counterparts of comparable complexity. For generic classical data without special structure, the established provable advantages remain largely confined to specially structured tasks: the known exponential separations rely on data generated by cryptographic problems, such as the discrete-logarithm classification task of Liu et al.~\cite{liu2021rigorous}, whose hidden algebraic structure a quantum learner can exploit but no classical learner can decode efficiently. Whether comparable gains carry over to typical workloads, which lack such engineered structure, remains to be confirmed empirically. The benefit of the quantum component depends on architecture, task, and training methodology, and advantages emerge primarily when data structure aligns with quantum circuit inductive
biases: periodic or oscillatory targets~\cite{Schuld_2021_Effect}, quantum-native data~\cite{Huang2022Science}, and data-scarce regimes~\cite{Caro_2022_Generalization}. These empirical limitations mirror concrete theoretical obstacles: variational circuits face \gls{BP} and cost concentration that impede training(\autoref{subsubsec:BP_QNN}), while broad classes of structurally restricted circuits admit efficient classical surrogation (\autoref{subsec:Dequantization}). Together, these results delineate the conditions for regimes in which a quantum layer is at once trainable and classically hard. Proposals have been made to bypass such problems: symmetry-aware and geometric ansatz, subspace-preserving layers (\autoref{subsec:Subspace_Preserving}), and reservoir-style models (\autoref{subsubsec:Quantum_Reservoir_Computing}). Whether these strategies can extend a quantum advantage beyond structured and quantum-native problems to broad impact on classical data remains genuinely open, and is the question this review is organized around.

\subsection{Scope and Structure of This Review}
\label{subsec:scope}

This review provides a comprehensive treatment of \glspl{HQNN}: the underlying theory, architectures, implementations, and empirical evidence. It covers both settings where quantum components have proven theoretical advantages and those where the evidence is primarily empirical. The sections are organized from theoretical foundations to empirical validation, so the reader can build the mathematical and architectural framework before assessing where hybrid models deliver measurable benefit. By presenting the main challenges and the most promising new architectures developed to address them, it is hoped that clear directions can be offered for future practitioners in the field.

Readers new to \gls{QML} may wish to start with
\autoref{sec:foundations} for background on classical and quantum neural network layers. Theoretical aspects for hybridization can be found in \autoref{sec:theory}. Practitioners focused on application evidence can find in \autoref{sec:empirical} consolidated results across domains reported from the literature. Those interested primarily in the theoretical case for hybridization should focus on \autoref{sec:theory}.

\section{Foundations of HQNNs}
\label{sec:foundations}
    \subsection{Classical Neural Network}

Classical neural networks serve as the fundamental building blocks of modern machine learning, providing the structural framework upon which hybrid architectures are constructed. At its core, a classical neural network layer transforms an input vector $x \in \mathbb{R}^{d_{in}}$ into an output representation $y \in \mathbb{R}^{d_{out}}$ through an affine transformation followed by a non-linear activation function. Mathematically, this is expressed as $y = \sigma(Wx + b)$, where $W \in \mathbb{R}^{d_{out} \times d_{in}}$ represents the weight matrix, $b \in \mathbb{R}^{d_{out}}$ is the bias vector, and $\sigma$ denotes a non-linear mapping such as the Rectified Linear Unit (ReLU), sigmoid, or hyperbolic tangent functions~\cite{goodfellow2016deep}. 
%Tensor representation of the linear layer is illustrated in \autoref{fig:classical-vs-quantum-layer}(a). 
The primary objective of these layers is to extract increasingly "abstract" features from raw data, a process motivated by the Universal Approximation Theorem \cite{hornik1989multilayer}, which posits that a sufficiently deep or wide network of such layers can approximate any continuous function to arbitrary precision. The diversity of classical layers, discussed below, allows for specialized processing of different data modalities, which is a critical consideration when designing the classical components of an HQNN. Fully connected (or linear) layers facilitate global information exchange between neurons, making them suitable for general-purpose feature processing. In contrast, \glspl{CNN} utilize localized kernels to exploit spatial correlations and translation equivariance in imagery, significantly reducing parameter counts compared to dense architectures~\cite{lecun1998mnist, he2016resnet}. For sequential data, recurrent layers such as \gls{LSTM} cells incorporate internal gating mechanisms to maintain temporal context and mitigate the vanishing gradient problem, which is a common challenge in training deep models~\cite{hochreiter1997lstm}. These specialized layers provide the inductive biases necessary for efficient learning in specific domains like computer vision and time-series forecasting.

Training these layers involves the iterative optimization of the parameters $W, b$ to minimize a task-specific loss function. This is typically achieved through gradient-based optimization algorithms, most notably \gls{SGD} and its variants like Adam, which rely on the backpropagation algorithm to efficiently compute gradients via the chain rule~\cite{goodfellow2016deep}. In the context of HQNNs, these classical layers are often integrated to pre-process the data and reduce its dimensionality before they are encoded into quantum states, or as post-processing units to map quantum measurement outcomes to final class probabilities or regression targets (see \autoref{fig:QNN}). This synergy allows the model to leverage the mature infrastructure of classical deep learning while strategically deploying quantum layers to explore complex feature spaces that remain computationally expensive for purely classical systems~\cite{Schuld_2021_Effect, Benedetti_2019_Parameterized}.

    \subsection{Quantum Neural Network}
\label{sec: Quantum Neural_Network}

QNNs are a particular application of variational algorithms \cite{cerezo_variational_2021}, which leverage the toolbox of classical optimization. The different elements of this toolbox can be categorized into several groups. In a standard supervised learning task, a parameterized function $f$, called a \emph{model}, is optimized to match the targets for a finite dataset. \emph{Quantum models} on $n$ qubits are defined as the family of parameterized functions $f : \mathcal{X} \times \Theta \rightarrow \mathbb{R}$ obtained by encoding the data into the qubits and subsequently estimating the expectation value of some Hermitian observable $O$ via repeated measurement, formally 
 \begin{equation}\label{eq:quantum_Model}
    f(x,\theta) = \langle O\rangle = \bra{0} U(x,\theta)^\dagger O U(x,\theta) \ket{0} 
 \end{equation}
where $U(x,\theta)$ is a $2^n$-dimensional unitary, $\theta \in \Theta$ is the vector of trainable parameters and $x= (x_1,\dots,x_D) \in \mathcal{X} \subset \mathbb{R}^D$ is the classical data vector, see \autoref{fig:QNN}. More generally, one measures several observables to obtain a vector valued quantum model $f(x,\theta) = \left(\langle O_1\rangle,\dots, \langle O_M\rangle \right)$. 

Formally, general numerical simulation of QNN relies on tensor multiplications with dimensions $N=2^n$ that grows exponentially with the number of qubits $n$, and becomes intractable for classical computation for large number of qubits, whereas on quantum hardware,  $f(x,\theta)$ is found as an output of the physical process involving these $n$ qubits.

        \subsubsection{Quantum states}\label{sec:quantum-states}

The state of an isolated quantum system that can be in $d$ different, mutually exclusive states, can be described by a unit vector of size $d$, noted $\ket{\psi}$ in a complex Hilbert space $\mathcal{H}$ of dimension $d$, that is, a complete inner-product space over $\mathbb{C}$. The inner product $\braket{\varphi}{\psi} \in \mathbb{C}$ induces the norm $\lVert \psi \rVert = \sqrt{\braket{\psi}{\psi}}$. The elementary building block of quantum computation is the \emph{qubit}, a two-level system whose Hilbert space is $\mathcal{H}_1 = \mathbb{C}^2$. Fixing the computational basis $\{\ket{0},\ket{1}\}$, any pure single-qubit state reads
\begin{equation}
    \ket{\psi} = \alpha \ket{0} + \beta \ket{1}, \qquad \alpha,\beta \in \mathbb{C}, \quad |\alpha|^2 + |\beta|^2 = 1,
    \label{eq:single-qubit}
\end{equation}
By contrast with a classical bit, the coefficients $\alpha,\beta$ allow for arbitrary coherent superpositions of the two basis states. The Hilbert space of a composite system is obtained as the \emph{tensor product} of the Hilbert spaces of its subsystems. For an $n$-qubit register, $\mathcal{H} = \mathcal{H}_1^{\otimes n}$ with computational basis $\{\ket{x_1 x_2 \cdots x_n} \mid x_i \in \{0,1\}\}$, a shorthand for the product states $\ket{x_1}\otimes\ket{x_2}\otimes \cdots \otimes\ket{x_n}$. The dimension of $\mathcal{H}$ grows \emph{exponentially} with the number of qubits, and a generic state is a superposition of product states
\begin{equation}\label{eq: state psi}
    \ket{\psi} = \sum_{x \in \{0,1\}^n} c_x \ket{x}, \qquad \sum_x |c_x|^2 = 1,
\end{equation}
specified by $2^n$ complex amplitudes, where states equal up to a global phase are physically equivalent. This exponential state space is the resource that quantum machine learning seeks to exploit. Crucially, general vectors in $\mathcal{H}$ cannot be written as a tensor product $\ket{\psi_1}\otimes\cdots\otimes\ket{\psi_n}$ of single-qubit states; such non-factorisable (pure) states are called \emph{entangled} \cite{dewolf2023quantumcomputinglecturenotes,horodecki2009quantum}.  Entanglement is often viewed are a necessary condition for quantum models to offer an advantage over classical counterparts. Notice however, that it is also known that entanglement is not sufficient (see \autoref{subsec:Dequantization}).

When the system is not perfectly isolated, or when only a subsystem of a larger composite system is accessible, its state is described more generally by a density operator $\rho$ satisfying $\rho = \rho^{\dagger}$, $\rho \succeq 0$, and $\Tr \rho = 1$. Pure states correspond to rank-one projectors $\rho = \ket{\psi}\bra{\psi}$, while $\Tr(\rho^2) < 1$ signals a statistical mixture. The reduced state of a subsystem $A$ of a bipartite system $AB$ is recovered through the partial trace, considering a measurement operation as described in \autoref{subsubsec:Output_Readout}.

\subsubsection{Gates, operations, and unitaries}
        \label{subsubsec:gates-unitaries}

Between measurements, a closed quantum system evolves according to the Schr\"odinger equation,
\begin{equation}
    i \, \frac{\mathrm{d}}{\mathrm{d}t} \ket{\psi(t)} = H(t)\, \ket{\psi(t)},
    \label{eq:schrodinger}
\end{equation}
where the Hamiltonian $H(t)$ is a Hermitian operator on $\mathcal{H}$. Integrating \autoref{eq:schrodinger} yields the propagator
\begin{equation}
    U = \mathcal{T}\exp\!\left(- i\int_0^t H(s)\,\mathrm{d}s\right),
\end{equation}
which reduces to $U= \exp(-iHt)$ for time-independent $H$. In the following we will absorb the time $t$ in the Hamiltonian, and simply write $U= \exp(-iH)$.  Because $H = H^{\dagger}$ is Hermitian, the propagator is unitary $U^{\dagger}U = U U^{\dagger} = \mathbb{I}$, which ensures that the norm $\braket{\psi}{\psi}$, and hence the total probability, is preserved by the evolution \cite{Nielsen2011Quantum,dewolf2023quantumcomputinglecturenotes}. A unitary that acts on a small number of qubits is called a \emph{quantum gate}. For example, the single-qubit Pauli operators are given by
\begin{equation}\label{eq: pauli}
    X = \begin{pmatrix} 0 & 1 \\ 1 & 0 \end{pmatrix}, \quad
    Y = \begin{pmatrix} 0 & -i \\ i & 0 \end{pmatrix}, \quad
    Z = \begin{pmatrix} 1 & 0 \\ 0 & -1 \end{pmatrix},
\end{equation}
and are both unitary and Hermitian, and can thus be used both as gates and Hamiltonians.

A set of gates is said to be \emph{universal}, if any unitary on $n$ qubits can be approximated to arbitrary precision by a finite product of these elements. A \emph{quantum circuit} is a sequence of quantum gates applied in a specific order. \emph{Parametrized gates} commonly defined as $e^{-i\theta H}$ make the link with \autoref{eq:schrodinger}, and are at the core of \gls{VQC} used in HQNNs. While, some quantum processing units offer universal sets of single and two qubit gates, others allow the use of global parametric Hamiltonians acting on the entire system \cite{bluvstein_fault-tolerant_2026}.

        \subsubsection{Input Encoding}\label{subsubsec:Encoding_Classical_Information}

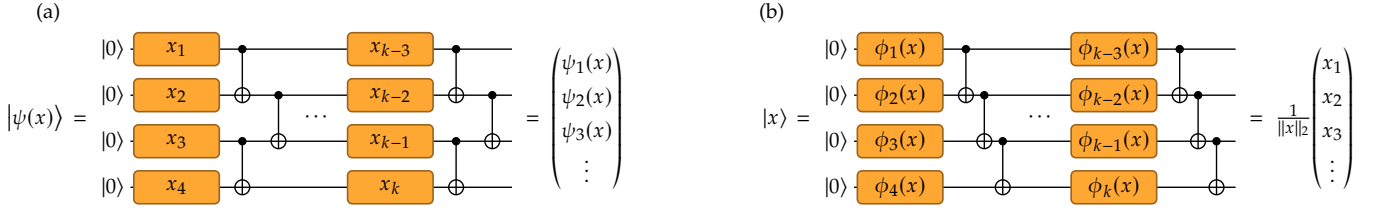
\begin{figure*}[!t]
    \centering
    % Fig 4: angle encoding (a) vs amplitude encoding (b), drawn as |state> = circuit
% acting on |0...0>. Both panels share the SAME circuit ansatz; (a) feeds the raw
% features x_i directly into the gates (angle encoding), while (b) feeds precomputed
% angles phi_i(x) that load x into the amplitudes (amplitude encoding).
% Two full circuits side by side, scaled to \linewidth.
\begingroup
% Palette from Fig 8: orange = encoding/data gates
\definecolor{encFill}{RGB}{255,167,51}
\tikzset{
  qwire/.style={line width=0.6pt},
  bgate/.style={draw=encFill!60!black, line width=0.7pt, fill=encFill, rounded corners=2pt,
                minimum width=1.3cm, minimum height=0.5cm, inner sep=1.5pt},
}
\resizebox{\linewidth}{!}{%
\begin{tikzpicture}[x=1cm,y=1cm]
% ============================== (a) angle encoding ==============================
\begin{scope}[xshift=0cm]
\node[font=\rmfamily] at (0.0, 0.55) {(a)};
\node[anchor=east] at (0.7,-1.05) {$\ket{\psi(x)}\,=$};
\foreach \i in {0,...,3}{
  \node at (0.98, -0.7*\i) {$\ket{0}$};
  \draw[qwire] (1.25, -0.7*\i) -- (7.05, -0.7*\i);
}
\node at (4.1,-1.05) {$\cdots$};
\foreach \i/\y in {1/0, 2/-0.7, 3/-1.4, 4/-2.1}{
  \node[bgate] at (1.95, \y) {$x_{\i}$};
}
\foreach \x/\ya/\yb in {2.95/0/-0.7, 2.95/-1.4/-2.1, 3.5/-0.7/-1.4}{
  \filldraw (\x,\ya) circle (1.6pt);
  \draw[qwire] (\x,\ya) -- (\x,\yb);
  \draw[qwire, fill=white] (\x,\yb) circle (3.2pt);
  \draw[qwire] (\x,{\yb-0.113}) -- (\x,{\yb+0.113});
  \draw[qwire] ({\x-0.113},\yb) -- ({\x+0.113},\yb);
}
\foreach \i/\y in {k-3/0, k-2/-0.7, k-1/-1.4, k/-2.1}{
  \node[bgate] at (5.2, \y) {$x_{\i}$};
}
\foreach \x/\ya/\yb in {6.2/0/-0.7, 6.2/-1.4/-2.1, 6.75/-0.7/-1.4}{
  \filldraw (\x,\ya) circle (1.6pt);
  \draw[qwire] (\x,\ya) -- (\x,\yb);
  \draw[qwire, fill=white] (\x,\yb) circle (3.2pt);
  \draw[qwire] (\x,{\yb-0.113}) -- (\x,{\yb+0.113});
  \draw[qwire] ({\x-0.113},\yb) -- ({\x+0.113},\yb);
}
\node[anchor=west] at (7.15,-1.05)
  {$=\;\begin{pmatrix} \psi_1(x)\\ \psi_2(x)\\ \psi_3(x)\\ \vdots \end{pmatrix}$};
\end{scope}
% ============================ (b) amplitude encoding ============================
\begin{scope}[xshift=11cm]
\node[font=\rmfamily] at (0.0, 0.55) {(b)};
\node[anchor=east] at (0.7,-1.05) {$\ket{x}\,=$};
\foreach \i in {0,...,3}{
  \node at (0.98, -0.7*\i) {$\ket{0}$};
  \draw[qwire] (1.25, -0.7*\i) -- (7.05, -0.7*\i);
}
\node at (4.1,-1.05) {$\cdots$};
\foreach \i/\y in {1/0, 2/-0.7, 3/-1.4, 4/-2.1}{
  \node[bgate] at (1.95, \y) {$\phi_{\i}(x)$};
}
\foreach \x/\ya/\yb in {2.95/0/-0.7, 3.23/-0.7/-1.4, 3.51/-1.4/-2.1}{
  \filldraw (\x,\ya) circle (1.6pt);
  \draw[qwire] (\x,\ya) -- (\x,\yb);
  \draw[qwire, fill=white] (\x,\yb) circle (3.2pt);
  \draw[qwire] (\x,{\yb-0.113}) -- (\x,{\yb+0.113});
  \draw[qwire] ({\x-0.113},\yb) -- ({\x+0.113},\yb);
}
\foreach \i/\y in {k-3/0, k-2/-0.7, k-1/-1.4, k/-2.1}{
  \node[bgate] at (5.2, \y) {$\phi_{\i}(x)$};
}
\foreach \x/\ya/\yb in {6.2/0/-0.7, 6.48/-0.7/-1.4, 6.76/-1.4/-2.1}{
  \filldraw (\x,\ya) circle (1.6pt);
  \draw[qwire] (\x,\ya) -- (\x,\yb);
  \draw[qwire, fill=white] (\x,\yb) circle (3.2pt);
  \draw[qwire] (\x,{\yb-0.113}) -- (\x,{\yb+0.113});
  \draw[qwire] ({\x-0.113},\yb) -- ({\x+0.113},\yb);
}
\node[anchor=west] at (7.15,-1.05)
  {$=\;\frac{1}{\lVert x\rVert_2}\!\begin{pmatrix} x_1\\ x_2\\ x_3\\ \vdots \end{pmatrix}$};
\end{scope}
\end{tikzpicture}%
}%
\endgroup
    \caption{The trade-off between the two main data-encoding strategies, illustrated on $n=4$ qubits. (a) Angle encoding: 
    parameters are trivially encoded into angles of a circuit, populating all $2^n$ basis states with amplitudes that are nonlinear (trigonometric) products of the data. (b) Amplitude encoding: circuit is derived numerically or analytically to reproduce a known state vector $\ket{x} = \sum_{i} \frac{x_i}{||x||_2} \ket{e_i}$ such that the data sits directly in the amplitudes.}
    \label{fig:encoding}
\end{figure*}

The encoding part of a QNN maps classical input data $x \in \mathbb{R}^d$ into a quantum state via a VQC.  A natural approach called \emph{Hamiltonian encoding} consists of evolving the system with an Hamiltonian parametrized with the classical data features $U(x)=e^{- i H(x)}$. A commonly used example~\cite{Schuld2021_QML_book} is the \emph{angle encoding}, where a single feature is encoded in the quantum system via a parametric gate $e^{-i x H}$, and the  Hamiltonian is usually chosen to be one of the Pauli operators in \autoref{eq: pauli}, see \autoref{fig:encoding}(a). 
One can choose how such encoding gates for different features are placed within the quantum circuit: they can be applied on different qubits and interjected with fixed or trainable quantum gates parametrized by $\theta$. In particular, data re-uploading \cite{PerezSalinas2020datareuploading, Schuld_2021_Effect} consist of placing several encoding gates parametrized by the same feature insider the circuit and can be used to enrich the model feature map (see \autoref{subsec:VQC_as_Fourier_Models}); while a large number of features can be injected into a fixed number of qubits by interjecting the data encoding layers for successive features with trainable entangling layers \cite{sagingalieva2023hybrid, lusnig2024hybrid, sagingalieva2025photovoltaic, periyasamy2022incremental}; the two ideas can also be combined~\cite{periyasamy2023bcqq}.

Alternatively one can approach the input encoding from top-to-bottom, as a global mapping between the feature space and the Hilbert space. Here, the canonical example is the \emph{amplitude encoding}, \autoref{fig:encoding}(b), where the data is represented in the amplitudes of a quantum state:
\begin{equation}\label{eq:amplitude_encoding}
    \forall x \in \mathbb{R}^d\, \quad |x\rangle := \sum_{i=1}^d
    \frac{x_i}{||x||_2} \ket{e_i} \, ,
\end{equation}
with $\{ e_i \}_{i=1}^d$ an orthonormal basis of states, such that the QNN model in \autoref{eq:quantum_Model} reads as
 \begin{equation}
    f(x,\theta) = \langle x| U(\theta)^\dagger O U(\theta) |x\rangle. 
 \end{equation}
 Such encoding allows for a more linear-algebraic approach, where the input vector $x$ is directly multiplied by a trainable unitary matrix $U(\theta)$.
 Amplitude encoding can accommodate up to a number of features that scales exponentially with the number of qubits~\cite{long_efficient_2001, plesch_quantum-state_2011}, but requires sophisticated circuits with the number of single and two-qubit gates scaling together with the number of features. 
Thus, while amplitude encoding appears as a building block in various long-term \gls{QML} algorithms, such as the quantum linear system solver procedure \cite{harrow_quantum_2009, morales_quantum_2025}, its practical implementation is very challenging~\cite{Aaronson2015Complexity}. Amplitude encoding on a polynomial subspace as a near-term approach has been particularly developed recently on Hamming-weight preserving quantum circuits \cite{farias_quantum_2025, johri_nearest_2021, cherrat_quantum_2024,
monbroussou_trainability_2025}, see also \autoref{subsec:Subspace_Preserving}. Moreover, it was noticed that natural data can be well compressed by tensor networks~\cite{dilip2022data} that can be then efficiently encoded into quantum amplitudes using tensor-network circuits~\cite{iaconis2023tensor}. In addition, one can represent the classical data directly by constructing an Hamiltonian evolution based on the classical data $e^{-iH(x)}$. This approach is native for several hardware, including adiabatic quantum computation \cite{albash_adiabatic_2018} or neutral atoms \cite{henriet_quantum_2020}.

\subsubsection{Output Readout}\label{subsubsec:Output_Readout}

Classical description of quantum states is not directly accessible, and no copy of an arbitrary unknown state can be made due to the no cloning theorem. Retrieving classical information from the quantum realm can be done through repetitive probabilistic \emph{measurements}, each of which leading to the
collapse of the quantum state. A common measurement used in QML is described by a \gls{POVM}, which
is a set of positive semi-definite observables $\{ E_i \}$ that sum to the identity operator $\sum_i E_i =I$.  Each element corresponds to an outcome, and the probability of observing it for a state $\rho$ is given by $\Pr[i] = \Tr(E_i \rho)$.  The most common example is the computational basis measurement. It is composed of the projectors on all states in the computational basis $\{E_x = \ketbra{x}\}$. For this measuremtn and a general pure state $\ket{\psi }$ in Eq.~\eqref{eq: state psi}, the probability to observe an outcome $x$ is given by 
\begin{equation}
    \Pr[x] = \Tr(E_x \ketbra{\psi}) = |\braket{x}{\psi}|^2 = |c_x|^2.
\end{equation}
Note that the most general way to extract classical information from a quantum system is provided by quantum instruments~\cite{Nielsen2011Quantum}, which output both a classical outcome $i$ and the post-measurement quantum state. However, quantum instruments are rarely considered in the QML literature.

The expected value $\langle O \rangle$ of any observable, output by the quantum model in \autoref{eq:quantum_Model}, can be estimated by combining the results of one measurement $\langle O \rangle = \sum_i o_i \Tr( E_i \rho)$, or several measurements $\{ E_{i|1} \},\dots,\{ E_{i|j} \},\dots$ performed with some probability $p_j$ 
\begin{equation}
\langle O \rangle = \sum_{i,j} p_j \, o_{ij} \Tr( E_{i|j} \rho).
\end{equation}
While the estimation of the expected values of several observables can often be done simultaneously, this procedure requires to repeat the same input encoding and output measurements. We will further discuss this {\it sampling cost} in \autoref{subsec:trainability_QNN}.

Consider a state $\rho$ defined on a bipartite Hilbert space $\mathcal{H} = \mathcal{H}_A \otimes \mathcal{H}_B$. Measuring subsystem $A$ with a POVM $\{E_i\}$ (with $E_i \geq 0$ and $\sum_i E_i = I_A$) returns outcome $i$ with probability $p_i = \Tr\!\left[(E_i \otimes I_B)\,\rho\right]$ and leaves subsystem $B$ in the conditional state:
\begin{equation}
    \rho_{B|i} = \frac{1}{p_i}\,\Tr_A\!\left[ (E_i \otimes I_B)\,\rho \right] \; ,
\end{equation}
with $\Tr_A$ the partial trace over $\mathcal{H}_A$.

Extracting properties of complex, large-scale quantum systems is an important question in quantum computing. Indeed, the number of parameters needed to describe a quantum system scales exponentially with the number of its constituents. These parameters cannot be accessed directly, but must
be estimated by measuring the system. To avoid the use of many samples to predict a quantum state, techniques have been developed including
classical shadow \cite{aaronson_shadow_2018, huang_predicting_2020, thomas_shedding_2025}.

\subsubsection{Training}

A classical optimizer can be employed to iteratively tune the parameters of the quantum circuit in order to minimize the cost function and thus find the desired optimal solution. Classical gradient based techniques can be used to train a \gls{VQC}, but computing the gradient is not a trivial task. \textbf{Parameter Shift rule} is a technique developed to provide an unbiased estimation of the derivatives of a quantum model, by evaluating the function on a set of distant points. Initially developed for quantum gates based on an Hamiltonian with two unique eigenvalues \cite{Mitarai_2018_QClearning, Schuld_2019_ParamShift}, it has been generalized for multi-parametrized gate \cite{wierichs_general_2022}.

For example, consider an Hermitian matrix $H$ with two distinct eigenvalues $\pm \lambda$, corresponding to a parametrized quantum gate $G(\theta) = e^{-i\theta H}$. It has been shown \cite{Schuld_2019_ParamShift} that, using the Fourier representation of the quantum model (see \autoref{subsec:VQC_as_Fourier_Models}), the following property holds for a quantum model as defined in \autoref{eq:quantum_Model}:
\begin{equation}
    \partial_{\theta} f = \lambda \left(f(\theta + \frac{\pi}{4 \lambda}) - f(\theta - \frac{\pi}{4 \lambda})\right) \, .  
\end{equation}

For \gls{VQC} that are not based on parametric single-qubit gates, such as variational photonic quantum ones, method similar to the parameter-shift rule have been recently proposed and tested experimentally \cite{hoch_variational_2025, facelli_exact_2024}.

The parameter-shift rule yields exact gradients, but its cost scales linearly with the number of parameters $p$: each partial derivative requires two dedicated circuit evaluations, i.e.\ $2p$ evaluations per optimization step. 

A popular gradient-free alternative is the \gls{SPSA} \cite{spall1992multivariate, spall1998implementation}, a zeroth-order method that relies only on evaluations of the cost function. At each iteration all $p$ parameters are perturbed \emph{simultaneously} along a single random direction, typically chosen from the discrete ensemble $\Delta_k\in\{-1,1\}^p$. Two evaluations of the function $f(\theta_k \pm c_k \Delta_k)$, with $c_k$ the perturbation scheduled at step $k$, yield a stochastic estimate of the gradient that is used to update the parameters. The cost per step is therefore two circuit evaluations, \emph{independently} of the number of parameters $p$. Each estimate is very noisy but nearly unbiased, and the iteration implicitly averages out zero-mean fluctuations, including the shot noise of quantum measurements, so that convergence to a local minimum is guaranteed under mild conditions \cite{spall1992multivariate}. This built-in noise averaging makes SPSA particularly attractive for training variational circuits on noisy hardware \cite{Kandala2017HardwareefficientVQ}. Its behavior is governed by a few hyperparameters controlling the perturbation and step-size sequences; these must be tuned to the problem and can have a decisive impact on convergence \cite{spall1998implementation, bonetmonroig2023performance}. In the QML setting, systematic benchmarks have compared SPSA and SPSA-based gradient estimates against parameter-shift and other optimizers on variational circuits \cite{ wiedmann2023empirical}.

\subsection{Hybrid Quantum Neural Network} 

\glspl{HQNN} represent a synergistic integration of classical deep learning modules and parameterized quantum circuits within a unified software stack~\cite{Mitarai_2018_QClearning, mari2020transfer, Chen2020HybridNN, meyer2024qiskit}. While classical neural networks serve as the fundamental building blocks of modern machine learning, HQNNs also use quantum layers, e.g., as illustrated in \autoref{fig:QNN}, to combine complementary capabilities of both paradigms. By integrating these classical modules as pre-processing units, HQNNs can efficiently reduce the dimensionality of high-dimensional inputs before they are encoded into quantum states~\cite{mari2020transfer}. Subsequently, a quantum layer acts as a highly expressive transformation, exploring complex feature spaces that remain computationally expensive for purely classical systems~\cite{Schuld_2021_Effect, Benedetti_2019_Parameterized}. Finally, classical post-processing units map the resulting quantum measurement outcomes to final class probabilities or regression targets~\cite{Chen2020HybridNN, Sharma_2020_Noiseresilience}. Mathematically, this hybrid pipeline maps an input vector $x \in \mathbb{R}^{d_{in}}$ to an output prediction $y \in \mathbb{R}^{d_{out}}$ through a sequence of alternating classical and quantum functions. These hybrid models leverage the mature infrastructure of classical optimization. During the training phase, the HQNN undergoes iterative parameter updates to minimize a task-specific loss function. This is typically achieved through gradient-based optimization algorithms. Classical parameters are updated via standard backpropagation techniques using the chain rule, while quantum gradients are evaluated using hardware-compatible analytic methods, such as the parameter-shift rule~\cite{Schuld_2019_ParamShift, Mitarai_2018_QClearning, periyasamy2024guided}. This cohesive architecture ensures that the distinct inductive biases of both the classical components and the quantum circuits, such as their Fourier spectral structure~\cite{Schuld_2021_Effect, wiedmann2025fourier}, unitarity constraints, and entanglement, are jointly optimized.

    \section{Theoretical Aspects of HQNN}
    \label{sec:theory}

In this Section, we present some important theoretical aspects that help to understand and design correctly \glspl{HQNN}. In particular, one must pay a particular attention to the capacity of training, the expressivity figure of merits and the potential de-quantization techniques that can affect a new model \cite{monbroussou_quantum_2025}.

        \subsection{Variational Models as Fourier
        Series}\label{subsec:VQC_as_Fourier_Models}

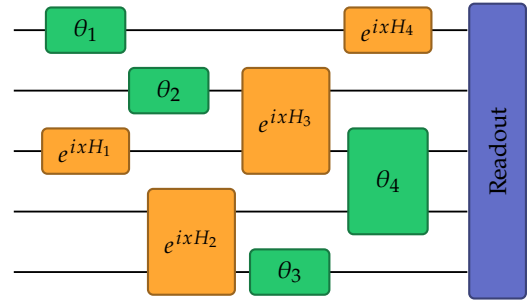
\begin{figure}[h!]
    \centering
    % Fig 5: QNN as truncated Fourier series. Green = trainable blocks -> Fourier
% coefficients; orange = data-encoding blocks -> frequency spectrum. Formula
% colors are saturated so the correspondence is visible (Pavel's comment).
% Drawn at natural size (no resizebox) so all text matches the body font.
\begingroup
% Palette from Fig 8: green = trainable, orange = encoding, violet = readout
\definecolor{trainFill}{RGB}{38,200,110}
\definecolor{encFill}{RGB}{255,167,51}
\definecolor{readFill}{RGB}{99,110,200}
\tikzset{
  qwire/.style={line width=0.7pt},
  tgate/.style={draw=trainFill!60!black, line width=0.8pt, rounded corners=2pt, fill=trainFill,
                minimum width=1.05cm, minimum height=0.6cm, inner sep=2pt},
  egate/.style={draw=encFill!60!black, line width=0.8pt, rounded corners=2pt, fill=encFill,
                minimum width=1.15cm, minimum height=0.6cm, inner sep=2pt},
}
\begin{tikzpicture}[x=1cm,y=1cm]
% wires
\foreach \y in {0,-0.8,-1.6,-2.4,-3.2}{
  \draw[qwire] (0,\y) -- (6.0,\y);
}
% trainable (green) blocks
\node[tgate] at (0.95, 0)    {$\theta_1$};
\node[tgate] at (2.05,-0.8)  {$\theta_2$};
\node[tgate] at (3.65,-3.2)  {$\theta_3$};
\node[tgate, minimum height=1.4cm] at (4.95,-2.0) {$\theta_4$};
% encoding (orange) blocks
\node[egate] at (0.95,-1.6)  {$e^{ixH_1}$};
\node[egate, minimum height=1.4cm] at (2.35,-2.8) {$e^{ixH_2}$};
\node[egate, minimum height=1.4cm] at (3.6,-1.2) {$e^{ixH_3}$};
\node[egate] at (4.95, 0)    {$e^{ixH_4}$};
% readout
\node[draw=readFill!60!black, line width=0.8pt, rounded corners=2pt, fill=readFill,
      minimum width=3.9cm, minimum height=0.75cm, rotate=90, font=\rmfamily]
      at (6.4,-1.6) {Readout};
\end{tikzpicture}%
\endgroup
    \caption{QNN as a truncated Fourier series, \autoref{eq:quantum_Fourier_Model}. The trainable blocks (green) determine the Fourier coefficients $c_\omega(\theta)$, while the data-encoding blocks (orange) determine the accessible frequency spectrum $\Omega$.}
    \label{fig:VQC_Fourier}
\end{figure}

From the seminal work \cite{Schuld_2021_Effect}, it is known that if the angle encoding strategy is considered (see \autoref{subsubsec:Encoding_Classical_Information}), the quantum model generated  can be written as a Fourier series \footnote{a classical preprocessing of the data still lead to truncated series expansion in another basis as explained in \cite{thabet_when_2026}}. Its spectrum $\Omega$ depends on the eigenvalues of the encoding Hamiltonians, and the associated Fourier
coefficients depend mainly on the parameterized unitaries. Under these assumptions, the obtained model is called a \emph{Quantum Fourier Model}
(QFM)~\cite{Schuld_2021_Effect, mhiri_constrained_2025}, which is defined as follows:
\definecolor{coefC}{RGB}{0,115,60}
\definecolor{freqC}{RGB}{179,117,36}
\begin{equation}\label{eq:quantum_Fourier_Model}
    f(x,\theta) = \sum_{\textcolor{freqC}{\omega} \in \textcolor{freqC}{\Omega}} \textcolor{coefC}{c_\omega(\theta)}\, e^{i \textcolor{freqC}{\omega}^T x} \, ,
\end{equation}
where the frequency spectrum $\Omega$ collects the accessible frequencies (differences of the encoding-Hamiltonian eigenvalues) and the coefficients $c_\omega(\theta)$ are set by the trainable unitaries, see \autoref{fig:VQC_Fourier} for the color coding. A QFM is thus fully characterized by its Fourier spectrum $\Omega$ and the parametrized Fourier feature vector ${\bf c}(\theta) \in \mathbb{C}^{|\Omega|}$.

Using re-uploading layers \cite{PerezSalinas2020datareuploading} or a classical pre-processing of the data \cite{peters_generalization_2023, kordzanganeh2023exponentially, shin2023exponential, Jaderberg2024LetQNN} increase the size of the Fourier spectrum  $|\Omega|$ leading to more complex quantum model. While the impact of the Fourier model structure is still an open field of research \cite{strobl_fourier_2025}, it has been shown that the average distribution of the quantum Fourier coefficients $c_\omega(\theta)$ is governed by the \emph{redundancies} of the frequencies $\omega$~\cite{mhiri_constrained_2025, campbell_circuit_2026}.

\begin{figure}[h!]
    \centering
    % Fig 6: Random Feature surrogate. Green = quantum Fourier coefficients |c_w|
% over the full spectrum Omega (orange); red = surrogate trained on the
% truncated sub-spectrum Omega_RF. Palette matches Fig5 (coefC/freqC).
% Drawn at natural size (no resizebox) so all text matches the body font.
\begingroup
% Palette from Fig 8 (green/orange); red kept as surrogate accent (no red in Fig 8)
\definecolor{coefC}{RGB}{0,115,60}
\definecolor{freqC}{RGB}{179,117,36}
\definecolor{rfC}{RGB}{205,60,60}
\definecolor{barGreen}{RGB}{136,225,175}
\definecolor{barRed}{RGB}{205,60,60}
\begin{tikzpicture}[x=1cm,y=1cm]
% bars: full spectrum (green), surrogate sub-spectrum overlaid (red, first 5)
\foreach \i/\h in {1/2.75, 2/2.25, 3/1.95, 4/1.45, 5/1.05, 6/0.55, 7/0.33, 8/0.22, 9/0.15, 10/0.11}{
  \fill[barGreen, rounded corners=1pt] ({0.15+(\i-1)*0.55},0)
    rectangle ({0.15+(\i-1)*0.55+0.36},\h);
}
\foreach \i/\h in {1/2.75, 2/2.25, 3/1.95, 4/1.45, 5/1.05}{
  \fill[barRed, rounded corners=1pt] ({0.15+(\i-1)*0.55+0.20},0)
    rectangle ({0.15+(\i-1)*0.55+0.38},\h);
}
% axes
\draw[line width=1.1pt, -{Stealth[length=2.8mm]}] (0,0) -- (0,3.5);
\draw[line width=1.1pt, -{Stealth[length=2.8mm]}] (0,0) -- (7.0,0);
\node[text=coefC, anchor=south west] at (-0.05,3.55) {$|c_\omega|$};
\node[anchor=south] at (6.75,0.18) {$\omega$};
% formulas
\node at (3.6,3.1)
  {$f_Q(x,\Theta) \,=\, \displaystyle\sum_{\textcolor{freqC}{\omega}\in\textcolor{freqC}{\Omega}}
   \textcolor{coefC}{c_{\omega}}\, e^{i\textcolor{freqC}{\omega}x}$};
\node at (4.8,1.8)
  {$f_{RF}(x,\beta) \,=\, \displaystyle\sum_{\textcolor{rfC}{\omega}\in\textcolor{rfC}{\Omega_{RF}}}
   \textcolor{rfC}{\beta_{\omega}}\, e^{i\textcolor{rfC}{\omega} x}$};
\draw[line width=1.1pt, rfC, -{Stealth[length=2.2mm]}]
  (2.0,2.6) to[bend right=40] (2.8,1.9);
% spectrum extents below the axis
\draw[line width=1.1pt, freqC, {Stealth[length=2.2mm]}-{Stealth[length=2.2mm]}]
  (0.05,-0.32) -- (5.55,-0.32);
\node[text=freqC] at (4.0,-0.72) {$\Omega$};
\draw[line width=1.1pt, rfC, {Stealth[length=2.2mm]}-{Stealth[length=2.2mm]}]
  (0.05,-0.88) -- (2.75,-0.88);
\node[text=rfC] at (1.4,-1.28) {$\Omega_{RF}$};
\end{tikzpicture}%
\endgroup
    \caption{Illustration of Random Feature Surrogate. The classical surrogate $f_{RF}$ retains only a polynomial sub-spectrum $\Omega_{RF} \subset \Omega$ (red) carrying the most important Fourier coefficients $|c_\omega|$ (green) of the quantum model's full spectrum $\Omega$ (orange).}
    \label{fig:RFF}
\end{figure}
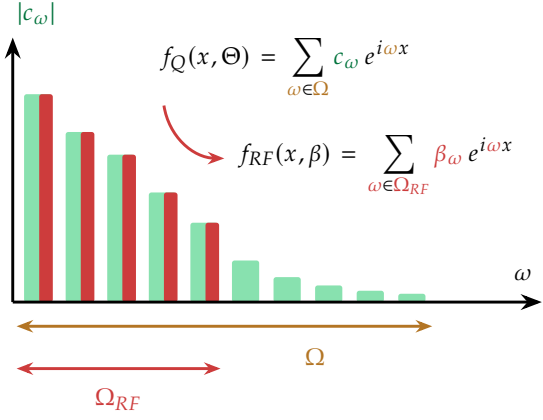

The size of the spectrum has been proposed as an upper-bound on the expressivity power of quantum models \cite{xiong_fundamental_2025}, and many results have pointed out the crucial role of the Fourier spectrum to avoid classical approximation techniques. First, a spectrum of polynomial size with respect to the number of qubits, allows to create a surrogate model (see \autoref{subsec:Dequantization}), which uses the same feature map and trains classically in polynomial time \cite{hofmann_kernel_2008}, ensuring the absence of exponential running time advantage. In the case where the Fourier spectrum is exponentially large, studies \cite{landman_classically_2022, sweke_potential_2025, sahebi_dequantization_2025} have shown that surrogate models on a polynomial sub-spectrum can still be defined on the most important features, based on Random Feature techniques \cite{rahimi_uniform_2008}, see \autoref{fig:RFF}.

However, other studies have showed provable quantum advantage from learning tasks, based on discrete inputs and cryptography primitives \cite{gyurik_exponential_2024, molteni_exponential_2026, liu2021rigorous, jerbi_shadows_2024}, meaning in practice that a quantum learner can classify data whose labels are hidden behind a cryptographic function, such as the discrete logarithm, by internally running a quantum subroutine like Shor's algorithm, a task no efficient classical learner can solve under standard hardness assumptions. This separation between a quantum model and a classical surrogate is due to the specific inductive bias ${\bf c}(\theta)$ that quantum models can realize efficiently~\cite{jerbi_quantum_2023}. Recent work \cite{thabet_when_2026} has also shown how a high norm of the Fourier feature vector, is a necessary condition to avoid classical approximation through Random Feature techniques: when the spectral weight is concentrated on a polynomial number of frequencies, a random-feature surrogate captures the model efficiently, see Fig.~\ref{fig:RFF}

        \subsection{Trainability of QNN}\label{subsec:trainability_QNN}

            \subsubsection{Sample Complexity}

In variational quantum neural networks, the notion of sample complexity has two distinct meanings. First, as in classical learning theory, one may ask how many training examples are required to achieve good generalization. Second, and more specifically to quantum models, one must account for the number of quantum measurement shots, i.e., repeated executions of the same circuit, required to estimate output expectation values and their gradients during training \cite{Benedetti_2019_Parameterized, cerezo_variational_2021}.

While the generalization behavior of quantum models is still an ongoing field of research \cite{gil-fuster_understanding_2024}, it has been shown that such capacity can be connected with encoding techniques \cite{caro_pseudo-dimension_2020, Caro_2021_encoding_genaralisation}, the structure of the hypothesis class \cite{banchi_generalization_2021}, and the number of trainable gates \cite{Caro_2022_Generalization}. Studies have focus on metrics of expressivity to quantify the power of a quantum neural network, such as effective dimension \cite{abbas2021power} (see \autoref{subsec:controllability_QNN}), equivalent Fourier spectrum \cite{xiong_fundamental_2025, schuld2019quantum} (see \autoref{subsec:VQC_as_Fourier_Models}), or the distribution of the equivalent unitary \cite{holmes_connecting_2022}.

The main bottleneck is the number of samples required to estimate the outcome with good accuracy, which can be exacerbated by \textbf{Barren Plateaus} \cite{McClean_2018, larocca_barren_2025} (see \autoref{subsubsec:BP_QNN}), a vanishing gradient phenomena where resolving a meaningful descent direction may itself require an exponentially large number of measurement shots. This difficulty is strengthen by hardware noise \cite{wang_noise-induced_2021, kubler_adaptive_2020}.

\begin{figure*}[t!]
    \centering
    \input{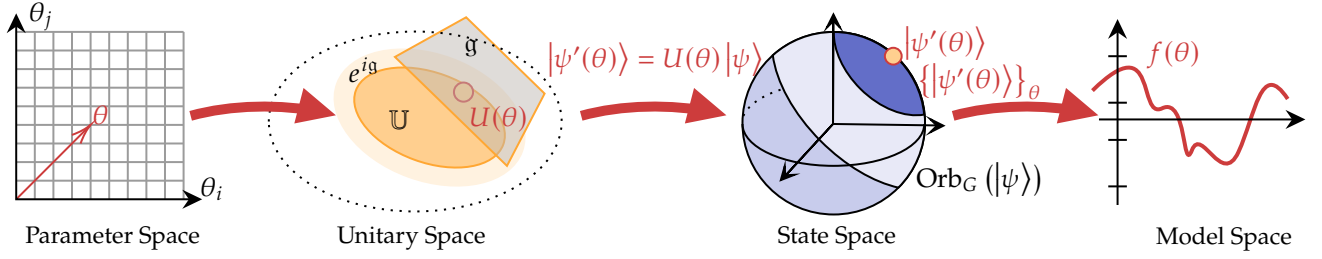}
    \caption{\textbf{The quantum model as a composition of maps, inspired by \cite{Larocca_2023}}. A choice of variational parameters $\theta$ determines the action of the QNN, represented as unitary transformation $U(\theta)$. Given an input $\ket{\psi}$ in a Hilbert space, the QNN produces an output $\ket{\psi'(\theta)}=U(\theta) \ket{\psi}$. Finally, by estimating the expectation value of some observable, one computes a quantum model $f(\theta)$.}
    \label{fig:Control_Spaces}
\end{figure*}

            \subsubsection{Barren Plateaus and Concentration}\label{subsubsec:BP_QNN}

One of the difficulties comes from vanishing gradient phenomenon, due to exponential concentration of the model, which in QML is referred to as \gls{BP} \cite{McClean_2018, larocca_barren_2025}, a
vanishing gradient phenomenon. Vanishing gradients and model concentration imply the requirement of a exponential number of samples with respect to the system size to train a QNN. 
In addition, it has been shown that using gradient free optimizers \cite{arrasmith_effect_2021}, and optimizers that use high order derivatives information \cite{cerezo_higher_2021} do not allow to avoid \gls{BP}s.

\begin{definition}[Barren Plateau]\label{def:Barren_Plateau}
    The cost function $\mathcal{C}(\theta)$ landscape of a $n$-qubit VQC is
    said to exhibit a \gls{BP} if for all $\lambda$:
    \begin{equation}
        \mathbb{E}_{\theta}[\partial_{\theta_\lambda} \mathcal{C}(\theta)] =
        0, \;\;\; \mathrm{Var}_{\theta}[\partial_{\theta_\lambda}
        \mathcal{C}(\theta)] = O\left(\frac{1}{b^n}\right),
    \end{equation}
    with $b > 1$.
\end{definition}

Diagnosing \gls{BP}s in VQCs is therefore important, and multiple methods have been used to do so, based on specific assumptions to allow an analytical study. First, by considering the set of achievable unitary close to Haar random distributed (forming at least a 2-design \cite{dankert_exact_2009}) over a Lie Group of exponentially large dimension, one can use Weingarten calculus \cite{mele_introduction_2024} or Representation theory tools to derive the variance of the expectation value of an observable \cite{McClean_2018, larocca_diagnosing_2022, fontana_characterizing_2024, ragone_lie_2024, ortiz_marrero_entanglement-induced_2021, mhiri_boson_2026, monbroussou_classical_2026}. Such hypothesis on the distribution is strong, especially for shallow circuits. Studies have considered the use of local, independent, and uncorrelated $2$-design gates \cite{Cerezo_2021_cost_BP, uvarov_barren_2021, pesah2021absence,
liu_presence_2022}. 

Recent works \cite{letcher_tight_2024} have pointed out this caveat in the usual theoretical framework, and offered new theoretical tools to study QNN gradients with realistic circuit assumptions, that can also be efficiently evaluated with classical resources. Theoretical frameworks have been proposed to avoid BPs, such as warm start strategies \cite{puig_variational_2025, mhiri_unifying_2025, shen_characterizing_2026}, but proof of dequantization matching the used hypothesis have proposed \cite{lerch_efficient_2024}. While theoretical studies have shown a connection between the proof of absence of BP and classical simulation of VQC \cite{cerezo_does_2025}, new results have shown heuristically that BP can be avoided using \gls{IQP} circuits \cite{recio-armengol_train_2025}, or quantum graph neural networks \cite{raj_scalable_2026}, with no known classical simulation method. Finally, existing theoretical framework have shown that avoiding BP can be done while ensuring a polynomial advantage which would be of particular interest for QPU with high repetition rate such as those based on photonics (see \autoref{subsec:Subspace_Preserving}).

\subsection{Controllability in Quantum Neural Network}\label{subsec:controllability_QNN}

A central challenge in designing quantum circuits for learning and optimization tasks is ensuring that the circuit is controllable in a meaningful sense: that its parameters can steer the output state across a sufficiently rich region of the relevant Hilbert space. Controllability, in this context, is not a binary property but a graded one, tightly bound to the notions of expressibility and trainability that determine the practical utility of a variational architecture. We refer to \cite{dalessandro_introduction_2021} for a more thorough study of quantum controllability. This notion provides a fundamental geometric perspective on a set of parameterized quantum circuits, and thus on an HQNN, viewed as a composition of maps from variational parameters to a unitary, the prepared state, and finally the measured model output (\autoref{fig:Control_Spaces}).

\subsubsection{Controllability in the unitary space}

On a general level a \gls{VQC} can be associated with its {\it reachable gateset}, i.e. the set of all unitary transformations that it is capable to realize
\begin{equation}
    \mathbb{U}=\{U(\theta)\}_\theta \subseteq SU(d),
\end{equation}
which is a subset of the special unitary group of the same dimension $d$ (usually, $d=2^n$ the global Hilbert space size). Several aspects are important to capture the properties of the variational anzatz. First, one can study the distribution of $U(\theta)$ inside the reachable gateset, induced by a natural distribution of the parameters $\theta$, for instance asking if it approximates a unitary $k$-design~\cite{holmes_connecting_2022}. Second, one can ask how close is the reachable gateset to the full group $SU(d)$ as a set, or, more generally, what is the dimension of $\mathbb{U}$. Controllability is typically concerned with the latter question.

Characterizing the reachable gateset of a variational circuit is in general intractable. Nevertheless,
a rigorous framework for analyzing controllability is provided by the \gls{DLA} of the circuit. Given the set of all gate generators $\{H_1, \dots, H_D\}$ appearing in the circuit, i.e.~Hamiltonian matrices used in the unitary gates $e^{-i \theta H_1}$ (see \autoref{subsubsec:gates-unitaries}), the DLA is defined as the Lie closure of the circuit’s generators $\mathfrak{g} ={\rm Lie}( H_1, \dots, H_D )$. The Lie group generated by the DLA, $G=e^{i\mathfrak{g}}$, gives an upper-bound on the reachable gateset $\mathbb{U}\subseteq G$.
% The DLA gives an upper-bound on the maximal set of achievable circuit equivalent Hamiltonian matrices. The Lie group, defined through $G=e^{i\mathfrak{g}}$ represent the maximal set of achievable circuit equivalent unitary matrices. maximal set of achievable circuit equivalent Hamiltonian matrices. The Lie group, defined through $G=e^{i\mathfrak{g}}$ represent the maximal set of achievable circuit equivalent unitary matrices. 
As a result, the DLA dimension gives an upperbound on the controllability of a quantum circuit in the unitary space. Obviously, this controllability is in practice also bounded by the number of independent parameters used in the circuit, its architecture, and the fact that some gates are usually not parametric.

\subsubsection{Controllability in the state space}

A natural extension to the DLA in the state space is the notion of orbit. Orbits of states \cite{dalessandro_introduction_2021, mamon_orbit_2025} $\ket{\psi}$ or $\rho$ under a unitary group $G$ are defined as $\text{Orb}_G(\ket{\psi}) := \{U \ket{\psi} | U \in G \}$ and $\text{Orb}_G(\rho) := \{U \rho U^\dagger | U \in G \}$. The dimension of the orbit is bounded by the dimension of the DLA previously defined, as the state is simply a projection of the Lie group onto a specific initial state, and by the dimension of the state space. As for the DLA, the orbit dimension gives an upperbound on the controllability of a quantum circuit in the state space that can be consider in addition with the number of independent parameters used in the circuit.

The DLA gives a global top-to-bottom view on the controllability of a VQC and the reachable state orbit. A complementary local perspective is offered by the \gls{QFIM}  $\mathcal{F}(\theta)$~\cite{liu_quantum_2019, bengtsson2017geometry, haug2024generalization}. The latter describes the geometry of the quantum states $\ket{\psi(\theta)}= U(\theta)\ket{\psi}$ around a specific parameter value $\theta$. Formally, the QFIM is given by
\begin{equation}\label{eq:QFIM}
\mathcal{F}_{ij}(\theta) = 4 \, {\rm Re}\Big( \langle \partial_i \psi | \partial_j \psi\rangle - \braket{\partial_i \psi}{\psi}\braket{\psi}{\partial_j \psi}\Big),
\end{equation}
with $\ket{\partial_i \psi} =\frac{\partial}{\partial \theta_i} \ket{\psi(\theta)}$, and defines the so called Fubini-Study metric on the manifold of pure quantum states. QFIM relates to parameter estimation via the quantum Cram\'er-Rao bound~\cite{giovannetti2011advances}, and extends the idea of natural gradients to the quantum realm~\cite{stokes2020quantum, mcardle_variational_2019, gacon2021simultaneous}. In particular, the rank of the QFIM gives the dimension of the local patch of the state orbit reachable by perturbing the parameter value around $\theta$. In addition, it has been established~\cite{Larocca_2023} that the regime where the rank of the QFIM is saturated, referred to as overparametrization, corresponds to a computational phase transition where the QNN trainability is greatly improved. Finally, the \gls{DQFIM}~\cite{haug2024generalization} extends this geometric perspective to include an explicit dependence on the training data available to the quantum model.
% In order to consider the controlability of a quantum state $\ket{\psi}$ for a fixed set of parameters $\theta$, and not an upperbound, one can consider the Quantum Fisher Information matrix \cite{liu_quantum_2019, bengtsson2017geometry, haug2024generalization} (QFIM) $\mathcal{F}(\theta)$ rank:
% \begin{equation}\label{eq:QFIM}
% \mathcal{F}_{ij}(\theta) = 4 \, {\rm Re}\Big( \langle \partial_i \psi | \partial_j \psi\rangle - \braket{\partial_i \psi}{\psi}\braket{\psi}{\partial_j \psi}\Big),
% \end{equation}
% with $\ket{\partial_i \psi} =\frac{\partial}{\partial \theta_i} \ket{\psi(\theta)}$, and defines the so called Fubini-Study metric. The QFIM eigenstates give the different direction the considered state can take. Related to parameter estimation via the quantum Cram\'er-Rao bound~\cite{giovannetti2011advances}, the QFIM offers an information-geometric view of the set of quantum states prepared by a variational quantum circuit extends the idea of natural gradients to the quantum realm~\cite{stokes2020quantum, mcardle_variational_2019, gacon2021simultaneous}. In addition, it has been established \cite{Larocca_2023} that the regime where the rank of the QFIM is saturated, referred as overparametrization, corresponds to a computational phase transition where the QNN trainability is greatly improved.

\subsubsection{Controllability in the model space}

The question of controllability is also natural to pose for the quantum model, obtained by performing a measurement on the quantum system. At the basic level the quantum model is described by the parametric probability distribution $p(y|\theta) = \bra{\psi(\theta)} E_y \ket{\psi(\theta)}$ of measurement outputs, for the measurement $\{E_y\}$ performed on the state  $\ket{\psi(\theta)}=U(\theta)\ket{\psi}$ prepared by the VQC.

In particular, the sensitively of the model to small parameter variations is characterized by the Fisher Information Matrix (FIM), given by 
\begin{equation}
\begin{split}
    \mathcal{I}_{ij}(\theta) & = \mathbb{E} \left[ (\partial_i  \log p(Y|\theta)) (\partial_j  \log p(Y|\theta))^\top \right] \; ,
\end{split}
\end{equation}
where the expectation value is taken over the random variable $Y\sim p(y|\theta)$. The FIM defines a Riemannian metric on the set of probability distributions, which is naturally related to their statistical properties~\cite{kay1993statistical}. In particular, the Cram\'er-Rao bound~\cite{kay1993statistical} states that for any unbiased estimator $\hat{\theta}$ of the the parameters $\theta$, the covariance matrix satisfies $\text{Cov}[\hat{\theta}-\theta] \geq \mathcal{I}(\theta)^{-1}$, thus providing a fundamental lower bound on estimation precision. Furthermore by monotonicity the FIM satisfies $ \mathcal{I}(\theta)\preceq  \mathcal{F}(\theta)$ for all possible measurements of the quantum state.

% characterizes  This geometric viewpoint is crucial for understanding the trainability of HQNNs, as the Fisher information provides the natural information-geometric metrics  on the set of probability distributions and quantum states.\\

% In the former case, the \gls{FIM} $\mathcal{I}(\theta)$ associated with a parameterized probability distribution $p(y|\theta)$ is defined as
% \begin{equation}
% \begin{split}
%     \mathcal{I}(\theta) & = \mathbb{E} \left[ (\nabla_\theta  \log p(Y|\theta)) (\nabla_\theta  \log p(Y|\theta))^\top \right] \; ,
% \end{split}
% \end{equation}
% where the expectation value is taken over the random variable $Y\sim p(y|\theta)$, and satisfies ${\rm QFIM}\succeq {\rm FIM}$ for all possible measurements of the quantum state. The Fisher information defines a Riemannian metric on the set of probability distributions, which is naturally related to their statistical properties~\cite{kay1993statistical}. In particular, the Cram\'er-Rao bound~\cite{kay1993statistical} states that for any unbiased estimator $\hat{\theta}$ of the the parameters $\theta$, the covariance matrix satisfies $\text{Cov}[\hat{\theta}-\theta] \geq \mathcal{I}(\theta)^{-1}$, thus providing a fundamental lower bound on estimation precision.\\

The FIM generalizes to parametrized probabilistic functions $f: (x , \theta) \mapsto Y\simeq p(y|x,\theta)$ describing the distribution of measurement outputs of a parametrized quantum circuit for the input data $x$~\cite{abbas2021power}. For a dataset $ \mathcal{X}$, one then defines the joint random variable 
\begin{equation}\label{eq: XYtogether}
(X,Y) \sim p(x,y|\theta):=\frac{1}{|\mathcal{X}|} p(y|x, \theta) \; ,
\end{equation}
capturing the parametric joint distributions of the input data and outputs of the function. For the quantum model the FIM $\mathcal{I}(\theta)$ of this distribution is the equivalent of the data QFIM for the quantum states. It has been shown that this FIM is a valuable analytical tool to study trainability, expressivity and generalization of the learning algorithm~\cite{abbas2021power}. For instance, regions of parameter space with vanishingly small norm of the  FIM  correspond to plateaus where gradient-based optimization becomes inefficient, as measurements provide little information about parameter updates.

\subsection{De-quantization}\label{subsec:Dequantization}

While promising near-term quantum algorithms are developed, the research community is challenging them by also working on the dequantization of quantum algorithms, making sometimes difficult the identification of quantum advantage for HQNN, and variational quantum algorithms in general. De-quantization techniques \cite{masot-llima_prospects_2025} allow one to perform a \textbf{(approximate) classical simulation or emulation} of quantum functions such as defined in \autoref{eq:quantum_Model}. Such a technique is referred to as \emph{efficient} if it can be run in polynomial time with respect to the system size (for e.g., number of qubits) and the inverse precision. De-quantization is an active field of research, and this section does not pretend to be exhaustive. We present in the following some of the most important techiques regarding HQNN.

    \subsubsection*{Exact simulations}

\noindent The most direct approach to classically simulating a quantum computation is \textbf{state vector} (or Schr\"odinger-style) simulation, in which the full $2^n$-dimensional amplitude vector of an $n$-qubit register is stored in memory and updated gate by gate through sparse matrix--vector multiplications. While exact and applicable to arbitrary circuits, its memory footprint grows exponentially with the number of qubits, limiting even the largest supercomputer implementations to roughly $45$-$50$ qubits~\cite{de_raedt_massively_2019, haner_05_2017}. Beyond this regime, a variety of specialized techniques trade generality for efficiency, and we give some of them in the following:

\textbf{Stabilizer methods} have been proposed to simulate Clifford-dominated circuits in polynomial time, with a cost growing exponentially only in the number of non-Clifford (magic) resources \cite{aaronson2004improved, bravyi_improved_2016}.

\textbf{Tensor networks} can simulate exactly quantum states with low entanglement, by representing the amplitudes as a network of contractions between low-rank tensors whose bond dimension bounds the entanglement across a cut \cite{berezutskii_tensor_2025}. When the entanglement generated by the circuit stays bounded, e.g. for shallow circuits of $\mathcal{O}(\log n)$ depth in a one-dimensional geometry, a matrix product state (MPS) of polynomial bond dimension stores the state and evaluates local observables in polynomial time.

\textbf{Lie algebraic simulation} is a technique \cite{goh_lie-algebraic_2025, barligea_enabling_2026}, sometimes referred as $\mathfrak{g}$-sim, based on the Lie algebraic structure of the dynamical process. When the set of Hamiltonian matrices used in the computation generates a Lie algebra of polynomial size, one can use such the basis of this Lie algebra to simulate the computation in polynomial time if the observable is part of Lie algebra. This hypothesis matches to the one that avoid \gls{BP}s with Lie Algebraic ansatz (see \autoref{subsubsec:BP_QNN}), contributing to the trade-off between trainability and classical simulation \cite{cerezo_does_2025}.

\textbf{Specific simulation techniques for non-universal ansatz} can be constructed, by using their specific mathematical structure. For example, circuits made of matchgates \cite{jozsa_matchgates_2008}, noninteracting fermionic circuits \cite{terhal_classical_2002}, or bosonic circuit \cite{bartlett_efficient_2002, clifford_classical_2018, quesada_exact_2020, quesada_simulating_2019, chabaud_classical_2021, bourassa_fast_2021, lim_classical_2025, seron_efficient_2024} offer specific simulation techniques.  

    \subsubsection*{Approximate simulations}

\noindent Beyond the regimes where exact simulation is efficient, several methods approximate a quantum computation up to a controlled error $\epsilon$, and remain efficient when this precision can be reached in polynomial time.

\textbf{Tensor networks} also provide approximate simulations beyond the shallow-depth circuits, by controllably truncating the bond dimension of a tensor-network ansatz to some finite value $\chi$ that discards the smallest Schmidt coefficients at each cut \cite{berezutskii_tensor_2025} while keeping error under given target $\epsilon$. Circuit noise typically increases experimental error and thereby extends the reach of the tensor-network approximation.

\textbf{Pauli Propagation}  \cite{rudolph_classical_2023, rudolph_pauli_2025, Begusic2024fast, shao_simulating_2024, angrisani_classically_2025} is a technique based on the representation of a quantum algorithm by computing the evolution of Pauli strings under a chosen quantum dynamics or quantum circuit. It can be used to perform  a back propagation of the observable (see \autoref{subsubsec:Output_Readout}) in the Heisenberg picture where the observable is represented with its coefficient in the basis of Pauli operators. As the Pauli basis has $4^n$ elements, this representation is efficient only when the evolved observable can be well approximated by a truncation to $\mathrm{poly}(n)$ terms, which holds for Clifford-dominated circuits and, more generally, on average over ensembles of noisy or random circuits \cite{angrisani_classically_2025}. Such Pauli-path methods are limited by the system’s "magic", which is upperbounded by the degree of non-Clifford operation in the quantum computation \cite{lin_utility-scale_2026}.

    \subsubsection*{Surrogate modelling}

\noindent Surrogate models are classical models that span the same basis functions as the corresponding quantum model. The weights of these basis functions are trained classically, without the restrictions the circuit structure imposes on the coefficients of the quantum output function, so a surrogate can match or exceed the performance of its quantum counterpart. Surrogate modelling therefore provides a natural benchmark that a quantum model must outperform to claim an advantage. Evading it requires that the basis functions be inaccessible to the surrogate, or that their effective number grow exponentially with the system size.

\textbf{Tensor networks} can be used to build such surrogates, by training a tensor-network representation of the quantum model directly on the classical feature space \cite{shin_dequantizing_2024}.

\textbf{Pauli Propagation}  can also be used to construct surrogates, learning the relevant Pauli coefficients of the model once and reusing them classically at inference \cite{bermejo_quantum_2026}.

\textbf{Random Feature Techniques} can be used to design surrogate models based on the decomposition of quantum models in feature maps of interest. The most common example is for quantum Fourier models \cite{Schuld_2021_Effect, schuld2019quantum, sweke_potential_2025, landman_classically_2022} and discussed more in details in \autoref{subsec:VQC_as_Fourier_Models}.
        
\section{Architectures of HQNN}
\label{sec:architectures}

The design space of Hybrid Quantum Neural Networks includes many architectural patterns that differ fundamentally in how quantum and classical components exchange information, how data is encoded into quantum states, and what computational primitives drive the learning process. In this Section, we first discuss how variational quantum layers can be used in a classical machine learning pipeline. Then, we present some computational paradigm of interest that are promising for future applications of HQNN.

%%%

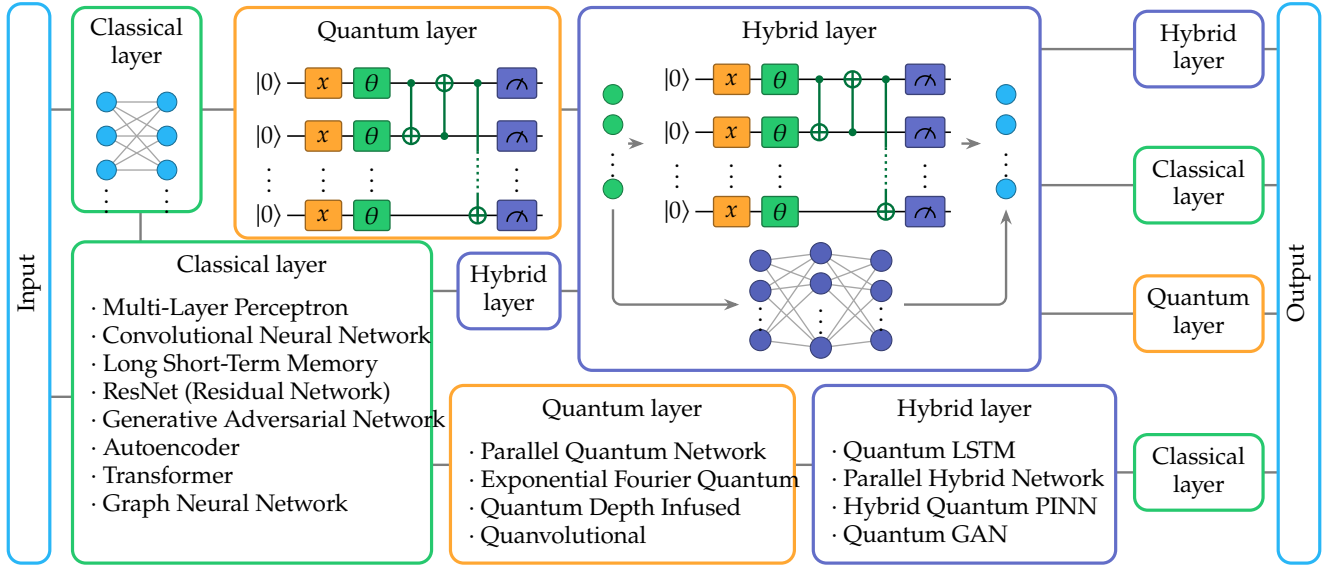
\begin{figure*}[t!]
    \centering
    % Fig 8: HQNN composition map (TikZ redraw of Fig8.pdf, original colors/style).
% Body font via \rmfamily. Total width 17.35cm <= \textwidth (494.5pt).
% Uniform inner padding ~0.13cm; uniform inter-box gaps ~0.25cm.
\begingroup
\definecolor{cyanB}{RGB}{41,182,246}
\definecolor{greenB}{RGB}{38,200,110}
\definecolor{orangeB}{RGB}{255,167,51}
\definecolor{violetB}{RGB}{99,110,200}
\definecolor{dkG}{RGB}{0,115,60}
\definecolor{nnV}{RGB}{85,98,185}
\definecolor{conn}{RGB}{125,125,125}
\tikzset{
  lbox/.style={draw, line width=1.2pt, rounded corners=5pt, fill=white},
  gateX/.style={draw=orangeB!60!black, line width=0.5pt, fill=orangeB, rounded corners=1pt,
                minimum width=0.48cm, minimum height=0.4cm, inner sep=1pt},
  gateW/.style={draw=greenB!60!black, line width=0.5pt, fill=greenB, rounded corners=1pt,
                minimum width=0.48cm, minimum height=0.4cm, inner sep=1pt},
  meterbox/.style={draw=violetB!60!black, line width=0.5pt, fill=violetB, rounded corners=1pt,
                minimum width=0.52cm, minimum height=0.4cm, inner sep=1pt},
  cline/.style={conn, line width=0.9pt},
  carrow/.style={conn, line width=0.9pt, -{Stealth[length=2mm]}, rounded corners=3pt},
  nnedge/.style={black!35, line width=0.5pt},
  qcirc/.pic={
    % three-wire circuit: |0> - x - theta - entanglers - meter ; width 3.62
    \foreach \y in {0,-0.7,-1.75}{
      \node[font=\rmfamily] at (0,\y) {$\ket{0}$};
      \draw[line width=0.6pt] (0.24,\y) -- (3.62,\y);
    }
    \node at (0,-1.18) {$\vdots$};
    \foreach \y in {0,-0.7,-1.75}{
      \node[gateX] at (0.72,\y) {$x$};
      \node[gateW] at (1.36,\y) {$\theta$};
    }
    \node at (0.72,-1.18) {$\vdots$};
    \node at (1.36,-1.18) {$\vdots$};
    % entanglers in dark green
    \fill[dkG] (1.88,0) circle (1.5pt);
    \draw[dkG, line width=0.9pt] (1.88,0) -- (1.88,-0.7);
    \draw[dkG, line width=0.9pt] (1.88,-0.7) circle (3pt);
    \draw[dkG, line width=0.9pt] (1.88,-0.805) -- (1.88,-0.595);
    \draw[dkG, line width=0.9pt] (1.775,-0.7) -- (1.985,-0.7);
    \fill[dkG] (2.32,-0.7) circle (1.5pt);
    \draw[dkG, line width=0.9pt] (2.32,-0.7) -- (2.32,0);
    \draw[dkG, line width=0.9pt] (2.32,0) circle (3pt);
    \draw[dkG, line width=0.9pt] (2.32,-0.105) -- (2.32,0.105);
    \draw[dkG, line width=0.9pt] (2.215,0) -- (2.425,0);
    \fill[dkG] (2.76,0) circle (1.5pt);
    \draw[dkG, line width=0.9pt] (2.76,0) -- (2.76,-0.9);
    \draw[dkG, line width=0.9pt, dash pattern=on 1pt off 1.6pt] (2.76,-0.9) -- (2.76,-1.45);
    \draw[dkG, line width=0.9pt] (2.76,-1.45) -- (2.76,-1.75);
    \draw[dkG, line width=0.9pt] (2.76,-1.75) circle (3pt);
    \draw[dkG, line width=0.9pt] (2.76,-1.855) -- (2.76,-1.645);
    \draw[dkG, line width=0.9pt] (2.655,-1.75) -- (2.865,-1.75);
    % meters
    \foreach \y in {0,-0.7,-1.75}{
      \node[meterbox] at (3.28,\y) {};
      \draw[black, line width=0.6pt] (3.15,\y-0.085) arc[start angle=180, end angle=0, radius=0.13];
      \draw[black, line width=0.6pt] (3.28,\y-0.085) -- (3.355,\y+0.065);
    }
    \node at (3.28,-1.18) {$\vdots$};
  },
}
\begin{tikzpicture}[x=1cm,y=1cm, every node/.style={font=\rmfamily}]
% ------------------------------------------------------------- connectors
\draw[cline] (0.6,6.2) -- (0.9,6.2);
\draw[cline] (0.6,2.4) -- (0.9,2.4);
\draw[cline] (1.8,4.85) -- (1.8,4.45);
\draw[cline] (2.6,6.2) -- (3.05,6.2);
\draw[cline] (7.35,6.2) -- (7.6,6.2);
\draw[cline] (13.7,7.0) -- (14.95,7.0);
\draw[cline] (16.65,7.0) -- (16.85,7.0);
\draw[cline] (13.7,5.2) -- (14.95,5.2);
\draw[cline] (16.65,5.2) -- (16.85,5.2);
\draw[cline] (13.7,3.5) -- (14.95,3.5);
\draw[cline] (16.65,3.5) -- (16.85,3.5);
\draw[cline] (5.65,3.8) -- (6.0,3.8);
\draw[cline] (7.35,3.8) -- (7.6,3.8);
\draw[cline] (5.65,1.5) -- (5.9,1.5);
\draw[cline] (10.45,1.5) -- (10.7,1.5);
\draw[cline] (14.7,1.4) -- (14.95,1.4);
\draw[cline] (16.65,1.4) -- (16.85,1.4);
% ------------------------------------------------------------- title
\node[font=\rmfamily\bfseries, text=dkG] at (8.7,8.0) {HYBRID QUANTUM NEURAL NETWORK};
% ------------------------------------------------------------- input / output
\draw[lbox, draw=cyanB] (0.05,0.2) rectangle (0.6,7.6);
\node[rotate=90] at (0.325,3.9) {Input};
\draw[lbox, draw=cyanB] (16.85,0.2) rectangle (17.4,7.6);
\node[rotate=90] at (17.125,3.9) {Output};
% ------------------------------------------------------------- top-left classical
\draw[lbox, draw=greenB] (0.9,4.85) rectangle (2.6,7.55);
\node[align=center] at (1.75,7.07) {Classical\\layer};
\foreach \ya in {6.3,5.85,5.4}{ \foreach \yb in {6.3,5.85,5.4}{
  \draw[nnedge] (1.35,\ya) -- (2.15,\yb); }}
\foreach \y in {6.3,5.85,5.4}{
  \fill[cyanB, draw=cyanB!60!black, line width=0.4pt] (1.35,\y) circle (0.13);
  \fill[cyanB, draw=cyanB!60!black, line width=0.4pt] (2.15,\y) circle (0.13); }
\node at (1.35,5.09) {$\vdots$};
\node at (2.15,5.09) {$\vdots$};
% ------------------------------------------------------------- quantum layer (top)
\draw[lbox, draw=orangeB] (3.05,4.5) rectangle (7.35,7.55);
\node at (5.2,7.22) {Quantum layer};
\pic at (3.5,6.55) {qcirc};
% ------------------------------------------------------------- hybrid layer (top right)
\draw[lbox, draw=violetB] (7.6,2.75) rectangle (13.7,7.55);
\node at (10.65,7.22) {Hybrid layer};
\foreach \y in {6.4,6.0,5.15}{
  \fill[greenB, draw=greenB!60!black, line width=0.4pt] (8.05,\y) circle (0.13); }
\node at (8.05,5.57) {$\vdots$};
\draw[carrow] (8.25,5.75) -- (8.46,5.75);
\pic at (8.9,6.6) {qcirc};
\foreach \y in {6.4,6.0,5.15}{
  \fill[cyanB, draw=cyanB!60!black, line width=0.4pt] (13.25,\y) circle (0.13); }
\node at (13.25,5.57) {$\vdots$};
\draw[carrow] (12.66,5.75) -- (12.92,5.75);
% small classical NN directly below the circuit (circuit bottom wire y=4.7)
\foreach \ya in {4.2,3.8,3.15}{ \foreach \yb in {4.3,3.9,3.05}{
  \draw[nnedge] (10.0,\ya) -- (10.8,\yb); }}
\foreach \ya in {4.3,3.9,3.05}{ \foreach \yb in {4.2,3.8,3.15}{
  \draw[nnedge] (10.8,\ya) -- (11.6,\yb); }}
\foreach \y in {4.2,3.8,3.15}{
  \fill[nnV, draw=nnV!60!black, line width=0.4pt] (10.0,\y) circle (0.14);
  \fill[nnV, draw=nnV!60!black, line width=0.4pt] (11.6,\y) circle (0.14); }
\foreach \y in {4.3,3.9,3.05}{
  \fill[nnV, draw=nnV!60!black, line width=0.4pt] (10.8,\y) circle (0.14); }
\node at (10.0,3.53) {$\vdots$};
\node at (10.8,3.5) {$\vdots$};
\node at (11.6,3.53) {$\vdots$};
\draw[carrow] (8.05,4.88) -- (8.05,3.62) -- (9.72,3.62);
\draw[carrow] (11.9,3.62) -- (13.25,3.62) -- (13.25,4.88);
% ------------------------------------------------------------- right column
\draw[lbox, draw=violetB] (14.95,6.5) rectangle (16.65,7.5);
\node[align=center] at (15.8,7.0) {Hybrid\\layer};
\draw[lbox, draw=greenB] (14.95,4.7) rectangle (16.65,5.7);
\node[align=center] at (15.8,5.2) {Classical\\layer};
\draw[lbox, draw=orangeB] (14.95,3.0) rectangle (16.65,4.0);
\node[align=center] at (15.8,3.5) {Quantum\\layer};
\draw[lbox, draw=greenB] (14.95,0.9) rectangle (16.65,1.9);
\node[align=center] at (15.8,1.4) {Classical\\layer};
% ------------------------------------------------------------- big classical (bottom left)
\draw[lbox, draw=greenB] (0.9,0.2) rectangle (5.65,4.45);
\node at (3.275,4.12) {Classical layer};
\node[align=left, anchor=north west] at (1.03,3.82)
  {$\cdot$ Multi-Layer Perceptron\\
   $\cdot$ Convolutional Neural Network\\
   $\cdot$ Long Short-Term Memory\\
   $\cdot$ ResNet (Residual Network)\\
   $\cdot$ Generative Adversarial Network\\
   $\cdot$ Autoencoder\\
   $\cdot$ Transformer\\
   $\cdot$ Graph Neural Network};
% ------------------------------------------------------------- hybrid small (middle)
\draw[lbox, draw=violetB] (6.0,3.3) rectangle (7.35,4.3);
\node[align=center] at (6.675,3.8) {Hybrid\\layer};
% ------------------------------------------------------------- quantum list (bottom)
\draw[lbox, draw=orangeB] (5.9,0.2) rectangle (10.45,2.55);
\node at (8.175,2.22) {Quantum layer};
\node[align=left, anchor=north west] at (6.03,1.92)
  {$\cdot$ Parallel Quantum Network\\
   $\cdot$ Exponential Fourier Quantum\\
   $\cdot$ Quantum Depth Infused\\
   $\cdot$ Quanvolutional};
% ------------------------------------------------------------- hybrid list (bottom)
\draw[lbox, draw=violetB] (10.7,0.2) rectangle (14.7,2.55);
\node at (12.7,2.22) {Hybrid layer};
\node[align=left, anchor=north west] at (10.83,1.92)
  {$\cdot$ Quantum LSTM\\
   $\cdot$ Parallel Hybrid Network\\
   $\cdot$ Hybrid Quantum PINN\\
   $\cdot$ Quantum GAN};
\end{tikzpicture}%
\endgroup
    \caption{Hybrid Quantum Neural Network consists of different types of layers: quantum (orange boxes), classical (green boxes) and hybrid (violet boxes). Three top boxes give examples of schematic representations of the layers of each type. Three bottom boxes include non-exhaustive lists of possible representatives for each
    type.}
    \label{fig:enter-label}
\end{figure*}

The design space of \glspl{HQNN} spans many architectural patterns that differ in three respects: how the quantum and classical components exchange information, how classical data are encoded into the quantum state, and which computational paradigm the quantum layer realizes. The corresponding classical, quantum, and hybrid layer types are summarized in \autoref{fig:enter-label}/ \autoref{subsec:circuits_pipeline} describes how a \gls{VQC} is placed within an otherwise classical machine-learning pipeline and the functional roles it can take there. \autoref{subsec:topology_info} treats the information-flow topology, that is, how the modules are wired together, and why this choice carries different weight in the \gls{NISQ} and \gls{FTQC} regimes. Finally, we present the computational paradigms of the quantum layer that are of interest for future \gls{HQNN} applications.

\subsection{Quantum Circuits in a ML Pipeline}\label{subsec:circuits_pipeline}

An \gls{HQNN} is assembled from three kinds of building block, shown in \autoref{fig:enter-label}: classical layers, quantum layers, and hybrid layers that combine the two. The quantum layer is a \gls{VQC}, a parameterized map that composes with classical layers into a single differentiable pipeline (\autoref{sec:foundations}).

Within this pipeline a \gls{VQC} can occupy one of several functional roles. Its parameters may be trained jointly with the classical weights; it may be held fixed as a feature map, as in quantum kernel methods and quantum reservoir computing (\autoref{subsubsec:Quantum_Reservoir_Computing}); or it may be trained in classical simulation and executed on quantum hardware only at inference. These roles differ in which parameters are learned rather than in the arrangement of layers.

\subsection{Topology of information}\label{subsec:topology_info}

The information-flow topology is the path data follow between the classical and quantum modules, and two arrangements are distinguished~\cite{Mitarai_2018_QClearning, mari2020transfer, Chen2020HybridNN, kordzanganeh2023parallel}. In a \emph{sequential} topology the modules form a single chain, with the \gls{VQC} placed between a classical pre- and post-processing block, as in the dressed quantum circuit~\cite{mari2020transfer}; a post-processing block trained on the circuit's measurement outputs can also absorb the systematic part of the device's readout errors, a form of learned error mitigation~\cite{Chen2020HybridNN}. In a \emph{parallel} topology the same input feeds a classical and a quantum branch that are evaluated independently and merged at the output, as in the \gls{PHN}~\cite{kordzanganeh2023parallel}.

The topology fixes the functional role of the \gls{VQC}: in a sequential chain it transforms a representation produced by the classical part, whereas in a parallel branch it contributes an independently learned function combined only at the output. How far this choice determines the model depends on the hardware regime. On \gls{NISQ} devices, where decoherence is the binding constraint, the topology is limited to what a shallow, low-qubit circuit can execute and is fixed empirically. Under \gls{FTQC}, where logical qubits behave essentially ideally and the cost is the deterministic synthesis budget (\autoref{subsec:FTQC_challenges}), it is instead governed by theoretical properties of the resulting model such as controllability (\autoref{subsec:controllability_QNN}) and trainability (\autoref{subsec:trainability_QNN}).

\subsection{Computational paradigm}

The topology of information flow dictates how the quantum layer is wired to the classical part and how data enters it; however, another critical design constraint on an \gls*{HQNN} pipeline is the computational paradigm of the quantum layer, that is, which model of quantum computation it realizes.
The default choice, a universal circuit of freely parameterized gates that can in principle reach any unitary, is the most flexible but is generically hard to train, since such circuits typically suffer from \gls{BP} (\autoref{subsec:trainability_QNN}). This motivates paradigms that deliberately give up universality, or move the trainable parameters elsewhere, in exchange for trainability guarantees, classical simulability, or a more hardware-native implementation. Each case of the paradigm explained below sets a different balance between the class of functions the layer can represent, the difficulty of training it, the cost of simulating it classically, and how readily it maps onto available hardware.

\subsubsection{Subspace Preserving Ansatz}\label{subsec:Subspace_Preserving}

Variational circuit often suffer from \gls{BP} (see \autoref{subsec:trainability_QNN}), a vanishing gradient phenomena that prevent an efficient training of QNN layers. Formally, a $n$-qubit circuit is
subspace preserving if there exists a direct sum decomposition of its Hilbert
space $\mathcal{H} = \bigoplus_i \mathcal{H}_i$ where each $\mathcal{H}_i$ is
preserved by the action of the unitaries / channels. Theoretical guarantees
\cite{larocca_diagnosing_2022, ragone_lie_2024, fontana_characterizing_2024,
lee_progress_2021} on the training can be offered by restricting exponentially
large Hilbert space described by quantum computation to a subspace of
polynomial dimension. Such reduction correspond to a use of non universal
quantum operation. By enforcing symmetries in the computation, studies have
reduced the set of achievable equivalent unitary to a Lie group of polynomial
size, while offering new properties to the quantum computation. The use of symmetry in the computation is sometime refers as \Gls{GQML} \cite{larocca_group-invariant_2022, meyer_exploiting_2023,zheng_speeding_2023, west_provably_2024}.

Hamming-weight preserving circuits have been exploited to offer near-term
amplitude encoding techniques on a basis of fixed hamming-weight
\cite{farias_quantum_2025}, and tailored made algorithms that mimic classical
machine learning subroutine such as \textbf{vision transformers}
\cite{cherrat_quantum_2024}, \textbf{orthogonal layers}
\cite{landman_quantum_2022}, \textbf{quantum inspired adapters}
\cite{raj_hyper_2025}, \textbf{convolutional network}
\cite{monbroussou_subspace_2025}, \textbf{nearest centroid classifier}
\cite{johri_nearest_2021}, \textbf{Fourier neural operator}
\cite{jain_quantum_2024}, \textbf{quantum graph neural networks} \cite{raj_scalable_2026}, or algorithms based on \textbf{compound layers
architecture} \cite{kerenidis_quantum_2022}. The most common Hamming-weight
preserving gates used are the \gls{RBS} and \gls{FBS}, both inspired by bosonic beam splitters:
\begin{equation*}
    \begin{split}
    RBS(\theta) &= \begin{pmatrix}
            1 & 0 & 0 & 0 \\
            0 & \cos(\theta) & \sin(\theta) & 0 \\
            0 & -\sin(\theta) & \cos(\theta) & 0 \\
            0 & 0 & 0 & 1 \\
        \end{pmatrix} \quad \text{and} \\
    FBS(\theta) &= \begin{pmatrix}
            1 & 0 & 0 & 0 \\
            0 & \cos(\theta) & (-1)^{f}\sin(\theta) & 0 \\
            0 & (-1)^{f+1}\sin(\theta) & \cos(\theta) & 0 \\
            0 & 0 & 0 & 1 \\
        \end{pmatrix}\, ,        
    \end{split}
\end{equation*}
with $f = \sum_{i<l<j} s_l$ such that $i,j$ are the qubits of application and
$s_i \dots s_j \in \{0,1\}$ are the excitation of the qubits in between. Most
of Hamming-weight proposals can offer theoretical guarantees on their training
at the cost of classical simulation in polynomial time (see
\autoref{subsec:trainability_QNN}).

\begin{figure*}[t!]
    \centering
    % Linear optical circuit (a) + photonic homomorphism (b).
% TikZ redraw of Fig11.pdf keeping its colors/style; body font via rmfamily.
% Single column: total width <= 8.3cm; panel (b) stacked below (a).
\begingroup
% Palette from Fig 8 (cyan/violet/green); red kept for beam splitters (no red in Fig 8)
\definecolor{bsR}{RGB}{205,60,60}
\definecolor{psB}{RGB}{41,182,246}
\definecolor{frV}{RGB}{99,110,200}
\definecolor{dotG}{RGB}{38,200,110}
\begin{tikzpicture}[x=1cm,y=1cm, every node/.style={font=\rmfamily}]
\node at (0.25,0.45) {a)};
\draw[frV, line width=1.5pt, rounded corners=9pt] (0.66,0.35) rectangle (8.0,-4.55);
\fill[dotG] (0.36,-0.00) circle (0.11);
\fill[dotG] (0.36,-0.60) circle (0.11);
\node at (0.36,-1.20) {$\vdots$};
\fill[dotG] (0.36,-2.40) circle (0.11);
%\fill[dotG] (0.36,-3.00) circle (0.11);
%\fill[dotG] (0.36,-3.60) circle (0.11);
%\fill[dotG] (0.36,-4.20) circle (0.11);
\fill[bsR, rounded corners=1.5pt] (0.85,-0.44) rectangle (1.45,-0.16);
\fill[bsR, rounded corners=1.5pt] (0.85,-1.64) rectangle (1.45,-1.36);
\fill[bsR, rounded corners=1.5pt] (0.85,-2.84) rectangle (1.45,-2.56);
\fill[bsR, rounded corners=1.5pt] (0.85,-4.04) rectangle (1.45,-3.76);
\fill[bsR, rounded corners=1.5pt] (2.57,-0.44) rectangle (3.17,-0.16);
\fill[bsR, rounded corners=1.5pt] (2.57,-1.64) rectangle (3.17,-1.36);
\fill[bsR, rounded corners=1.5pt] (2.57,-2.84) rectangle (3.17,-2.56);
\fill[bsR, rounded corners=1.5pt] (2.57,-4.04) rectangle (3.17,-3.76);
\fill[bsR, rounded corners=1.5pt] (4.29,-0.44) rectangle (4.89,-0.16);
\fill[bsR, rounded corners=1.5pt] (4.29,-1.64) rectangle (4.89,-1.36);
\fill[bsR, rounded corners=1.5pt] (4.29,-2.84) rectangle (4.89,-2.56);
\fill[bsR, rounded corners=1.5pt] (4.29,-4.04) rectangle (4.89,-3.76);
\fill[bsR, rounded corners=1.5pt] (6.01,-0.44) rectangle (6.61,-0.16);
\fill[bsR, rounded corners=1.5pt] (6.01,-1.64) rectangle (6.61,-1.36);
\fill[bsR, rounded corners=1.5pt] (6.01,-2.84) rectangle (6.61,-2.56);
\fill[bsR, rounded corners=1.5pt] (6.01,-4.04) rectangle (6.61,-3.76);
\fill[bsR, rounded corners=1.5pt] (1.71,-1.04) rectangle (2.31,-0.76);
\fill[bsR, rounded corners=1.5pt] (1.71,-2.24) rectangle (2.31,-1.96);
\fill[bsR, rounded corners=1.5pt] (1.71,-3.44) rectangle (2.31,-3.16);
\fill[bsR, rounded corners=1.5pt] (3.43,-1.04) rectangle (4.03,-0.76);
\fill[bsR, rounded corners=1.5pt] (3.43,-2.24) rectangle (4.03,-1.96);
\fill[bsR, rounded corners=1.5pt] (3.43,-3.44) rectangle (4.03,-3.16);
\fill[bsR, rounded corners=1.5pt] (5.15,-1.04) rectangle (5.75,-0.76);
\fill[bsR, rounded corners=1.5pt] (5.15,-2.24) rectangle (5.75,-1.96);
\fill[bsR, rounded corners=1.5pt] (5.15,-3.44) rectangle (5.75,-3.16);
\fill[bsR, rounded corners=1.5pt] (6.87,-1.04) rectangle (7.47,-0.76);
\fill[bsR, rounded corners=1.5pt] (6.87,-2.24) rectangle (7.47,-1.96);
\fill[bsR, rounded corners=1.5pt] (6.87,-3.44) rectangle (7.47,-3.16);
\draw[line width=0.7pt] (0.66,-0.00) -- (0.84,-0.00) to[out=0,in=180] (1.15,-0.17) to[out=0,in=180] (1.46,-0.00) -- (2.56,-0.00) to[out=0,in=180] (2.87,-0.17) to[out=0,in=180] (3.18,-0.00) -- (4.28,-0.00) to[out=0,in=180] (4.59,-0.17) to[out=0,in=180] (4.90,-0.00) -- (6.00,-0.00) to[out=0,in=180] (6.31,-0.17) to[out=0,in=180] (6.62,-0.00) -- (8.0,-0.00);
\draw[line width=0.7pt] (0.66,-0.60) -- (0.84,-0.60) to[out=0,in=180] (1.15,-0.43) to[out=0,in=180] (1.46,-0.60) -- (1.70,-0.60) to[out=0,in=180] (2.01,-0.77) to[out=0,in=180] (2.32,-0.60) -- (2.56,-0.60) to[out=0,in=180] (2.87,-0.43) to[out=0,in=180] (3.18,-0.60) -- (3.42,-0.60) to[out=0,in=180] (3.73,-0.77) to[out=0,in=180] (4.04,-0.60) -- (4.28,-0.60) to[out=0,in=180] (4.59,-0.43) to[out=0,in=180] (4.90,-0.60) -- (5.14,-0.60) to[out=0,in=180] (5.45,-0.77) to[out=0,in=180] (5.76,-0.60) -- (6.00,-0.60) to[out=0,in=180] (6.31,-0.43) to[out=0,in=180] (6.62,-0.60) -- (6.86,-0.60) to[out=0,in=180] (7.17,-0.77) to[out=0,in=180] (7.48,-0.60) -- (8.0,-0.60);
\draw[line width=0.7pt] (0.66,-1.20) -- (0.84,-1.20) to[out=0,in=180] (1.15,-1.37) to[out=0,in=180] (1.46,-1.20) -- (1.70,-1.20) to[out=0,in=180] (2.01,-1.03) to[out=0,in=180] (2.32,-1.20) -- (2.56,-1.20) to[out=0,in=180] (2.87,-1.37) to[out=0,in=180] (3.18,-1.20) -- (3.42,-1.20) to[out=0,in=180] (3.73,-1.03) to[out=0,in=180] (4.04,-1.20) -- (4.28,-1.20) to[out=0,in=180] (4.59,-1.37) to[out=0,in=180] (4.90,-1.20) -- (5.14,-1.20) to[out=0,in=180] (5.45,-1.03) to[out=0,in=180] (5.76,-1.20) -- (6.00,-1.20) to[out=0,in=180] (6.31,-1.37) to[out=0,in=180] (6.62,-1.20) -- (6.86,-1.20) to[out=0,in=180] (7.17,-1.03) to[out=0,in=180] (7.48,-1.20) -- (8.0,-1.20);
\draw[line width=0.7pt] (0.66,-1.80) -- (0.84,-1.80) to[out=0,in=180] (1.15,-1.63) to[out=0,in=180] (1.46,-1.80) -- (1.70,-1.80) to[out=0,in=180] (2.01,-1.97) to[out=0,in=180] (2.32,-1.80) -- (2.56,-1.80) to[out=0,in=180] (2.87,-1.63) to[out=0,in=180] (3.18,-1.80) -- (3.42,-1.80) to[out=0,in=180] (3.73,-1.97) to[out=0,in=180] (4.04,-1.80) -- (4.28,-1.80) to[out=0,in=180] (4.59,-1.63) to[out=0,in=180] (4.90,-1.80) -- (5.14,-1.80) to[out=0,in=180] (5.45,-1.97) to[out=0,in=180] (5.76,-1.80) -- (6.00,-1.80) to[out=0,in=180] (6.31,-1.63) to[out=0,in=180] (6.62,-1.80) -- (6.86,-1.80) to[out=0,in=180] (7.17,-1.97) to[out=0,in=180] (7.48,-1.80) -- (8.0,-1.80);
\draw[line width=0.7pt] (0.66,-2.40) -- (0.84,-2.40) to[out=0,in=180] (1.15,-2.57) to[out=0,in=180] (1.46,-2.40) -- (1.70,-2.40) to[out=0,in=180] (2.01,-2.23) to[out=0,in=180] (2.32,-2.40) -- (2.56,-2.40) to[out=0,in=180] (2.87,-2.57) to[out=0,in=180] (3.18,-2.40) -- (3.42,-2.40) to[out=0,in=180] (3.73,-2.23) to[out=0,in=180] (4.04,-2.40) -- (4.28,-2.40) to[out=0,in=180] (4.59,-2.57) to[out=0,in=180] (4.90,-2.40) -- (5.14,-2.40) to[out=0,in=180] (5.45,-2.23) to[out=0,in=180] (5.76,-2.40) -- (6.00,-2.40) to[out=0,in=180] (6.31,-2.57) to[out=0,in=180] (6.62,-2.40) -- (6.86,-2.40) to[out=0,in=180] (7.17,-2.23) to[out=0,in=180] (7.48,-2.40) -- (8.0,-2.40);
\draw[line width=0.7pt] (0.66,-3.00) -- (0.84,-3.00) to[out=0,in=180] (1.15,-2.83) to[out=0,in=180] (1.46,-3.00) -- (1.70,-3.00) to[out=0,in=180] (2.01,-3.17) to[out=0,in=180] (2.32,-3.00) -- (2.56,-3.00) to[out=0,in=180] (2.87,-2.83) to[out=0,in=180] (3.18,-3.00) -- (3.42,-3.00) to[out=0,in=180] (3.73,-3.17) to[out=0,in=180] (4.04,-3.00) -- (4.28,-3.00) to[out=0,in=180] (4.59,-2.83) to[out=0,in=180] (4.90,-3.00) -- (5.14,-3.00) to[out=0,in=180] (5.45,-3.17) to[out=0,in=180] (5.76,-3.00) -- (6.00,-3.00) to[out=0,in=180] (6.31,-2.83) to[out=0,in=180] (6.62,-3.00) -- (6.86,-3.00) to[out=0,in=180] (7.17,-3.17) to[out=0,in=180] (7.48,-3.00) -- (8.0,-3.00);
\draw[line width=0.7pt] (0.66,-3.60) -- (0.84,-3.60) to[out=0,in=180] (1.15,-3.77) to[out=0,in=180] (1.46,-3.60) -- (1.70,-3.60) to[out=0,in=180] (2.01,-3.43) to[out=0,in=180] (2.32,-3.60) -- (2.56,-3.60) to[out=0,in=180] (2.87,-3.77) to[out=0,in=180] (3.18,-3.60) -- (3.42,-3.60) to[out=0,in=180] (3.73,-3.43) to[out=0,in=180] (4.04,-3.60) -- (4.28,-3.60) to[out=0,in=180] (4.59,-3.77) to[out=0,in=180] (4.90,-3.60) -- (5.14,-3.60) to[out=0,in=180] (5.45,-3.43) to[out=0,in=180] (5.76,-3.60) -- (6.00,-3.60) to[out=0,in=180] (6.31,-3.77) to[out=0,in=180] (6.62,-3.60) -- (6.86,-3.60) to[out=0,in=180] (7.17,-3.43) to[out=0,in=180] (7.48,-3.60) -- (8.0,-3.60);
\draw[line width=0.7pt] (0.66,-4.20) -- (0.84,-4.20) to[out=0,in=180] (1.15,-4.03) to[out=0,in=180] (1.46,-4.20) -- (2.56,-4.20) to[out=0,in=180] (2.87,-4.03) to[out=0,in=180] (3.18,-4.20) -- (4.28,-4.20) to[out=0,in=180] (4.59,-4.03) to[out=0,in=180] (4.90,-4.20) -- (6.00,-4.20) to[out=0,in=180] (6.31,-4.03) to[out=0,in=180] (6.62,-4.20) -- (8.0,-4.20);
\fill[psB] (1.53,-0.09) rectangle (1.62,0.09);
\fill[psB] (1.53,-1.29) rectangle (1.62,-1.11);
\fill[psB] (1.53,-2.49) rectangle (1.62,-2.31);
\fill[psB] (1.53,-3.69) rectangle (1.62,-3.51);
\fill[psB] (3.25,-0.09) rectangle (3.34,0.09);
\fill[psB] (3.25,-1.29) rectangle (3.34,-1.11);
\fill[psB] (3.25,-2.49) rectangle (3.34,-2.31);
\fill[psB] (3.25,-3.69) rectangle (3.34,-3.51);
\fill[psB] (4.97,-0.09) rectangle (5.06,0.09);
\fill[psB] (4.97,-1.29) rectangle (5.06,-1.11);
\fill[psB] (4.97,-2.49) rectangle (5.06,-2.31);
\fill[psB] (4.97,-3.69) rectangle (5.06,-3.51);
\fill[psB] (6.69,-0.09) rectangle (6.78,0.09);
\fill[psB] (6.69,-1.29) rectangle (6.78,-1.11);
\fill[psB] (6.69,-2.49) rectangle (6.78,-2.31);
\fill[psB] (6.69,-3.69) rectangle (6.78,-3.51);
\fill[psB] (2.39,-0.69) rectangle (2.48,-0.51);
\fill[psB] (2.39,-1.89) rectangle (2.48,-1.71);
\fill[psB] (2.39,-3.09) rectangle (2.48,-2.91);
\fill[psB] (4.11,-0.69) rectangle (4.20,-0.51);
\fill[psB] (4.11,-1.89) rectangle (4.20,-1.71);
\fill[psB] (4.11,-3.09) rectangle (4.20,-2.91);
\fill[psB] (5.83,-0.69) rectangle (5.92,-0.51);
\fill[psB] (5.83,-1.89) rectangle (5.92,-1.71);
\fill[psB] (5.83,-3.09) rectangle (5.92,-2.91);
\fill[psB] (7.55,-0.69) rectangle (7.64,-0.51);
\fill[psB] (7.55,-1.89) rectangle (7.64,-1.71);
\fill[psB] (7.55,-3.09) rectangle (7.64,-2.91);
\node at (8.75,0.45) {b)};
% enclosing matrix with diagonal subspace blocks
\draw[line width=0.9pt] (9.4,-0.1) rectangle (13.3,-4.0);
\fill[frV] (9.52,-0.22) rectangle (9.84,-0.54);
\draw[line width=0.7pt] (9.84,-0.54) rectangle (10.6,-1.3);
\draw[line width=0.7pt] (10.6,-1.3) rectangle (11.65,-2.35);
\draw[line width=0.7pt, dash pattern=on 0.8pt off 1.6pt] (11.71,-2.41) -- (11.97,-2.67);
\draw[line width=0.7pt] (12.03,-2.73) rectangle (13.18,-3.88);
\node at (12.6,-3.3) {$\varphi^{m}_{\textcolor{dotG}{n}}(\textcolor{frV}{U})$};
\draw[dotG, line width=1.7pt, -{Stealth[length=2.8mm]}] (9.97,-0.3) .. controls (11.9,-0.35) and (12.9,-1.2) .. (12.62,-2.62);
\node[align=center] at (10.35,-3.35) {n photon\\ input state};
\node at (11.35,-4.45) {$\varphi^{m}_{n} : U(m)\,\rightarrow\,U(\mathcal{H}_{m,n})$};
\end{tikzpicture}%
\endgroup
    \caption{Linear Optical Circuit (a) and equivalent unitary given by the photonic homomorphism (b), see the text.}
    \label{fig:linear_optical_circuit}
\end{figure*}
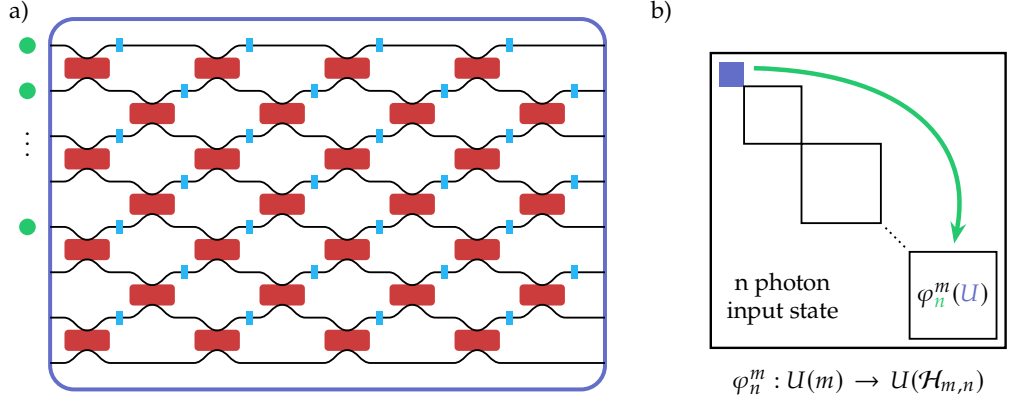

Particle-number preserving circuits such as Linear Optical Circuits \cite{Aaronson2015Complexity} (\autoref{fig:linear_optical_circuit}) have been used in several proposals for
HQNNs architecture. Such circuits fundamentally differ from the qubit based ones, even if photonic qubit architecture can be proposed based on measurement based technique \cite{knill_scheme_2001, bartolucci_fusion_based_2023, aghaee_rad_scaling_2025}, as linear optical circuits with $m$ modes preserve the number of particle $n$ in the computation. In addition, it has been shown \cite{aaronson_computational_2011} that such a linear interferometer is perfectly defined by its equivalent unitary in the subspace of a single particle (in $U(m)$), and its action considering $n$ particles is given by the photonic homomorphism:
\begin{align}
    \varphi_n^m: U(m) \rightarrow U(\mathcal{H}_{m,n}) \\
    \bra{s} \varphi_n^m(U(m)) \ket{t} = \frac{Per[ U_{s,t}(m)]}{\sqrt{s! \,t !}}  
\end{align}
with $\mathcal{H}_{m,n}$ of dimension $\binom{m+n-1}{n}$, $s, t \in \llbracket 0, n \rrbracket^m$ with $\sum_{i=1}^m s_i = \sum_{i=1}^m t_i = n$, and $U_{s,t}(m)$ a $n \times n$ matrix corresponding to the permutation of $U(m)$ as detailed in \cite{aaronson_computational_2011}. \newline

This specific structured has been exploited using linear optical circuit \cite{austin_hybrid_2025, killoran_continuous-variable_2019,
yin_experimental_2025, gan_fock_2022}, and adaptive linear optics \cite{chabaud_quantum_2021, monbroussou_toward_2025, monbroussou_photonic_2025, hoch_quantum_2025} as near-term candidates for HQNNs. In addition, recent proposals \cite{kolarovszki_generative_2026, gottlieb_efficient_2026} have been using linear optical platform due to the proven hardness of sampling from bosonic circuit \cite{aaronson_computational_2011, lund_quantum_2017, deshpande_quantum_2022, brod_complexity_2015}, but also for the potential energetic advantage \cite{soret_quantum_2026}.

\subsubsection{Quantum Reservoir Computing}\label{subsubsec:Quantum_Reservoir_Computing}

Reservoir computing (RC) is a framework based on \gls{RNN} that is well suited for temporal/sequential data processing. The reservoir is fixed, i.e., not trainable, and only the readout is trained with a simple method such as linear regression. The input data is encoded in a high-dimensional space by the reservoir, similarly to kernel methods. Instead of \glspl{RNN}, other nonlinear dynamical systems can be used as reservoirs. In particular, physical RC using reservoirs based on physical phenomena has recently attracted increasing interest in many research areas \cite{tanaka_recent_2019, chmielewski_quantum_2026}.

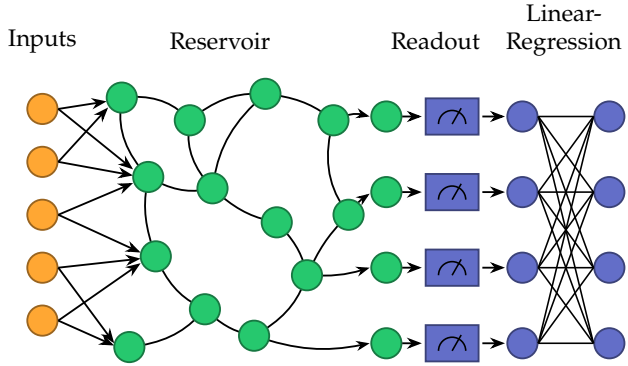
\begin{figure}[t!]
    \centering
    % Quantum reservoir computing (TikZ).
% Single-column layout; classical (linear-regression) nodes use the Fig 1
% classical-NN blue. Body font via rmfamily.
\begingroup
% Palette from Fig 8: orange = inputs, green = reservoir, violet = readout/classical
\definecolor{inO}{RGB}{255,167,51}
\definecolor{resG}{RGB}{38,200,110}
\definecolor{outB}{RGB}{99,110,200}
\definecolor{mtrV}{RGB}{99,110,200}
\tikzset{
  innode/.style={fill=inO, draw=inO!60!black, line width=0.7pt},
  resnode/.style={fill=resG, draw=resG!60!black, line width=0.7pt},
  bluenode/.style={fill=outB, draw=outB!60!black, line width=0.7pt},
  sig/.style={black, line width=0.7pt, -{Stealth[length=2mm]}},
  res/.style={black, line width=0.7pt, -{Stealth[length=1.8mm]}},
  mtr/.style={draw=mtrV!60!black, line width=0.7pt, fill=mtrV, minimum width=0.7cm,
              minimum height=0.48cm, inner sep=1pt},
}
\begin{tikzpicture}[x=1cm,y=1cm, every node/.style={font=\rmfamily}]
% headers
\node at (0.55,3.3) {Inputs};
\node at (2.9,3.3)  {Reservoir};
\node at (5.75,3.3) {Readout};
\node[align=center] at (7.45,3.45) {Linear-\\Regression};
% reservoir internal edges (curved, drawn first)
\draw[res] (1.6,2.55) to[bend left=25] (2.5,2.25);
\draw[res] (2.5,2.25) to[bend left=20] (3.5,2.6);
\draw[res] (3.5,2.6) to[bend left=15] (4.4,2.25);
\draw[res] (1.95,1.5) to[bend right=20] (2.8,1.35);
\draw[res] (2.8,1.35) to[bend left=25] (2.5,2.25);
\draw[res] (2.8,1.35) to[bend right=15] (3.65,0.9);
\draw[res] (2.05,0.45) to[bend right=20] (2.7,-0.25);
\draw[res] (2.7,-0.25) to[bend right=15] (3.35,-0.6);
\draw[res] (3.35,-0.6) to[bend right=20] (4.05,0.2);
\draw[res] (3.65,0.9) to[bend left=15] (4.05,0.2);
\draw[res] (4.05,0.2) to[bend left=10] (4.6,1.0);
\draw[res] (1.95,1.5) to[bend left=25] (1.6,2.55);
\draw[res] (2.05,0.45) to[bend left=15] (1.95,1.5);
\draw[res] (1.7,-0.75) to[bend right=20] (2.7,-0.25);
\draw[res] (3.5,2.6) to[bend right=20] (2.8,1.35);
\draw[res] (4.4,2.25) to[bend left=20] (4.88,2.3);
\draw[res] (4.6,1.0) to[bend left=10] (4.88,1.28);
\draw[res] (4.05,0.2) to[bend right=10] (4.88,0.3);
\draw[res] (3.35,-0.6) to[bend right=15] (4.88,-0.7);
\draw[res] (4.4,2.25) to[bend right=25] (4.6,1.0);
% inputs -> reservoir
\foreach \y in {2.4,1.7,1.0,0.3,-0.4}{
  \node[innode, circle, minimum size=0.4cm] at (0.55,\y) {}; }
\draw[sig] (0.76,2.4) -- (1.4,2.5);
\draw[sig] (0.76,2.4) -- (1.75,1.6);
\draw[sig] (0.76,1.7) -- (1.42,2.42);
\draw[sig] (0.76,1.7) -- (1.75,1.52);
\draw[sig] (0.76,1.0) -- (1.75,1.42);
\draw[sig] (0.76,1.0) -- (1.85,0.5);
\draw[sig] (0.76,0.3) -- (1.85,0.42);
\draw[sig] (0.76,0.3) -- (1.5,-0.7);
\draw[sig] (0.76,-0.4) -- (1.5,-0.73);
\draw[sig] (0.76,-0.4) -- (1.85,0.36);
% reservoir nodes
\foreach \x/\y in {1.6/2.55, 2.5/2.25, 3.5/2.6, 1.95/1.5, 2.8/1.35,
                   2.05/0.45, 2.7/-0.25, 3.65/0.9, 3.35/-0.6, 4.4/2.25,
                   4.05/0.2, 1.7/-0.75, 4.6/1.0}{
  \node[resnode, circle, minimum size=0.4cm] at (\x,\y) {}; }
% read-out chain: output nodes -> meters -> blue nodes
\foreach \y in {2.3,1.3,0.3,-0.7}{
  \node[resnode, circle, minimum size=0.4cm] at (5.1,\y) {};
  \draw[sig] (5.31,\y) -- (5.57,\y);
  \node[mtr] at (5.97,\y) {};
  \draw[black, line width=0.6pt] (5.79,\y-0.1) arc[start angle=180, end angle=0, radius=0.18];
  \draw[black, line width=0.6pt] (5.97,\y-0.1) -- (6.09,\y+0.11);
  \draw[sig] (6.37,\y) -- (6.67,\y);
  \node[bluenode, circle, minimum size=0.4cm] at (6.9,\y) {};
  \node[bluenode, circle, minimum size=0.4cm] at (8.05,\y) {};
}
% full bipartite linear regression
\foreach \ya in {2.3,1.3,0.3,-0.7}{ \foreach \yb in {2.3,1.3,0.3,-0.7}{
  \draw[black, line width=0.6pt] (7.11,\ya) -- (7.84,\yb); }}
\end{tikzpicture}%
\endgroup
    \caption{Principles of quantum reservoir computing. The reservoir (green) are measured , and a classical neural networks output the model.}
    \label{fig:QRC}
\end{figure}

Quantum reservoir computing (\autoref{fig:QRC}) is a quantum analogue of classical reservoir computing \cite{jaeger_harnessing_2004, maass_real-time_2002} in which an input signal drives a fixed high-dimensional dynamical system, while only the final readout layer is trained \cite{fujii_harnessing_2017, nakajima_boosting_2019, kutvonen_optimizing_2020}. This makes learning substantially simpler than in fully trainable recurrent models and is especially attractive for near-term quantum hardware. The exponentially large Hilbert space also acts as a temporal memory: because the input subsystem is refreshed at each step while the remaining degrees of freedom evolve unitarily, past inputs persist in the reservoir state and can be recovered by the readout, with the achievable short-term memory and nonlinearity quantified by the information processing capacity \cite{Mart_nez_Pe_a_2020, suzuki_natural_2022}. Proposals have been made using quantum oscillators \cite{dudas2023quantum, govia_quantum_2021}, on super-conducting qubits \cite{suzuki_natural_2022}, Rydberg atoms \cite{bravo_quantum_2022}

\subsubsection{Linear combination of unitaries}

\Gls{LCU} is the standard primitive for realizing non-unitary linear maps on a quantum processor, expressing a target operator $A$ as a linear combination of elementary unitaries $U_j$ with complex coefficients $\alpha_j$~\cite{childs2012hamiltonian, berry2015simulating}:

\begin{equation}
A = \sum_{j=1}^{M} \alpha_j U_j,
\end{equation}

To realize this operator, the circuit introduces an ancillary register to mediate a controlled unitary operation. Specifically, the ancilla is first prepared in a state with amplitudes proportional to $\sqrt{\alpha_j}$. This ancilla then controls the application of the respective unitaries $U_j$ on the system register. After reversing the state preparation, post-selecting on the all-zero outcome of the ancilla register implements a block-encoding of $A/\alpha$. This succeeds with probability $\|A\ket{\psi}\|^2 / \alpha^2$, in which case the input $\ket{\psi}$ is left on the system register in the normalized state $A\ket{\psi}/\|A\ket{\psi}\|$, and amplitude amplification can reduce the number of repetitions needed to reach this successful outcome to $\mathcal{O}(\alpha)$ queries of the underlying subroutines~\cite{gilyen2019quantum, low2019hamiltonian}. The construction is most valuable when the individual unitaries $U_j$ correspond to hardware-native gates or admit efficient circuit representations, so that their weighted superposition captures linear transformations that a single variational unitary cannot realize directly.

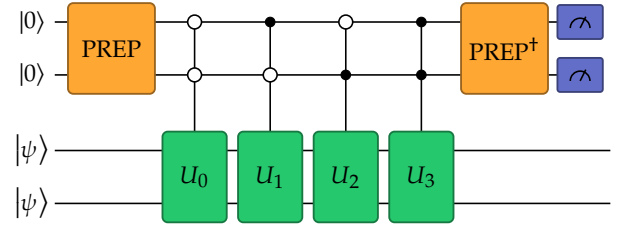
\begin{figure}[htb]
  \centering
  % Fig 13: LCU circuit (TikZ redraw of the former yquant version), drawn in the
% style/palette of Figs 1, 4, 8: orange = state preparation (PREP), green =
% unitaries, violet = measurement. Filled dot = control on |1>, open = on |0>.
\begingroup
\definecolor{prepC}{RGB}{255,167,51}
\definecolor{uC}{RGB}{38,200,110}
\definecolor{mC}{RGB}{99,110,200}
\tikzset{
  qwire/.style={line width=0.6pt},
  prepg/.style={draw=prepC!60!black, line width=0.7pt, fill=prepC, rounded corners=2pt,
                minimum width=1.15cm, minimum height=1.2cm, inner sep=1.5pt},
  ug/.style={draw=uC!60!black, line width=0.7pt, fill=uC, rounded corners=2pt,
             minimum width=0.85cm, minimum height=1.2cm, inner sep=1.5pt},
  mtr/.style={draw=mC!60!black, line width=0.7pt, fill=mC, rounded corners=1pt,
              minimum width=0.6cm, minimum height=0.45cm, inner sep=1pt},
}
\begin{tikzpicture}[x=1cm,y=1cm, every node/.style={font=\rmfamily}]
% wires: two ancillas (top), two system qubits (bottom)
\foreach \y in {0,-0.7}{
  \node at (0.2,\y) {$\ket{0}$};
  \draw[qwire] (0.5,\y) -- (7.45,\y);
}
\foreach \y in {-1.7,-2.4}{
  \node at (0.2,\y) {$\ket{\psi}$};
  \draw[qwire] (0.5,\y) -- (7.85,\y);
}
% control lines (drawn first, controls and boxes go on top)
\foreach \x in {2.35,3.35,4.35,5.35}{
  \draw[qwire] (\x,0) -- (\x,-1.45);
}
% PREP on the ancillas
\node[prepg] at (1.25,-0.35) {PREP};
% SELECT: U_i on the system, controlled on the ancilla state |i>
\node[ug] at (2.35,-2.05) {$U_0$};
\node[ug] at (3.35,-2.05) {$U_1$};
\node[ug] at (4.35,-2.05) {$U_2$};
\node[ug] at (5.35,-2.05) {$U_3$};
% controls: open circle = condition on |0>, filled dot = condition on |1>
% U_0 fires on |00>, U_1 on |01> (a0=1), U_2 on |10> (a1=1), U_3 on |11>
\foreach \x in {2.35,4.35}{ \draw[qwire, fill=white] (\x,0) circle (2.6pt); }
\foreach \x in {3.35,5.35}{ \filldraw (\x,0) circle (1.8pt); }
\foreach \x in {2.35,3.35}{ \draw[qwire, fill=white] (\x,-0.7) circle (2.6pt); }
\foreach \x in {4.35,5.35}{ \filldraw (\x,-0.7) circle (1.8pt); }
% PREP^dagger on the ancillas
\node[prepg] at (6.45,-0.35) {PREP$^{\dagger}$};
% measure ancillas; post-select on the all-zeros outcome
\foreach \y in {0,-0.7}{
  \node[mtr] at (7.45,\y) {};
  \draw[black, line width=0.6pt] (7.32,\y-0.08) arc[start angle=180, end angle=0, radius=0.13];
  \draw[black, line width=0.6pt] (7.45,\y-0.08) -- (7.525,\y+0.07);
}
\end{tikzpicture}%
\endgroup
\caption{Quantum circuit implementing a linear combination of unitaries (LCU): the PREP block loads the coefficients onto the ancilla qubits, the SELECT block applies each unitary conditioned on the ancilla state, and the inverse PREP followed by post-selecting the all-zeros measurement applies the target operator to the system register.}
  \label{fig:lcu}
\end{figure}

The practical realization of LCU within \gls{VQC}s follows the PREP-SELECT-PREP$^\dagger$ pattern: an ancilla register is prepared (PREP) in a superposition that loads the normalized coefficients, a controlled block (SELECT) applies $U_j$ when the ancilla is in state $\ket{j}$, and a final PREP$^\dagger$ followed by projection onto the all-zero outcome of the ancilla register recovers the action of $A/\alpha$~\cite{childs2012hamiltonian} as shown in \autoref{fig:lcu}. The same template underpins Hamiltonian simulation, quantum linear algebra subroutines, and the quantum singular value transformation, where matrix functions and non-unitary operators arise naturally~\cite{harrow_quantum_2009, berry2015simulating, gilyen2019quantum}.

In an HQNN setting the input data are first encoded into the system state $\ket{\psi(x)}$, and the PREP circuit that loads the coefficients $\alpha_j$ becomes the trainable component while the basis $\{U_j\}$ stays fixed, so the layer realizes a parameterized and generally non-unitary operator $A_{\theta} = \sum_j \alpha_j(\theta)\,U_j$ whose coefficients and relative phases are the learnable parameters. Running PREP, SELECT and PREP$^\dagger$ leaves the system and ancilla in $\ket{0}_{\mathrm{anc}} \otimes (A_{\theta}/\alpha)\ket{\psi(x)} + \ket{\Phi^{\perp}}$, where $\ket{\Phi^{\perp}}$ has no overlap with the all-zero ancilla, so post-selecting on that outcome collapses the system to the normalized state $A_{\theta}\ket{\psi(x)}/\lVert A_{\theta}\ket{\psi(x)}\rVert$, which is the actual output of the layer. A prediction is then read out as the expectation value of a chosen observable $O$ on this conditional state, 

\begin{equation}
    f(x,\theta) = \frac{\langle\psi(x)|A_{\theta}^{\dagger} O A_{\theta}|\psi(x)\rangle}{\langle\psi(x)|A_{\theta}^{\dagger} A_{\theta}|\psi(x)\rangle},
\end{equation} 
so that training adjusts $\theta$ to shape the non-unitary map $A_{\theta}$ applied to the encoded input. The state-dependent normalization in this expression, which arises precisely because the output is conditioned on successful post-selection, which introduces a state-dependent nonlinearity of a kind that a single unitary layer cannot produce through unitary evolution alone, and it is this learnable non-unitary map, rather than the block-encoding seen as a fixed subroutine, that gives LCU its value as a network layer.

Such architectures have been proposed as a route toward improved expressivity with comparatively few trainable parameters, at the cost of increased ancilla overhead and post-selection inefficiency. In practice, LCU-based layers remain a largely proof-of-concept tool in the near-term \gls{HQNN} literature, since every added unitary increases the complexity of the SELECT and PREP circuits and may increase the ancilla requirements logarithmically with the number of terms, and may increase the coefficient $1$-norm $\alpha$ depending on the chosen decomposition, thereby reducing the post-selection success probability. The approach is therefore best suited to problem classes with a natural operator decomposition or a concrete need for fine-grained control over non-unitary transformations.

\subsubsection{Quantum Generative models}\label{subsec:QGAN}

Quantum generative learning models have been proposed as an quantum extension or alternative of classical generative learning models \cite{tian_recent_2023}. The goal is to simulate the process that produced the data of interest, and to reproduce its distribution, allowing for example to generate new samples. The first proposal is \textbf{quantum Boltzmann machines} \cite{amin_quantum_2018}, inspired by the use of Boltzmann distribution in classical machine learning \cite{cheng_information_2018}, based on quantum Boltzmann distribution of a quantum Hamiltonian. Empirical results suggest that, under certain conditions, quantum annealing hardware samples from a Gibbs or a Boltzmann distribution \cite{Benedetti2018Unsupervised, amin_quantum_2018, benedetti_estimation_2016, amin_searching_2015} which can be useful for machine learning purposes.

In quantum circuit Born machines \cite{liu_differentiable_2018, coyle_born_2020, benedetti_generative_2019}, one trains a parameterized quantum circuit so that its output probability distribution matches a distribution of interest. In particular, recent works have focus on the use of quantum circuits for generative models in regimes that are classically trainable, but with a quantum deployment for inference. In particular, studies have used the hardness of sampling from random bosonic \cite{kolarovszki_generative_2026, gottlieb_efficient_2026}, fermionic \cite{bako_fermionic_2025, kerenidis_scalable_2026}, or \gls{IQP} \cite{kasture_protocols_2023, recio-armengol_train_2025} circuit distributions. The hardness is usually only proven for random circuit, and is not for pretrained ansatz, but it motivates the use of such circuits from a complexity perspective. While the consideration of an advantage through quantum sampling for the deployment of the models is of particular interest regarding the usual trade-off between classical simulation and trainability of quantum models (see \autoref{subsec:trainability_QNN} and \cite{cerezo_does_2025}), recent results have shown \cite{recio-armengol_train_2025, shen_characterizing_2026} that training \gls{IQP} circuit is possible without \gls{BP} and direct de-quantization technique.

\Glspl{QGAN} are the adversarial instance of the \gls{HQNN} design space introduced above, in which one
or both networks of a classical
\gls{GAN}~\cite{goodfellow2014gan, creswell_generative_2018} are replaced by a variational quantum circuit~\cite{lloyd2018quantum, dallaire2018quantum}. The generator learns to produce synthetic samples while the discriminator learns to tell them apart from real data, and the two are trained against one another in a minimax game.

In the most general hybrid setting, a quantum generator $G_\theta$
parameterized by quantum circuit angles $\theta$ maps a latent random
variable $z \sim p_z$ to a generated data distribution either through
direct measurement of its output state or through the expectation values
of observables, while a classical or quantum discriminator $D_\phi$
with parameters $\phi$ outputs the probability that a given sample
originated from the true data distribution rather than the generator.
A common training objective follows the standard \gls{GAN} formulation~\cite{goodfellow2014gan}:
 
\begin{equation}
\begin{split}
    \min_\theta \max_\phi V(D, G) = & \,\mathbb{E}_{x \sim p_{\text{data}}(x)}[\log D_\phi(x)] \\
     & + \mathbb{E}_{z \sim p_z(z)}[\log(1 - D_\phi(G_\theta(z)))],
\end{split}
\end{equation}
 
where $p_{\text{data}}$ denotes the true data distribution.
Alternatively, the generator may be classical while the discriminator is
quantum, or both components may be quantum circuits acting on different
registers. The choice of configuration depends on the specific problem
structure, available quantum resources, and the desired inductive biases.

QGANs have a handful of properties that set them apart from generic
variational architectures. First, recent work has identified classes of QGAN cost functions and circuit architectures that can provably avoid barren plateaus under certain realistic assumptions, which is a notable exception to the vanishing-gradient issues discussed in \autoref{subsubsec:BP_QNN}~\cite{letcher_tight_2024,
Cerezo_2021_cost_BP}. Second, quantum generators can in principle represent distributions associated with quantum circuits that are conjectured to be classically hard to sample from, especially those with nontrivial entanglement or arising from quantum data~\cite{arute2019quantum}, while quantum discriminators
can in principle exploit the exponentially large Hilbert space for
feature extraction, though concrete task-specific advantages are still
under active investigation. Gradient updates for the quantum components
follow the standard parameter-shift rule introduced in
\autoref{subsec:trainability_QNN}, and the hybrid structure makes
QGANs natural candidates for the \gls{NISQ} era, since most of the
heavy computation can be offloaded to classical components. Reported
applications span distribution loading and quantum state
preparation~\cite{zoufal2019quantum}, anomaly detection on classical
datasets, and small-scale generative modeling~\cite{Benedetti2018Unsupervised}, showing a path toward using quantum processors for unsupervised and semi-supervised tasks beyond supervised classification, albeit mostly at proof-of-concept scale so far.

\subsubsection{Fault-tolerant quantum machine learning}\label{subsubsec:FTQC_QML}

The paradigms above are tailored to near-term hardware, curtailing circuit depth or expressivity to remain trainable under noise. A separate paradigm instead assumes fault tolerance, where logical error rates are low enough to execute the deep, coherent circuits that those constraints preclude. The quantum layer is then not a shallow trained ansatz but an algorithmic subroutine with a provable complexity, drawn from the quantum linear-algebra primitives that predate the variational era: the quantum linear-systems algorithm~\cite{harrow_quantum_2009}, quantum principal component analysis~\cite{Lloyd2014QuantumUnsupervised}, quantum support vector machines~\cite{rebentrost_quantum_2014}, and quantum recommendation systems~\cite{kerenidis_recommendation_2017}. Many are unified by the quantum singular value transformation that also underlies the \gls{LCU} layer above~\cite{gilyen2019quantum} and are accelerated by quantum amplitude estimation~\cite{brassard_quantum_2002}. These routines act on the linear-algebraic core of learning, such as matrix inversion, eigendecomposition, and inner-product estimation, and offer up to exponential speedups in the input dimension, conditioned on coherent data access through a quantum random-access memory~\cite{giovannetti_quantum_2008}.

Two facts qualify this paradigm. First, the speedups rest on strong assumptions, including efficient state preparation, well-conditioned operators, and a readout limited to summary statistics; several have since been matched by quantum-inspired classical algorithms under analogous sampling access~\cite{tang_quantum_2021, chia_sampling-based_2020}, so a genuine separation is expected only for inputs with the right structure (\autoref{subsec:Dequantization}). Second, executing these deep circuits on logical qubits incurs the deterministic synthesis and magic-state cost of \autoref{subsec:FTQC_challenges}, which the variational paradigms largely avoid. Within an \gls{HQNN}, such a subroutine therefore enters as a fixed, provably efficient block rather than a trained layer (\autoref{subsec:circuits_pipeline}), and the guarantees on the model follow from algorithmic complexity rather than from empirical tuning.

\section{Implementation on Classical and Quantum Hardware and HQNN Stack}
\label{sec:implementation}

Turning an architecture from \autoref{sec:architectures} into a running experiment requires a software and hardware stack that spans several layers of abstraction: a hybrid program, a classical or quantum execution backend, and the heterogeneous devices underneath. This section maps that stack and reviews the practical trade-offs an \gls{HQNN} practitioner faces today. \autoref{sec:5_stack} introduces the stack as a whole. \autoref{sec:5_classical} surveys the differentiable software frameworks that host the majority of current \gls{HQNN} results by simulating the quantum layer on classical hardware. \autoref{sec:5_quantum} reviews the leading quantum hardware platforms, the regimes in which each is competitive, and the operational trade-offs that shape platform choice.

\subsection{The HQNN software stack}
\label{sec:5_stack}

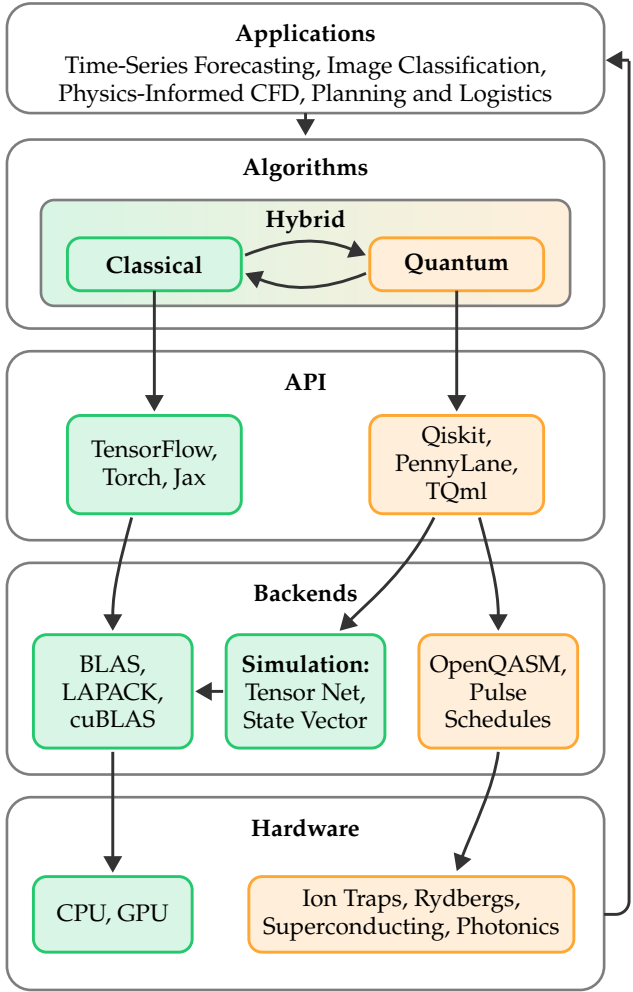
\begin{figure}[htbp!]
    \centering
    % HQNN software stack (TikZ).
% Body-font text (rmfamily, 10pt, upright), uniform box sizes per layer,
% consistent borders, triangular arrow tips. Width <= 8.33cm (single column).
\begingroup
% Palette from Fig 8: classical = green, quantum = orange (matches Fig 8 semantics)
\definecolor{layG}{RGB}{125,125,125}
\definecolor{clO}{RGB}{38,200,110}
\definecolor{clOf}{RGB}{216,245,229}
\definecolor{qnG}{RGB}{255,167,51}
\definecolor{qnGf}{RGB}{255,239,218}
\tikzset{
  layer/.style={draw=layG, line width=1.0pt, rounded corners=9pt, fill=white},
  obox/.style={draw=clO, line width=1.1pt, rounded corners=5pt, fill=clOf,
               align=center, inner sep=3pt, font=\rmfamily},
  gbox/.style={draw=qnG, line width=1.1pt, rounded corners=5pt, fill=qnGf,
               align=center, inner sep=3pt, font=\rmfamily},
  flow/.style={black!80, line width=1.1pt, -{Triangle[length=2.2mm,width=2.4mm]}},
}
\begin{tikzpicture}[x=1cm,y=1cm, every node/.style={font=\rmfamily}]
% ---------------------------------------------------------------- Applications
\draw[layer] (0.05,11.3) rectangle (7.95,12.75);
\node[font=\rmfamily\bfseries] at (4.0,12.33) {Applications};
\node[align=center, font=\rmfamily\small] at (4.0,11.72) {Time-Series Forecasting, Image Classification,\\ Physics-Informed CFD, Planning and Logistics};
\draw[flow] (4.0,11.3) -- (4.0,11.0);
% ---------------------------------------------------------------- Algorithms
\draw[layer] (0.05,8.45) rectangle (7.95,10.95);
\node[font=\rmfamily\bfseries] at (4.0,10.55) {Algorithms};
\shade[left color=clOf, right color=qnGf, rounded corners=5pt, draw=layG,
       line width=1.0pt] (0.5,8.75) rectangle (7.5,10.15);
\node[font=\rmfamily\bfseries] at (4.0,9.87) {Hybrid};
\node[obox, minimum width=2.3cm, minimum height=0.7cm, font=\rmfamily\bfseries]
  at (2.0,9.3) {Classical};
\node[gbox, minimum width=2.3cm, minimum height=0.7cm, font=\rmfamily\bfseries]
  at (6.0,9.3) {Quantum};
\draw[flow] (3.2,9.42) to[bend left=22] (4.8,9.42);
\draw[flow] (4.8,9.18) to[bend left=22] (3.2,9.18);
% ---------------------------------------------------------------- API
\draw[layer] (0.05,5.65) rectangle (7.95,8.15);
\node[font=\rmfamily\bfseries] at (4.0,7.75) {API};
\node[obox, minimum width=2.3cm, minimum height=1.3cm]
  at (2.0,6.65) {TensorFlow,\\Torch, Jax};
\node[gbox, minimum width=2.3cm, minimum height=1.3cm]
  at (6.0,6.65) {Qiskit,\\PennyLane,\\TQml};
\draw[flow] (2.0,8.95) -- (2.0,7.35);
\draw[flow] (6.0,8.95) -- (6.0,7.35);
% ---------------------------------------------------------------- Backends
\draw[layer] (0.05,2.55) rectangle (7.95,5.35);
\node[font=\rmfamily\bfseries] at (4.0,4.95) {Backends};
\node[obox, minimum width=2.1cm, minimum height=1.5cm]
  at (1.45,3.65) {BLAS,\\LAPACK,\\cuBLAS};
\node[obox, minimum width=2.1cm, minimum height=1.5cm]
  at (4.0,3.65) {{\bfseries Simulation:}\\Tensor Net,\\State Vector};
\node[gbox, minimum width=2.1cm, minimum height=1.5cm]
  at (6.55,3.65) {OpenQASM,\\Pulse\\Schedules};
\draw[flow] (1.7,5.95) .. controls (1.55,5.4) and (1.45,5.0) .. (1.45,4.45);
\draw[flow] (5.7,5.95) .. controls (5.35,5.25) and (4.85,4.85) .. (4.45,4.45);
\draw[flow] (6.3,5.95) .. controls (6.45,5.4) and (6.55,5.0) .. (6.55,4.45);
\draw[flow] (2.91,3.65) -- (2.54,3.65);
% ---------------------------------------------------------------- Hardware
\draw[layer] (0.05,-0.3) rectangle (7.95,2.25);
\node[font=\rmfamily\bfseries] at (4.0,1.85) {Hardware};
\node[obox, minimum width=2.1cm, minimum height=1.0cm]
  at (1.45,0.7) {CPU, GPU};
\node[gbox, minimum width=4.3cm, minimum height=1.0cm]
  at (5.4,0.7) {Ion Traps, Rydbergs,\\Superconducting, Photonics};
\draw[flow] (1.45,2.85) -- (1.45,1.25);
\draw[flow] (6.55,2.85) .. controls (6.5,2.4) and (6.2,1.8) .. (6.0,1.27);
% ---------------------------------------------------------------- feedback loop
\draw[black!80, line width=1.1pt, rounded corners=5pt] (7.95,0.7) -- (8.28,0.7) -- (8.28,11.99);
\draw[flow] (8.28,12.0) -- (7.97,12.0);
\end{tikzpicture}%
\endgroup
    \caption{Layered architecture of the \gls{HQNN} software stack. \textbf{Applications} drive \textbf{hybrid algorithms} whose classical and quantum components exchange information through a shared training loop. The \textbf{API} layer exposes this loop to developers through classical deep-learning frameworks (PyTorch, TensorFlow, JAX) and quantum-circuit frameworks (Qiskit, PennyLane, TQml). \textbf{Backends} compile programs to executable form: linear-algebra libraries (BLAS, LAPACK, cuBLAS) on the classical side; state-vector and tensor-network simulators for classical simulation of quantum circuits; and OpenQASM or pulse schedules for quantum processors. The bottom \textbf{Hardware} layer comprises CPUs and GPUs for classical work and a range of \gls{QPU} modalities for quantum work.}
    \label{fig:hqnn_stack}
\end{figure}

\autoref{fig:hqnn_stack} organizes the tools and devices that a practitioner must navigate along five layers. \textbf{Applications} describe the problem domain the pipeline targets. \textbf{Algorithms} compose classical and quantum subroutines into a single hybrid program: the coupling between classical and quantum subroutines gives the stack its hybrid character and is what the lower layers must support efficiently. \textbf{APIs} are the user-facing abstractions. Quantum frameworks express the quantum layer, and where that layer is embedded in a classical surrounding model, classical frameworks define that part and drive its parameter optimization if needed. When such a classical component is present and the model is trained, the two interact through a shared optimization contract. A practical complication shapes the layer immediately below: running non-trivial circuits on today's \glspl{QPU} is constrained by noise, decoherence, limited qubit counts, and cloud-access latency and cost. For the vast majority of current \gls{HQNN} work, classical simulation of the quantum layer on CPUs or GPUs is therefore the current default execution path during training, with on-hardware execution reserved for end-to-end benchmarks and final validation. \textbf{Backends} translate the API abstractions into executable objects that support both routes: differentiable classical simulators running on CPUs and GPUs on the classical side, and OpenQASM programs and pulse schedules on the quantum side. \textbf{Hardware} is heterogeneous by design: CPUs and GPUs cover classical workloads and, for most published \gls{HQNN} results, also the simulated quantum layer; \glspl{QPU} come in several physical implementations reviewed in \autoref{sec:5_quantum}. A key practical value of the decomposition is modularity: a simulator, a device, or a compilation pass can be swapped without rewriting the model itself, provided the API contract is respected.

\subsection{Classical execution: differentiable simulation frameworks for HQNNs}
\label{sec:5_classical}

A large fraction of the \gls{HQNN} benchmark results reported in \autoref{sec:empirical} are obtained by simulating the quantum layer on classical hardware rather than executing it on a \gls{QPU}. Simulation offers three practical advantages that together explain its present dominance: gradient flow through the quantum layer is native, classical auto-differentiation composes with the surrounding deep-learning model, and iteration is cheap at the qubit counts typical of current architectures ($\lesssim 30$). The maturity of open-source libraries that realize this workflow is now a defining feature of the field; \autoref{tab:hqnn_frameworks} summarizes the main options reviewed below.

\textbf{PennyLane}~\cite{bergholm_pennylane_2020} is the most widely used entry point. It provides differentiable quantum nodes with PyTorch, TensorFlow, and JAX interfaces and supports analytic parameter-shift gradients~\cite{Mitarai_2018_QClearning, wierichs_general_2022}, adjoint differentiation, and back-propagation modes. Two sides of the same trade-off appear in practice: the reference \texttt{default.qubit} backend composes cleanly with Torch/JAX auto-grad, supports back-propagation, and vectorizes over input batches but becomes slow at higher qubit counts, while the C\texttt{++}-accelerated \texttt{lightning.qubit} and \texttt{lightning.gpu} backends reach much higher per-circuit throughput but neither implement back-propagation (they rely on adjoint differentiation) nor vectorize over a batch of inputs, instead looping over parameter sets internally. The canonical \gls{ML} pattern of vectorized mini-batch training is therefore awkward to express with these fast backends.

\textbf{TensorCircuit}~\cite{zhang_tensorcircuit_2023} provides an \emph{exact} general-purpose simulator: it represents ``circuit + observable'' as a tensor network and evaluates the final expectation value by searching for a cheap contraction order, without approximation. An MPS backend is always available alongside and approximates the state; it is practical when the circuit has low entanglement. Just-in-time compilation over JAX or TensorFlow and first-class GPU support make it well suited to mid-range \gls{HQNN} training runs.

\textbf{TorchQuantum}~\cite{wang_torchquantum_2022} integrates quantum layers as native PyTorch \texttt{nn.Module} objects so that an \gls{HQNN} can be assembled and trained without a framework boundary between classical and quantum components. \textbf{TensorFlow Quantum}~\cite{broughton_tfq_2020} bridges Cirq~\cite{cirq_developers_cirq_2024} and TensorFlow and was an early influence on the design of later libraries, though active development has slowed. \textbf{Qiskit Machine Learning}~\cite{javadi_qiskit_2024} provides PyTorch connectors on top of Qiskit's primitives interface and shares a compilation toolchain with IBM hardware. \textbf{NVIDIA cuQuantum}~\cite{bayraktar_cuquantum_2023} accelerates state-vector and tensor-network simulation on GPUs and is increasingly adopted as a backend by the higher-level libraries above. Sitting one layer up on the same stack, \textbf{NVIDIA CUDA-Q}~\cite{the_cuda-q_development_team_cuda-q_2025} is a kernel-based programming model and compiler for hybrid quantum-classical programs: it executes GPU-accelerated simulation (via cuQuantum) and physical \glspl{QPU} through a common interface and supports parameter-shift gradients for variational workloads, making it a vehicle for the hybrid execution path itself rather than a drop-in autodiff layer for the deep-learning frameworks above.

\textbf{Photonic and continuous-variable simulators} form a separate branch, because circuits on photonic hardware are not naturally expressed in the qubit gate model. Differentiability is not uniform across this branch. \textbf{Strawberry Fields}~\cite{killoran_continuous-variable_2019} was an early demonstration that photonic quantum layers can be trained by back-propagation, through its TensorFlow-backed Fock-space simulator, and \textbf{Piquasso}~\cite{kolarovszki_piquasso_2025} likewise supports gradient-based training via TensorFlow and JAX backends across Gaussian and Fock representations. \textbf{Perceval}~\cite{heurtel_perceval_2023}, by contrast, targets discrete-variable linear-optical circuits and is built around forward simulation with gradient-free parameter optimization, so end-to-end back-propagation relies on a separate companion layer rather than the core library. A recurring bottleneck is the cost of evaluating output probabilities: exact simulation of Gaussian boson sampling reduces to computing hafnians, for which dedicated high-performance routines such as \textbf{The Walrus}~\cite{gupt_walrus_2019} are required, and the best known classical algorithms scale exponentially in the photon number~\cite{quesada_exact_2020, clifford_classical_2018}. Photonic simulation therefore remains practical only at modest mode and photon counts, which constrains the scale of photonic \glspl{HQNN} that can be trained classically today.

\textbf{TQml Simulator}~\cite{kuzmin_tqml_2025} specializes in the layered ansatzes typical of \gls{HQNN} applications rather than arbitrary-circuit coverage. Its simulator is \emph{exact} and, for each layer of gates, selects among several mathematically equivalent evaluation strategies, choosing whichever is cheapest for the gate and register size at hand. Running on a single CPU thread in double precision, this layer-wise specialization yields up to an order-of-magnitude speed-up over a general-purpose reference simulator (PennyLane's \texttt{default.qubit}) on the ansatz families it targets, while remaining differentiable by standard back-propagation.

A cross-cutting theme across these libraries is \gls{JIT} compilation of the entire hybrid program, not only the quantum circuit, to remove framework-boundary overhead that would otherwise dominate short training steps. PennyLane's Catalyst, TensorCircuit's JIT mode, and Qiboml~\cite{robbiati_qiboml_2025} provide this capability through their respective autodiff tracing mechanisms.

\begin{table*}[htbp!]
\centering
\caption{Differentiable software frameworks for building and training \glspl{HQNN}. ``Interface'' lists the classical deep-learning frameworks whose auto-grad the library integrates with. ``Gradient'' records whether analytic parameter-shift (PSR), back-propagation through the simulator, or both are supported for quantum gradients. ``Simulation'' lists the main classical backends shipped with the library: state-vector (SV) and tensor-network (TN). Hardware-backend lists are representative, not exhaustive.}
\label{tab:hqnn_frameworks}
\resizebox{\linewidth}{!}{ 
\begin{tabular}{@{}lllll@{}}
\toprule
Framework & Interface & Gradient & Simulation & Hardware backends \\
\midrule
PennyLane~\cite{bergholm_pennylane_2020}          & PyTorch, TF, JAX  & PSR, backprop, adjoint$^{\dagger}$ & SV, TN                & IBM, IonQ, Rigetti, Xanadu, \ldots \\
TensorCircuit~\cite{zhang_tensorcircuit_2023}     & JAX, TF, PyTorch  & backprop               & SV, TN (exact + MPS)  & plug-ins for major cloud QPUs \\
TorchQuantum~\cite{wang_torchquantum_2022}        & PyTorch           & backprop, PSR          & SV                    & IBM via Qiskit \\
TFQ~\cite{broughton_tfq_2020}                     & TensorFlow        & PSR                    & SV (Cirq)             & Cirq-targeted QPUs \\
Qiskit ML~\cite{javadi_qiskit_2024}               & PyTorch           & PSR                    & SV, stabilizer (Aer)  & IBM \\
cuQuantum~\cite{bayraktar_cuquantum_2023}         & (backend library) & ---                    & SV, TN (GPU-only)     & used as a backend \\
TQml Simulator~\cite{kuzmin_tqml_2025}            & PyTorch, JAX      & backprop               & SV                    & internal/cloud QPUs \\
\bottomrule
\end{tabular}}
\vspace{2pt}
\footnotesize$^{\dagger}$ The C\texttt{++}-accelerated \texttt{lightning.*} backends use adjoint differentiation (no back-propagation) and loop over input batches internally; the slower \texttt{default.qubit} backend supports back-propagation and native batching.
\end{table*}

Beyond the choice of library, the underlying \emph{simulation method} (state-vector on CPU, state-vector on GPU, or tensor-network contraction; see the survey~\cite{cicero_simulation_2026} discussed above for the broader method landscape) sets how far the quantum layer can be pushed in qubits and depth.

\subsection{Quantum hardware platforms}
\label{sec:5_quantum}

\begin{figure}[htbp!]
    \centering
    \includegraphics[width=\linewidth]{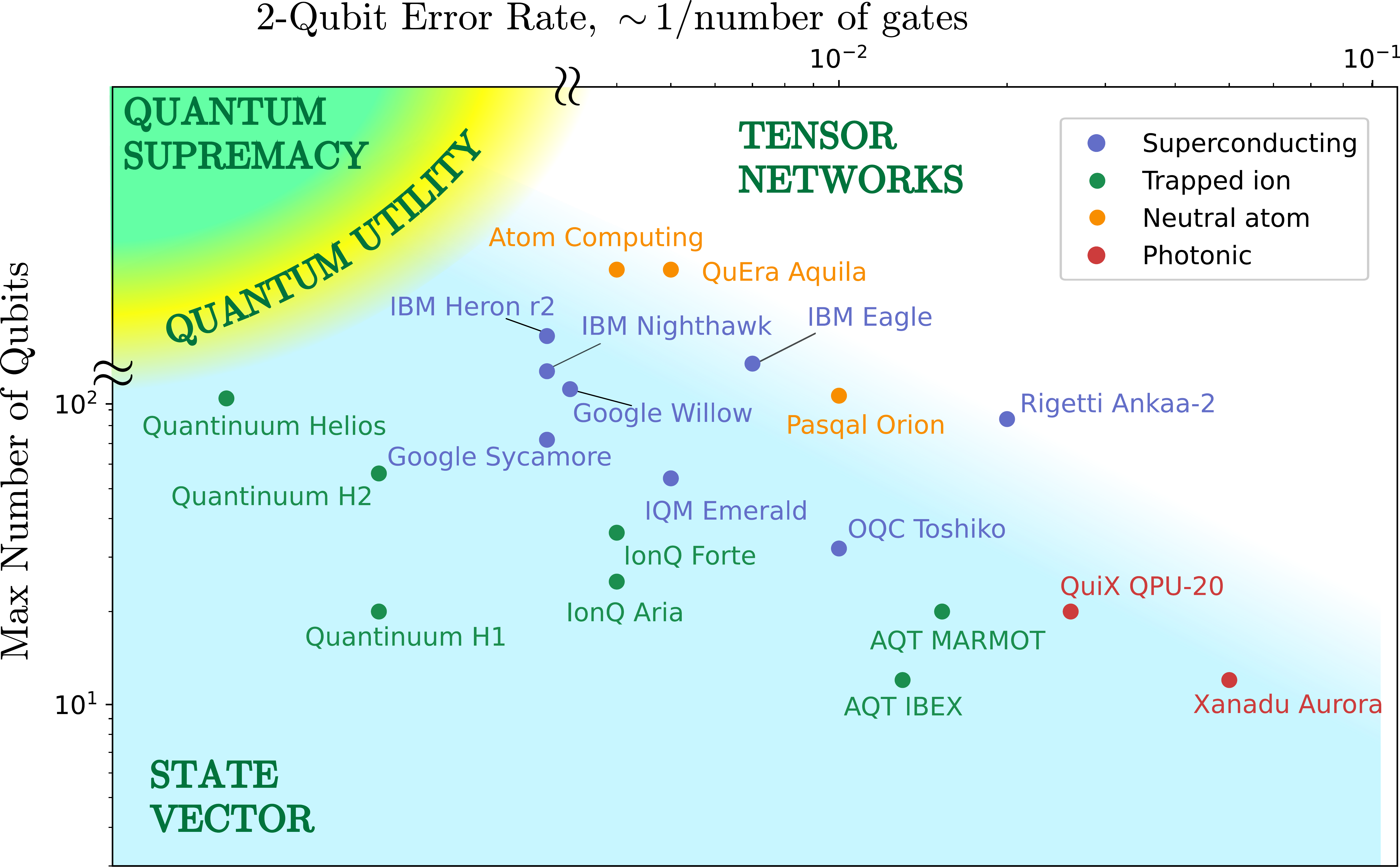}
    \caption{Placement of representative \glspl{QPU} in the plane of maximum qubit count versus two-qubit gate error (proportional to the inverse of achievable circuit depth). Shaded regions indicate where quantum supremacy, \emph{quantum utility}, tensor-network, and exact state-vector simulation regimes respectively apply.}
    \label{fig:hardware_plot_hqnn}
\end{figure}

A majority of gate-model quantum processor can be characterized by the number of controllable qubits, the fidelity of two-qubit operations, the qubit coherence time (which bounds how many gates can be applied before the state decoheres), the gate repetition rate (which together with fidelity fixes how deep a circuit can be trained in a given wall-clock budget), and the qubit connectivity graph. Beyond these performance figures, a platform is equally defined by its engineering context: the complexity and operational cost of keeping it running (cryogenic cooling, laser stabilization, ultra-high vacuum), the energy consumed per shot, the physical fragility of the system to vibration and electromagnetic noise, and its projected scaling path toward fault-tolerant sizes. These engineering factors frequently determine which platforms an organization can realistically deploy, independent of raw performance. For reference, \autoref{fig:hardware_plot_hqnn} maps the leading systems in the qubit-count/gate-error plane and indicates the regimes in which exact state-vector simulation, tensor-network surrogation, or hardware-only operation are respectively advantageous, and \autoref{tab:hqnn_qpus} gives order-of-magnitude numbers for each of the four modalities that dominate current \gls{HQNN} deployments and are discussed below.

\textbf{Superconducting circuits}~\cite{kim2023evidence} (IBM, Google, Rigetti, IQM) share fabrication techniques with semiconductor industry processes, operate at millikelvin temperatures, and offer fast two-qubit gates on the order of tens of nanoseconds. Connectivity is typically restricted to a nearest-neighbor lattice, which adds SWAP overhead for algorithms with long-range interactions but is offset by high repetition rate. Current devices span $\sim\!10^{2}$--$10^{3}$ qubits with two-qubit error rates in the $10^{-3}$ range, pushing them out of the tensor-network-simulable region and toward the quantum-utility regime of \autoref{fig:hardware_plot_hqnn}.

\textbf{Trapped ions}~\cite{decross2024computational} (Quantinuum, IonQ, AQT) currently lead on fidelity and coherence, with two-qubit errors below $10^{-3}$, long coherence times, and effectively all-to-all connectivity via ion shuttling in QCCD-style architectures~\cite{kielpinski_architecture_2002,pino_qccd_2021} or shared phonon buses. Their characteristic weakness is gate clock: two-qubit operations (typically M{\o}lmer--S{\o}rensen gates~\cite{sorensen_molmer_2000}) are orders of magnitude slower than their superconducting counterparts, which makes training loops that rely on many parameter-shift evaluations expensive in wall-clock time despite the high per-shot fidelity. Current devices approach the quantum-utility regime of \autoref{fig:hardware_plot_hqnn} from the state-vector-simulable side: qubit counts remain modest (tens to $\sim\!100$), but low error rates and high connectivity mean that circuits within reach are often still classically tractable.

\textbf{Neutral atoms}~\cite{balewski2024engineering, bluvstein2024logical} (QuEra, Pasqal, Atom Computing, planqc) use optically trapped atoms with Rydberg-mediated interactions. Recent experiments have demonstrated reconfigurable connectivity via atom shuttling~\cite{bluvstein2024logical} alongside early logical-qubit operations and fault-tolerant error-correction primitives~\cite{bluvstein2024logical}, placing this platform at the connectivity and error-correction frontiers simultaneously. Optical tweezer arrays are straightforward to scale to thousands of sites, but tweezer reconfiguration and atom shuttling add non-negligible latency between gate layers, so the effective clock of a deep circuit is set by how often the geometry needs to change. Gate times lie between those of ion traps and superconductors; qubit counts scale naturally with the size of the trap array. The platform is also a natural match for \emph{analog quantum reservoir computing}: the native many-body Rydberg dynamics provide a high-dimensional nonlinear feature map that has been exploited both theoretically~\cite{bravo_quantum_2022} and in large-scale experiments on QuEra Aquila~\cite{kornjavca2024large}.

\textbf{Photonic processors}~\cite{bourassa2021blueprint, larsen_integrated_2025, alexander_manufacturable_2025, taballione_20-mode_2023, aghaee_rad_scaling_2025, maring_versatile_2024} (Xanadu, PsiQuantum, ORCA, QuiX, Quandela, AegiQ) encode information in photonic modes and operate most components at room temperature. They are a natural match for continuous-variable and linear-optical encodings and have been used for end-to-end \gls{HQNN} demonstrations~\cite{austin_hybrid_2025, yin_experimental_2025,paparelle_experimental_2026, hoch_variational_2025, sedrakyan_photonic_2024, monbroussou_photonic_2025}. Fiber-optic interconnects enable native networking between chips and rack-scale modular scaling, as recently demonstrated by an 84-squeezer / 36-detector system networked over optical fiber~\cite{aghaee_rad_scaling_2025}. Modulation and switching operate on nanosecond timescales and the passive nature of waveguides, beam splitters, and phase shifters gives the platform a favorable energy-per-operation profile compared with cryogenic architectures. The room-temperature picture is only partial, however: high-fidelity photon-number-resolving detectors are typically superconducting nanowires that themselves require sub-kelvin cryogenics, which contributes substantially to the system-level power budget. A structural advantage for \gls{HQNN} workloads is that linear-optical circuits acting on a fixed number of photons implement unitaries on a particle-number-preserving subspace by construction, which is exactly the $k$-sector structure exploited by subspace-preserving \gls{HQNN} ansatzes (see~\autoref{subsec:Subspace_Preserving}). Present limitations remain photon loss and the difficulty of efficient deterministic single-photon generation.

\begin{table*}[htbp!]
\centering
\caption{Representative current-generation values for the four leading \gls{QPU} modalities used in \gls{HQNN} research: Superconducting~\cite{kim2023evidence}, Trapped ion~\cite{decross2024computational}, Neutral atom~\cite{balewski2024engineering, bluvstein2024logical}, and Photonic~\cite{bourassa2021blueprint}. Values evolve quickly; they are given as order-of-magnitude anchors rather than definitive specifications, and vendor-specific roadmaps should be consulted for up-to-date numbers.}
\label{tab:hqnn_qpus}
{\rmfamily
\begin{tabular}{@{}llllll@{}}
\toprule
Platform & Qubit count & 2-qubit error & Coherence time & 2-qubit gate time & Connectivity \\
\midrule
Superconducting                             & $10^{2}$--$10^{3}$       & $\sim\!10^{-3}$       & $\sim\!10^{2}$ $\mu$s   & 10--100 ns            & nearest-neighbor lattice \\
Trapped ion                       & $10$--$10^{2}$           & $\lesssim\!10^{-3}$   & $\gtrsim\!1$ s          & $\sim\!10^{2}$ $\mu$s  & all-to-all \\
Neutral atom  & $10^{2}$--$10^{3}$       & $10^{-3}$--$10^{-2}$  & $\sim\!1$ s             & $\sim\!1$ $\mu$s       & reconfigurable \\
Photonic                              & modes: $10$--$10^{2}$    & loss-dominated        & loss-limited            & sub-ns                 & native networking \\
\bottomrule
\end{tabular}}
\end{table*}

\subsection{Challenges for QML on FTQC}\label{subsec:FTQC_challenges}

As hardware moves from \gls{NISQ} prototypes toward \gls{FTQC}, the resource bottleneck for \gls{HQNN} changes character. Instead of fighting stochastic decoherence, one pays a deterministic cost to synthesize continuously parametrized operations on logical qubits. Variational layers describe each transformation as a sequence of single-qubit rotations whose phases are updated every training step, but logical-qubit hardware exposes only a discrete fault-tolerant gate set, most commonly the Clifford+$T$ set~\cite{Nielsen2011Quantum}. The Clifford group (generated by $H$, $S$, and $CNOT$) is cheap to implement fault-tolerantly but is not universal and can be simulated efficiently classically; adjoining the non-Clifford $T$ gate yields a discrete, universal set, at the cost that $T$ becomes the expensive resource (as detailed below). Every continuous rotation must therefore be approximated by a discrete sequence of these gates. The Solovay--Kitaev theorem~\cite{dawson_solovay-kitaev_2006} guarantees an approximation within tolerance $\varepsilon$ using $O(\log^c(1/\varepsilon))$ gates, and modern optimal $T$-count synthesis~\cite{ross_optimal_2016, kliuchnikov_synthesis_2013} reaches $\varepsilon \sim 10^{-6}$ with a few tens of $T$-gates per rotation. A nominally depth-one parametrized layer can therefore expand into hundreds of $T$-gates. Training amplifies the issue: parameter-shift gradient evaluations (\autoref{subsec:trainability_QNN}) require re-running the circuit at shifted parameter values, so every optimization step also triggers a re-synthesis of each modified rotation. Classical compilation latency, which was negligible in the \gls{NISQ} regime, becomes a non-trivial part of the training loop~\cite{ross_optimal_2016}.

Fault tolerance is achieved by encoding each logical qubit into many physical qubits using a quantum error-correcting code, so that errors can be detected and corrected faster than they accumulate. The surface code and its variants~\cite{fowler_surface_2012} are the leading candidates for near-term hardware, owing to its high error threshold and reliance only on nearest-neighbor interactions; its code distance $d$ sets how many physical qubits encode one logical qubit and how strongly errors are suppressed. The dominant resource cost of this discretization is the $T$-gate budget. In the surface code, as in most topological codes, Clifford operations are transversal (applied qubit-by-qubit across code blocks, so a single physical fault cannot spread into an uncorrectable error within a block) but the $T$-gate is not; it is implemented by injecting a high-fidelity so-called magic state prepared through an iterative distillation protocol~\cite{bravyi_universal_2005, fowler_surface_2012}. Each distilled magic state consumes many noisy ancillas and several rounds of stabilizer measurement, and realistic resource estimates~\cite{litinski_game_2019, beverland_resource_2022} are accordingly dominated by the throughput of dedicated magic-state factories that occupy a substantial fraction of the device. The qubit budget compounds the problem. At code distance $d$ each logical qubit costs $O(d^2)$ physical qubits, and $d \sim 15$--$25$ is typically needed to keep logical errors below the total $T$-count of a trained \gls{VQC}. A modest \gls{HQNN} with around $50$ logical qubits and a few hundred parametrized rotations per layer therefore translates into $10^{4}$--$10^{5}$ physical qubits, plus several factories running in parallel, a regime far removed from the gate-count discussion of \autoref{sec:5_quantum}. Several mitigations specific to \gls{HQNN} workloads are currently being explored. These include compiling whole ansatz layers rather than individual rotations~\cite{Vandaele2024Hadamard, vandeWetering2025Optimal}, sharing $T$-states across structurally identical parameter-shift evaluations~\cite{Sayginel2023FTVQA}, replacing finely discretized rotations by fixed discrete-parameter ansatzes~\cite{Ravi2023CAFQA}, and and substituting ansatzes built predominantly from cheap Clifford gates, with only a bounded amount of non-Clifford ($T$) content, for instance the subspace-preserving circuits of \autoref{subsec:Subspace_Preserving}, so that the dominant $T$-cost is capped by design. None of these design choices have a direct \gls{NISQ}-era counterpart.

\subsubsection{Hardware-modality dependence}
\textbf{Superconducting} processors are the main target for surface-code architectures. Their fast gate clock partially absorbs the inflated logical-circuit depth, but the $O(d^2)$ overhead is felt directly, since cryogenic wiring and control electronics already constrain physical-qubit growth, and routing on a nearest-neighbor lattice forces additional lattice-surgery operations between logical patches~\cite{litinski_game_2019}. \textbf{Trapped-ion} systems pay a different price. Their per-shot fidelities are high enough that smaller code distances suffice, and reconfigurable QCCD shuttling enables transversal logical gates between code blocks~\cite{decross2024computational}. The slow physical clock, however, means that $T$-heavy parametrized layers translate into very long wall-clock training steps. \textbf{Neutral-atom} arrays exploit reconfigurability and Rydberg-mediated multi-qubit gates to implement transversal logical operations on small codes, and recent experiments have demonstrated magic-state distillation and other fault-tolerant primitives in this geometry~\cite{bluvstein2024logical}; the binding constraint here is the latency of atom shuttling between data and distillation regions. 
\textbf{Photonic} platforms come at fault tolerance from a different angle. Rather than synthesizing magic states on the fly during the computation, several leading photonic architectures~\cite{bourassa2021blueprint, bartolucci_fusion_based_2023} build the required non-Clifford resources directly into how the photonic states are prepared beforehand. The dominant cost therefore shifts away from $T$-state throughput: the binding constraints become photon loss and the probabilistic nature of the entangling measurements on which these architectures rely. Whichever platform an \gls{HQNN} designer targets, the binding budget is then less the noisy gate count and more the rate and latency at which the fault-tolerant architecture can supply $T$-states per training step.

\section{Empirical Studies of HQNNs}
\label{sec:empirical}

This section presents practical applications of HQNN across multiple domains. We start by offering an overview of HQNN proposals for several use cases in \autoref{subsec:Domain_App_Results}: time series forecasting, image classification, computational fluid dynamics, and planning and logistics. Then, we discuss recent recent from benchmarking studies on heuristic proposals of HQNNs. Finally, we discussed alternative figure of merits for HQNNs, and how other benefits should be considered along with model running time and accuracy.

\subsection{Domain Applications and Results}\label{subsec:Domain_App_Results}

This subsection surveys HQNN applications across different domains, and highlight how different architectures can be used for different use cases.

\paragraph{Time Series Forecasting:} Learning from time series underpins demand and energy-generation forecasting, equipment condition monitoring, financial prediction, and anomaly detection, settings in which observations are typically noisy and non-stationary while exhibiting both short and long range dependencies. Classical approaches \cite{hochreiter1997lstm, eren_comprehensive_2024, serradilla_deep_2022, sezer_financial_2020, zamanzadeh_darban_deep_2024} address these tasks by modeling sequential dependencies via recurrence or attention mechanisms, but can strugle with complex or limited training data. 

\noindent\textbf{Variational quantum circuits} have been studied in machine learning architecture \cite{chen2022qlstm, sagingalieva2025photovoltaic, siemaszko2023cvqrnn, khan2024qlstm_solar, schetakis2025traffic, kurkin2025forecasting, lee2025predictive,su2025blsqlstm, thakkar2024financial}, and demonstrated a reduction in the required number of parameters, and faster convergence for small problem instances. However, benchmarks on larger \cite{fellner_quantum_2026, bowles2024better} indicate these gains do not yet carry over at scale for generic ansatze, underscoring the value of the structured architectures of \autoref{sec:architectures}.

instances\cite{fellner_quantum_2026} benchmarked the use of \gls{VQC} for time series and concluded that large instances tend to be in defavor of quantum as described in \cite{bowles2024better}.

\noindent\textbf{Reservoir computing} architectures (see \autoref{subsubsec:Quantum_Reservoir_Computing}) are promising candidates for time-series forecasting \cite{suzuki_natural_2022, mujal_time-series_2023}, and have been compared with RNNs in \cite{vlachas_backpropagation_2020}.

\noindent\textbf{\glspl{RNN} extension with continuous variable ressources} have been proposed \cite{siemaszko2023cvqrnn} to decrease the running time and energy consumption of these classical algorithms.

\paragraph{Image classification}

Image classification has long served as the canonical benchmark for validating novel ML architectures. Classical approaches rely on convolutional neural networks and, more recently, vision transformers \cite{lecun1998mnist, vaswani_attention_2017} to capture these hierarchies.

\noindent\textbf{Variational quantum circuits} have been explored as drop-in replacements or complements to classical layers \cite{senokosov2024quantum, shi_quantum_2024, elhag_quantum_2025, liu2025lcqhnn, henderson2020quanv, he2016resnet, liu2021qccnn, long2025qcqcnn, bokhan2022qcnn_multiclass, fan2024qccnn, vu2024quanv, kawase2024dqnn, houssein2022covid, acar2022breast}, typically reporting competitive accuracy on MNIST and FashionMNIST-scale tasks with substantially fewer trainable parameters. However, \cite{bowles2024better} benchmarked twelve QML models over 160 datasets and reported that classical baselines match or exceed them, with entanglement often being neutral or detrimental at these scales, indicating that generic circuits alone do not confer an advantage and motivating the symmetry-aware architectures discussed next.

\noindent\textbf{Architectures mimicking classical ML}, especially taking into account symmetries in the computation have been used, including algorithms with convolutionnal layers \cite{kerenidis_quantum_2020, cong2019quantum, monbroussou_subspace_2025, monbroussou_photonic_2025}, attention mechanisms \cite{cherrat_quantum_2024}, and orthogonal layers \cite{landman_quantum_2022} based on subspace preserving circuits. Based on a rich body of work between of quantum theory and symmetry, \textbf{Geometric Quantum Machine Learning} (GQML) \cite{perrier_quantum_2020,  wiersema_geometric_2025,tuysuz_symmetry_2024, ragone_representation_2023} have been consider to create model with inductive biases encoding the symmetries of a problem, leading to performance improvement.

\paragraph{Computational Fluid Dynamics:} \Gls{CFD} presents substantial computational challenges due to the need to solve the Navier--Stokes equations across complex geometries and wide ranges of scales. Physics-informed approaches incorporating governing equations into the loss function have emerged as a promising paradigm, and quantum enhancements offer potential improvements in parameter efficiency~\cite{gaitan2020navier, amaral2026qmlcfdreview,
ye2024hybridcfd}.

\noindent\textbf{Hybrid quantum classical PINNs} have been considered in several studies \cite{xiao2024piqnn, trahan2024qpinn, siegl2024piqc,farea2025qcpinn, song2025nisqcfd,berger2025teqpinn, marcandelli2025phqfno}  for canonical \gls{PDE} benchmarks, highlighting parameters reduction with similar performances and similar performances for small problem instances. More application-oriented architectures have been proposed for laminar flow in complex mixer geometries \cite{sedykh2024hybrid}.

\paragraph{Planning and Logistics:} Planning and logistics optimization involves combinatorial problems where the solution space grows exponentially with problem size. HQNNs have been applied to scheduling, routing, and resource allocation.

\noindent\textbf{Quantum attention-head layers} have been proposed to replace their classical counterparts in reinforcement learning agents with variational circuits \cite{sanches2022short, correll2023qnn}.

\noindent\textbf{Variational quantum circuits} have been used in reinforcement learning pipeline to project and train classical data in exponentially large Hilbert space \cite{haboury2023supervised, rainjonneau2023quantum, jahin2023qamplifynet, princy2025hqerl}.

\begin{table*}[t!]
\centering
\caption{Consolidated empirical results for HQNNs across application domains.
    "Param.\ Red." indicates parameter reduction relative to classical
    baselines achieving comparable accuracy. "Sim." denotes classical
    simulator; "QPU" denotes execution on quantum hardware. Dashes indicate
    metrics not reported.}
\label{tab:hqnn_results_expanded}
\scriptsize
\setlength{\tabcolsep}{2pt}
\renewcommand{\arraystretch}{1.0}
\begin{tabular}{|c|c|c|c|c|c|}
    \hline
    \multicolumn{1}{|@{}c@{}|}{\cellcolor{lightgray}\textbf{Task/Dataset}} & \multicolumn{1}{@{}c@{}|}{\cellcolor{lightgray}\textbf{Architecture}} & \multicolumn{1}{@{}c@{}|}{\cellcolor{lightgray}\textbf{Qubits}} & \multicolumn{1}{@{}c@{}|}{\cellcolor{lightgray}\textbf{Param.\ Red.}} & \multicolumn{1}{@{}c@{}|}{\cellcolor{lightgray}\textbf{Performance vs Classical}} & \multicolumn{1}{@{}c@{}|}{\cellcolor{lightgray}\textbf{Platform}} \\
    \hline
    \multicolumn{6}{|@{}c@{}|}{\cellcolor{blue!10}\textbf{Time-Series Forecasting}} \\
    \hline
    PV Power Forecasting~\protect\cite{sagingalieva2025photovoltaic} & Hybrid QLSTM with QDI & 4 & 61.2\% & $-$40\% MAE/MSE & Sim. \\
    Solar Forecasting~\protect\cite{khan2024qlstm_solar} & QLSTM & 2--8 & --- & Lower MAE; earlier convergence & Sim. \\
    Stock Index Prediction~\protect\cite{su2025blsqlstm} & BLS-QLSTM & --- & --- & Best on 6/6 metrics vs LSTM & Sim. \\
    Traffic Flow~\protect\cite{schetakis2025traffic} & QNN with re-uploading & 2--14 & --- & Outperforms classical at $\geq6$ qubits & Sim. \\
    Steam Mass Flow~\protect\cite{kurkin2025forecasting} & Parallel hybrid network & $2\times5$ & --- & Test MSE reduced by $>5.7\times$ & Sim. \\
    Blast Furnace Control~\protect\cite{lee2025predictive} & Hybrid QLSTM with QDI & 6 & --- & 25\% lower 1-h RMSE & Sim. \\
    Credit Risk~\protect\cite{thakkar2024financial} & OrthoResNN & 8 & 79.7\% & 54.29\% vs 54.20\% Gini (sim.) & Sim, IBM \\
    QLSTM Benchmark~\protect\cite{chen2022qlstm} & QLSTM & 4 & 12.0\% & Faster convergence; comparable & Sim. \\
    CV-QRNN~\protect\cite{siemaszko2023cvqrnn} & CV-Quantum RNN & 3 modes & --- & Fewer epochs to converge & Sim. \\
    \hline
    \multicolumn{6}{|@{}c@{}|}{\cellcolor{green!10}\textbf{Image Classification}} \\
    \hline
    MNIST (full)~\protect\cite{senokosov2024quantum} & HQNN-Parallel (PQN) & 4$\times$5 & 87.9\% & 99.21\% vs 98.71\% (+0.5\%) & Sim. \\
    Medical MNIST~\protect\cite{senokosov2024quantum} & HQNN-Parallel (PQN) & 4$\times$5 & 0.07\% & 99.97\% vs 99.96\% & Sim. \\
    CIFAR-10~\protect\cite{senokosov2024quantum} & HQNN-Parallel (PQN) & 4$\times$5 & 0.15\% & 82.78\% vs 82.64\% & Sim. \\
    Car Classification~\protect\cite{sagingalieva2023hyperparameter} & Hybrid Q-ResNet & 13 & 30.6\% & Grid search: 98.9\% vs 92.0\% & Sim. \\
    Ants vs Bees~\protect\cite{mari2020transfer} & Dressed Quantum Circuit & 4 & --- & 96.7\% sim.; 95\%/80\% QPU & Sim, IBM/Rigetti \\
    COVID-19 X-ray~\protect\cite{houssein2022covid} & Quanvolutional CNN & 4 & 76.2\% & 98.6\%; 99\% recall & Sim. \\
    Breast Cancer~\protect\cite{acar2022breast} & Q-Transfer Learning & 4 & --- & Competitive diagnostic & Sim, IBM \\
    MNIST(Quanv.)~\protect\cite{henderson2020quanv} & Random Quanvolutional & 9 & --- & Higher than classical CNN & Sim. \\
    Remote Sensing~\protect\cite{fan2024qccnn} & QC-CNN (Amplitude Enc.) & 10 & 7.5--11.5\% & Higher on 5/5 benchmarks & Sim. \\
    MNIST Multiclass~\protect\cite{bokhan2022qcnn_multiclass} & QCNN & 8+4 & 20.7\% & Successful 4-class; below classical CCNN & Sim. \\
    Hybrid QC-CNN~\protect\cite{liu2021qccnn} & Hybrid QCNN & 4 & --- & Higher learning accuracy than classical CNN & Sim. \\
    QCQ-CNN~\protect\cite{long2025qcqcnn} & Quantum convolutional & 4, 2 & --- & Improved vs single-stage & Sim. \\
    Quanv.\ Topology Study~\protect\cite{vu2024quanv} & Optimized Quanvolutional & 2--5 & --- & Improved vs classical CNN & Sim. \\
    Partitioned DQNN~\protect\cite{kawase2024dqnn} & Distributed QNNs & 7--8 per QNN & --- & Higher accuracy than single QNN & Sim. \\
    \hline
    \multicolumn{6}{|@{}c@{}|}{\cellcolor{orange!10}\textbf{Physics-Informed CFD}} \\
    \hline
    1D Burgers' Eq.~\protect\cite{siegl2024piqc} & Physics-Informed QC & 10 & --- & Comparable & Sim. \\
    5 PDE Benchmarks~\protect\cite{farea2025qcpinn} & QCPINN & 5 & 71.1--89.6\% & $-64\%$ to $+22\%$ $L_2$ error & Sim. \\
    Laminar Mixer Flow~\protect\cite{sedykh2024hybrid} & HQPINN with QDI & 3 & --- & 21\% lower loss & Sim. \\
    Navier--Stokes~\protect\cite{berger2025teqpinn} & TE-QPINN & 4 & --- & Qualitative agreement with OpenFOAM & Sim. \\
    Burgers/NS Flows~\protect\cite{marcandelli2025phqfno} & PH-QFNO & 12 & --- & Comparable; improved noise robustness & Sim. \\
    NISQ CFD Demo~\protect\cite{song2025nisqcfd} & VQE & 4 & --- & Readout 95--99\%; QPU CFD did not converge & Sim, IBM SC \\
    PI-QNN~\protect\cite{xiao2024piqnn} & Physics-Informed QNN & 3--10 & --- & 2--3 orders lower for trig.; worse on Burgers & Sim. \\
    Hybrid PINN~\protect\cite{trahan2024qpinn} & Hybrid QPINN & 5 & 57.7\% & Lower RMSE; fewer params & Sim. \\
    \hline
    \multicolumn{6}{|@{}c@{}|}{\cellcolor{purple!10}\textbf{Planning and Logistics}} \\
    \hline
    Vehicle Routing~\protect\cite{sanches2022short} & Quantum Attention & 4, 6 & --- & Competitive solution quality on simulation & Sim, Rigetti \\
    Automotive Supply Chain~\protect\cite{correll2023qnn} & Quantum multi-head attn. & 16, 8 & --- & 49\% demand satisfied on QPU & Sim, IonQ \\
    Satellite Mission~\protect\cite{rainjonneau2023quantum} & Hybrid Q-RL agent & 4 & --- & 98.5\% vs 75.8\%  & Sim. \\
    Evacuation Routing~\protect\cite{haboury2023supervised} & HQNN Classifier & 7 & --- & +7\% vs classical & Sim, IonQ\\
    Backorder Prediction~\protect\cite{jahin2023qamplifynet} & QAmplifyNet & 2 & --- & Improved prediction & Sim. \\
    \hline
\end{tabular}

\end{table*}

\subsection{Benchmarking of HQNN approaches}
\label{subsubsec:performance_summary}

Many proposals show significant improvements on small problem instances that can be simulated (see \autoref{tab:hqnn_results_expanded}). However, several studies have been proposed to discuss the use of HQNNs at scale. 

In particular, \cite{bowles2024better} systematically test $12$ popular quantum ML models on $160$ binary classification datasets and find that out-of-the-box classical models generally outperform the quantum classifiers. Notably, removing entanglement from quantum models often yields equal or better performance, questioning whether ``quantumness'' is the key ingredient at the scales tested. An open-source benchmarking package built on PennyLane \cite{bergholm_pennylane_2020} was offered to compare quantum solution on larger problem instances. For time series prediction,  \cite{fellner_quantum_2026} has beenchmarked variational QML architectures against classical methods, highlighting the difficulty of finding optimal ansatze. Two consistent lessons emerge: unstructured variational layers deliver their measured gains mainly at small scale, while architectures with theoretical guarantees on trainability and classical hardness (\autoref{sec:architectures}) are the credible candidates for advantage at scale. These benchmarks have helped move the field toward such structured designs.

In addition, simulation techniques and surrogate models (see \autoref{subsec:Dequantization}) have been proposed and tested to challenge HQNN models in QML \cite{schreiber_classical_2023, bermejo_quantum_2026, sweke_potential_2025, landman_classically_2022, shin_dequantizing_2024} and quantum algorithm in general \cite{tang_quantum_2021, chia_sampling-based_2020, tang_dequantizing_2022, gharibian_dequantizing_2022, fontana_classical_2025}. Coupled in studies offering solution to avoid dequantization \cite{thabet_when_2026, gil2024relation, jerbi_quantum_2023, angrisani_simulating_2026} and to achieve speedups in learning \cite{Huang2022Science, huang2021power, zhao_entanglement-induced_2025, martinez_efficient_2025, gonzalez-garcia_pauli_2025}, the scientific community is constantly proposing and challenging new ideas for future applications.

        \subsection{Figure of merits for quantum layers}
\label{subsec:runtime}

A complete assessment of HQNN must consider not only accuracy but also computational cost. This subsection analyzes runtime scaling for quantum layers executed on classical simulators and quantum hardware.

        \subsubsection{Running time and complexity}

The computational complexity of HQNNs is intimately tied to the hardness of simulating quantum circuits classically. Several sampling-based quantum models are believed to be classically intractable. For example, Boson Sampling \cite{aaronson_computational_2011} and \gls{IQP} circuits \cite{bremner_classical_2010} are conjectured to be classically hard. More broadly, large random quantum circuits are generally expected to define problems outside the reach of polynomial-time classical algorithms \cite{bouland_complexity_2019}. From a generalization perspective, HQNNs may offer advantages that go beyond mere computational hardness of evaluation. 
Classical hardness of learning has been established for tasks involving cryptographic functions, whose output can be estimated efficiently by a quantum circuit (e.g. using the Shor algoritm \cite{Shor1997Polynomial}) but not by any classical algorithm. It has been shown \cite{jerbi_quantum_2023, thabet_when_2026} that quantum learners can identify such functions in settings where classical learners provably cannot.

        \subsubsection{Energy Consumption}\label{subsubsec:Energy_Consumption}

The energy footprint of training and inference is an increasingly important metric, particularly as classical deep learning models grow to billions of parameters with associated carbon costs~\cite{strubell2019energy, patterson2021carbon}. Comparative energy analysis between quantum and classical computing for ML remains in its early stages, but several important observations can be made.

Energetic advantage asks a different question than runtime advantage: for a given computational task and target quality, can a quantum system deliver the result with lower total energy cost than the best classical alternative? This framing has been explicitly promoted by the Quantum Energy Initiative, which argues that energy should be treated as a core resource alongside qubits, gates, error rates, and wall-clock time when assessing scalable quantum technologies~\cite{auffeves2022quantum}. In practice, meaningful comparisons require full-stack accounting, since cooling and auxiliary subsystems can dominate total energy use in quantum computing architectures~\cite{fellous-asiani_optimizing_2023, mccollum_energy_2026}. Studies \cite{fellous-asiani_optimizing_2023, Woitzik2024} have proposed energetic footprint models for quantum algorithm building blocks such as gates or measurements (see \autoref{sec:foundations}), while more recent work \cite{carrasco-codina_energy_2026, soret_quantum_2026} have offered an hardware-aware study, by estimating the consumption of quantum computers using power consumption values from existing components and fully working setups.

From a theoretical standpoint, a rigorous framework for the energy-consumption advantage of quantum computation has recently been established, proving that quantum computers achieve an exponential energy advantage over classical ones for Simon's problem and proposing explicit experimental criteria for demonstrating this advantage~\cite{meier_energy-consumption_2025}. 

Preliminary estimates~\cite{jaschke2023quantum, fellous-asiani_optimizing_2023} suggest that current QPUs do not yet offer energy advantages over GPUs for typical ML workloads, primarily due to cryogenic overhead. This limitation is directly relevant to HQNN pipelines that invoke near-term superconducting or trapped-ion backends, where the cooling infrastructure dominates the total energy budget. However, fault-tolerant QPUs with higher gate counts, lower error rates, and improved cryogenic efficiency could shift this balance, particularly for problems where quantum algorithms provide superpolynomial speedups~\cite{meier_energy-consumption_2025}. Furthermore, HQNNs have been shown empirically to achieve comparable or better performance than classical neural networks with significantly fewer trainable parameters~\cite{bischof_hybrid_2025}, which translates into reduced gradient computation costs during training — a potentially meaningful energy saving even on classical simulation backends.

Photonic and neutral-atom architectures operating partially at or near room temperature represent the most promising near-term pathways toward energy-efficient QML, precisely because they avoid the dominant cryogenic cost that burdens superconducting platforms~\cite{soret_quantum_2026}. Quantum algorithms for generative \gls{AI} tasks, including training and inference of large models, are also beginning to be systematically evaluated for their potential to reduce the energy demands of the classical \gls{AI} pipeline~\cite{flother_accelerating_2025}

\section{Discussion, Challenges, and Perspectives}
    
    \label{sec:discussion}

The preceding sections have looked at \glspl{HQNN} from four complementary angles: their theoretical foundations (\autoref{sec:foundations}), the design space of architectures that have been proposed for them (\autoref{sec:architectures}), the software and hardware stack on which they are executed (\autoref{sec:implementation}), and the empirical results that the field has accumulated across application domains (\autoref{sec:empirical}). The picture that emerges is considerably sharper than it was previously: the field now has a much clearer view of where quantum layers help and why.  \glspl{HQNN} are a versatile design idiom and a serious testbed for thinking about the role of quantum resources inside machine-learning models. Theory has now established provable separations on specific constructed tasks, including some defined over large-scale classical data, yet the case for a robust, scale-invariant advantage across typical learning workloads is not yet settled. That these are proven separations rather than empirical observations is significant in its own right: they establish that the advantage is a real property of the models where it holds, not an artefact of insufficiently optimised classical methods. A recurring theme across these chapters is that the value of an \gls{HQNN} cannot be read off the size of its Hilbert space. The expressivity of a variational circuit can be studied through its accessible Fourier spectrum (\autoref{subsec:VQC_as_Fourier_Models}), its trainability is constrained by barren plateaus and concentration phenomena (\autoref{subsubsec:BP_QNN}), and a wide class of structurally restricted circuits admits efficient classical surrogation (\autoref{subsec:Dequantization}). These three observations interact in a way that is now understood with some precision: ansatzes that are provably trainable seem to admit classical simulation methods. Read constructively, this is less a barrier than a map: it marks the generic-circuit regime as unpromising and directs effort toward exploitable problem structure, as presented in \autoref{sec:architectures}. The architectural responses that have proved most informative therefore exploit problem structure rather than raw expressivity. Subspace-preserving layers (\autoref{subsec:Subspace_Preserving}), geometric and symmetry-aware ansatzes, and reservoir-style models that fix the quantum dynamics and train only a classical read-out (\autoref{subsubsec:Quantum_Reservoir_Computing}) all reduce the effective hypothesis class in ways that tend to improve trainability. The empirical record reviewed in \autoref{sec:empirical} is consistent with this theoretical picture: \glspl{HQNN} have demonstrated parameter reductions and competitive accuracy on small-scale instances of time-series forecasting, image classification, computational fluid dynamics, and planning problems, but systematic benchmarks against well-tuned classical baselines show that generic architectures do not yet carry these gains to scale. Rather than undermining the approach, this record identifies the regimes where an \gls{HQNN} is the informed choice: data-scarce tasks with exploitable symmetries, problems whose structure matches the inductive bias of a particular quantum layer, and settings where parameter efficiency or model size is the binding constraint.

\subsection{Open challenges}

Several open challenges run through the review and are worth flagging collectively. The first is trainability at scale. Although barren plateaus are now characterized analytically for a wide range of ansatz families, the same theoretical tools tend to imply classical simulability of the trainable regime, leaving open whether a strictly non-classically-simulable \gls{HQNN} can be trained reliably as the qubit count grows. Initialization strategies, warm starts, and data-dependent parameter distributions offer partial remedies, but no general prescription has emerged.

The second challenge is benchmarking. The community has converged on a small number of standard datasets that do not stress the regime in which quantum models are most likely to differ from classical ones. Benchmark suites that incorporate symmetry, data scarcity, ill-conditioning, and structured noise, together with transparent reporting of classical baselines and the resources consumed, would change the empirical conversation. The figure-of-merits discussion in \autoref{sec:empirical} argues for a multi-dimensional comparison rather than a single accuracy headline; adopting that practice as a community-wide convention would let the field separate compounding progress from results that depend on a particular favorable choice of benchmark.

The third challenge is the hardware path. Most \gls{HQNN} results today are obtained by simulating the quantum layer classically, with on-hardware execution reserved for final validation. As devices grow, the transition to fault-tolerant operation (\autoref{sec:5_quantum}) replaces stochastic noise with a deterministic resource cost of synthesizing continuously parametrized rotations on logical qubits. The design problem then shifts from controlling decoherence to managing $T$-gate budgets, magic-state throughput, and physical-to-logical qubit overheads. Ansatzes amenable to fault-tolerant compilation diverge substantially from those most extensively explored in the current \gls{NISQ} literature, and the \gls{QML} community has only recently begun to engage seriously with the use of early FTQC platform.

A fourth challenge is integration with the broader machine-learning stack. \glspl{HQNN} have, to date, been built largely by quantum specialists and have inherited the conventions of variational algorithms. Their adoption by a wider community will require differentiable software stacks (\autoref{sec:5_classical}) that compose cleanly with existing \gls{ML} component, and a shared vocabulary for explaining where and why a quantum layer is expected to help.

\subsection{Perspectives}

Looking forward, three directions appear to us most promising. The first is problem-aware architecture design. The strongest empirical results in \autoref{sec:empirical} converge on a common feature: the quantum layer encodes a known symmetry or conservation law of the underlying problem rather than acting as a generic function approximator. \gls{GQML}, subspace-preserving circuits, and reservoir architectures whose dynamics match the time-series structure of the task are all examples of this pattern. The second direction is hardware-aware design of quantum algorithms. The qualitative differences between the four \gls{QPU} modalities of \autoref{sec:5_quantum} are large enough that the same algorithmic idea can be ill-suited on one platform and a natural fit on another. Photonic processors paired with subspace-preserving \glspl{HQNN}, neutral-atom arrays paired with reservoir computing in the analog regime, trapped-ion devices paired with high-fidelity small-circuit training loops, and superconducting processors paired with depth-shallow ansatzes with structured data re-uploading are pairings that already show this. As fault-tolerant primitives mature on each modality, we expect the \gls{HQNN} design space to bifurcate by platform rather than converge.

The third direction concerns the figures of merit by which \glspl{HQNN} are judged. As classical simulation becomes more capable, the bar that a quantum layer has to clear in order to justify itself rises. Energy per gradient step, parameter count at fixed accuracy, sample complexity, and the latency cost of on-hardware execution are all metrics that the field will increasingly need to report alongside accuracy. Some of these favor quantum hardware in principle, since photonic and Rydberg processors operate at or near room temperature and avoid the cryogenic overhead that dominates the energy budget of superconducting platforms, but the gap between principle and practice depends on details that the current literature has not yet measured carefully.

What remains for the next phase of \gls{HQNN} research is a narrower empirical question than the broad "can quantum models outperform classical ones?" that has framed much of the literature so far. The pertinent question is which pairings of problem class, hardware modality, and scale make a quantum layer's structure worthwhile against a similarly resourced classical alternative. The body of theory and practice surveyed in this review already supplies much of what is needed to answer it, and points clearly to where the remaining work lies.

\bibliographystyle{apsrev4-2}
%\bibliography{references}
\bibliography{bibliography}

%\vskip3pt

\end{document}